\documentclass{aastex701}

\usepackage{amsmath}
\usepackage{amssymb,microtype,siunitx,booktabs}
\usepackage{graphicx}	
\usepackage{booktabs}
\usepackage{multirow}

\usepackage{algorithm}
\usepackage{algpseudocode}

\usepackage{subcaption} 
\usepackage{caption}     

\usepackage{todonotes}

\begin{document}

\title{Modeling the technical specifications for future multi-object spectroscopy instrumentation}

\author[orcid=0000-0002-0309-1599,sname='Ahmed']{Ummee T. Ahmed}
\affiliation{Australian Astronomical Optics, Macquarie University, 105 Delhi Rd, North Ryde, NSW 2113, Australia}
\email[show]{ummeetania.ahmed@mq.edu.au}

\author[orcid=0000-0002-6097-2747,gname=Andrew, sname='Hopkins']{Andrew M. Hopkins} 
\affiliation{School of Mathematical and Physical Sciences, 12 Wally’s Walk, Macquarie University, NSW 2109, Australia}
\email{andrew.hopkins@mq.edu.au}

\author[orcid=0000-0002-0742-379X,gname=Ellis, sname='Simon']{Simon C. Ellis} 
\affiliation{Australian Astronomical Optics, Macquarie University, 105 Delhi Rd, North Ryde, NSW 2113, Australia}
\email{simon.ellis@mq.edu.au}

\begin{abstract}

We investigate a proof-of-concept code designed to model the functionality of current generation astronomical fiber positioner instruments. This code accommodates an arbitrary number of fiber probes and astronomical targets. To establish the prerequisites for future fiber positioning technologies, we conduct a systematic analysis of both the observational and instrumentation parameter space (such as target density, fiber density, and exclusion and patrol radius).
The tool assesses the efficiency and completeness of such a system by modeling the allocation of fibers to targets over multiple visits or observations, enabling the quantification of performance based on the specific parameters. This tool holds significance in delineating the technical specifications for future multi-object spectroscopy instrumentation. This will inform the development of new technologies for the next generation of telescopes using novel instrumentation techniques or the extension of existing technologies.

\end{abstract}



\section{Introduction} 
\label{sec:intro}

Large scale observational surveys have attracted growing interest in astronomical research. These surveys have significantly advanced our understanding of the formation and evolution of the Universe \citep{2022arXiv220908049A}. Their remarkable success has inspired the development and construction of new multiplexed spectroscopic instruments. Currently, these surveys are mainly carried out at multi-object spectroscopy (MOS) facilities using large aperture telescopes \citep{2016SPIE.9908E..1SC}.
MOS has been pivotal in advancing the field of precision cosmology. Early surveys explored the large scale structure and evolution of the Universe, while current-generation surveys provide detailed maps across a wide range of redshifts. This technique has been essential for mapping the cosmic web and extracting increasingly precise cosmological information. It remains crucial for future studies aiming to map the evolving geometry of the Universe and structure throughout cosmic history. These efforts aim to achieve the most precise and comprehensive measurements of cosmological parameters \citep{2013MNRAS.428..447S,2016SPIE.9908E..1SC, 2017arXiv170101976E}.

MOS is a powerful technique that enables the acquisition of detailed spectra for numerous targets simultaneously. This is a notable improvement compared to single-object spectrographs. Technological advancements have significantly enhanced MOS capabilities over the past three decades. These improvements allow objects to be measured at greater depths and increase the number of targets observed simultaneously.
To achieve this, the two primary technologies \citep{massey2010astronomical} are multi-slit spectroscopy and multi-fiber spectroscopy. Multi-fiber systems in particular have been highly effective for studying galaxy evolution \citep[e.g.,][]{2011MNRAS.413..971D} and the large-scale structure of the Universe \citep[e.g.,][]{2016SPIE.9908E..1SC}, enabling the measurement of a wide range of cosmological parameters with increasing precision. The key features of MOS are multiplex and field of view (FoV). These features are essential for meeting the primary requirements of cosmological surveys, which are large sample sizes and significant survey volumes. They are also essential for galaxy evolution studies, enabling repeated observations to build up high sample completeness, and good signal-to-noise in the resulting spectra. Through the late 1990s into the 2000s these requirements often overlapped, but started to diverge in the late 2000s, as evidenced by the very different observational strategies adopted by surveys such as GAMA \citep[Galaxy And Mass Assembly, a galaxy evolution program][]{2011MNRAS.413..971D} and WiggleZ \citep[the WiggleZ Dark Energy Survey, a cosmology survey][]{2012PhRvD..86j3518P}, both conducted over a similar time period using the 2 degree Field (2dF) instrument on the Anglo-Australian Telescope (AAT).
Through the combination of wide-field sky coverage and high multiplex, MOS surveys continue to provide invaluable insights into the Universe \citep[e.g.,][]{2013MNRAS.428..447S}.

\begin{table*}[ht!]
    \centering
    \caption{Specifications of some of the existing fiber positioning systems. The table presents a comparative analysis of fiber positioner systems employed in multi-object spectroscopic instruments. This table shows differences in design approaches (e.g., tilting spine, dual-rotator, starbug, pick-and-place, and new design concept) along with their key instrumentation parameters and observational parameters, such as their FoV, fiber pitch, number of fibers, and the fiber density they can observe.}
    \label{tab:positioner_systems}
    \makebox[\textwidth]{%
    \resizebox{\textwidth}{!}{%
    \begin{tabular}{lccccccccc}
        \hline
        \textbf{Telescope} & \textbf{Positioner} & \textbf{FoV} & \textbf{Pitch} & \textbf{Pitch} & \textbf{Fiber Number} & \textbf{Fiber Density} & \textbf{References}\\
        & \textbf{Instruments} & (deg$^2$) &  (mm) & (arcmin) &  & (Fibers/deg$^2$) & \\
        \hline \hline
        & \multicolumn{5}{l} {\textbf{Tilting Spine Technology}} \\
        VISTA (4.1m) & 4MOST - AESOP Echidna & 4.2 & 9.542 & 2.68 & 2436 & 580 & 4MOST User Manual\footnote{https://www.4most.eu/cms/facility/overview/} \\
        Subaru (8m) & FMOS Echidna & 0.196 & 7.2 & 1.44 & 400 & 2040 & \citep{2010arXiv1006.3102K}\\
        MSE (11.25m) & Sphinx & 1.52 & 7.77 & 1.23 & 4332 & 2888 & \citep{2018SPIE10702E..1MS}\\
        \hline
        & \multicolumn{5}{l}{\textbf{Dual-rotator Technology}} \\
        Sloan (2.5m) & SDSS-V & 7.068 & 22.4 & 6.16 & 640 & 90.5 & {\citep{2020SPIE11447E..81P}}\\
        du Pont (2.5m)\footnote{Each instrument contains 500 robotic fiber positioners for optical and 298 for NIR spectroscopy.} & SDSS-V & 3.46 & 22.4 & 6.16 & 798 & 231.2 & {\citep{2020SPIE11447E..81P}}\\
        Mayall (4m) & DESI & {7.5} & 10.4 & 2.427 &  5000 & 707.35 & \citep{2024AJ....167...62D} \\
        LAMOST (4m) & LAMOST & 19.635 & 25.6 & 4.267 & 4000 & 203.72 & \citep{2012RAA....12.1197C}\\
        VLT (8.2m) & MOONS & 0.1389 & 25 & 0.72 & 1000 & 7199.42 & \citep{2022SPIE12189E..0VB}\\
        Subaru (8m) & PFS - Cobra & 1.33 &  8 & 1.439 & 2394 & 1804 & \citep{2014SPIE.9151E..1YF} \\
        \hline
        &  \multicolumn{5}{l}{\textbf{Starbug Technology}} \\
        UKST (1.2m) & TAIPAN & 28.27 & 10 & 11.13 &  150 & 5.3 & \citep{2022SPIE12184E..0YB} \\
        Keck II (10m) & FOBOS & 0.063 & 20 & 0.4584 & 1800 & 28\,571.43 & \citep{2020SPIE11447E..1DB, 2019BAAS...51g.198B} \\
        \hline
        & \multicolumn{5}{l}{\textbf{Pick-and-Place Technology}} \\
       WHT (4.2m) & WEAVE & 3.1416 &   &   & 960 & 305.57 & \citep{2012SPIE.8446E..0PD} \\
        AAT (3.9m) & 2dF  & 3.1416 &   &   & 400 & 127.32 & \citep{2002MNRAS.333..279L} \\
        \hline
       &  \multicolumn{5}{l}{\textbf{New Design Concept}} \\
        WST (12m) & FLEX & 3.1 &  7 & 0.75  & 20\,000 & 6451.61 & \citep{10.1117/12.3019907}\\
        \hline
    \end{tabular}
    }}
\end{table*}

Early examples of MOS systems, such as MEDUSA  \citep{hill1988history} and FLAIR \citep{1988ASPC....3..125W}, laid the groundwork for both the use of optical fibers in such systems, as well as advancements in multiplex.
Modern multi-object spectrographs are highly efficient tools. They enable observational programs targeting objects spread well beyond a single FoV of the telescope, through multiple pointings, by virtue of reconfigurable fiber or slit systems over a large FoV. Early robotic fiber systems include Hydra, on the WIYN Telescope \citep{1995SPIE.2476....2B}, and the Two-degree Field instrument on the Anglo-Australian Telescope \citep[2dF/AAOmega;][]{lewis2002anglo,saunders2004aaomega, smith2004aaomega, sharp2006performance}, while the contemporaneous Sloan Digital Sky Survey \citep[][]{2003AJ....126.2081A} continued to employ a manual fiber plugging system. The fiber Large Array Multi Element Spectrograph \citep[FLAMES;][]{pasquini2002installation} on the Very Large Telescope (VLT) and the fiber Multi-Object Spectrograph \citep[FMOS;][]{kimura2010fibre} on the Subaru Telescope are later examples. Other examples include Hectochelle \citep{szentgyorgyi2011hectochelle} and Hectospec \citep{fabricant2005hectospec} on the Multiple Mirror Telescope.
Building on the success of these instruments, a new generation of multi-object spectrographs has emerged, including 4MOST \citep{2022SPIE12184E..6NB}, DESI \citep{2024AJ....167...62D}, WEAVE \citep{2024MNRAS.530.2688J}, the Sloan Digital Sky Survey V \citep[SDSS-V;][]{2026AJ....171...52K}, MOONS \citep{2022SPIE12189E..0VB}, PFS \citep{2014SPIE.9151E..1YF}, and LAMOST \citep{2012RAA....12.1197C}.

These systems are designed to tackle increasingly complex scientific objectives. Each represents a significant advancement in both technology and design. They enable more efficient and expansive surveys than ever before. These instruments are tailored to address specific astrophysical challenges, highlighting the ongoing evolution of MOS. Moreover, with the rapid growth of imaging surveys, there is increasing interest in spectroscopic follow-up of the identified targets. This has led to substantial investment in upcoming MOS facilities  \citep[e.g.,][]{2023ConPh..64...47P, 2024arXiv240305398M}.

This study provides a detailed overview of the broad range of specifications potentially accessible to new fiber positioning systems. The focus is on identifying technologies with the highest potential impact in astronomy. A key aspect of this analysis is the evaluation of the performance of MOS fiber systems in terms of their basic specifications. Our proof-of-concept implementation systematically assigns fibers to targets while adhering to patrol and exclusion constraints. It uses a simple model for the allocation of fibers to targets, iterated through repeated visits, to achieve a required level of completeness, while tracking fiber allocation efficiency. Different existing fiber positioning systems are compared to assess the impact of their technical specifications on this observational performance. This approach is an important step in identifying areas where innovation could drive the most significant improvements. The investigation goes beyond a technology review by establishing an essential link between astronomical goals and technological capabilities. The study concludes with recommendations on the most promising areas for future development. We use the fourth and final data (DR4) release of the GAMA survey \citep{2022MNRAS.513..439D} in this work to investigate our proof-of-concept code, as explained in the details below.

The layout of this paper is as follows. We provide a brief introduction to the existing fiber positioning technologies in \S\,\ref{sec:existing_fiber}, followed by the dataset from the GAMA in \S~\ref{sec:data}. 
In \S~\ref{sec:code} we introduce the proof-of-concept code, introducing some basic parameters, defining fiber home position, selecting science targets, and details of the fiber to target allocation process.
The main results, including the details of our analysis on the main fiber to target simulation, and evaluating the existing fiber positioning systems are presented in \S~\ref{sec:results}. In \S~\ref{sec:discussion}, we summarize the key results and explore the future directions of our analysis, before presenting our conclusions in \S~\ref{sec:concl}.

\section{EXISTING FIBER POSITIONING SYSTEMS} 
\label{sec:existing_fiber}

In MOS fiber systems, light from multiple objects in a given FoV is collected using multiple optical fibers. 
Major spectroscopic surveys \citep[e.g.,][]{2001MNRAS.328.1039C, 2003AJ....126.2081A} typically assign a single fiber to each object, providing one measurement per galaxy. Each fiber directs the light to a spectrograph, where they are aligned to form a pseudo-slit. This setup ensures distinct spectra for each object. 
To achieve this outcome fiber positioner systems are essential for aligning fibers with targets in the FoV. Their design plays a critical role in the efficiency and adaptability of spectroscopic instruments. Robotic fiber positioner (RFP) arrays allow hundreds or thousands of fibers to be positioned simultaneously. RFP arrays can be divided into two broad types: (1)~``parallel positioners,'' which include tilting spines, positioners with two rotational axes 
(often called phi-theta or theta-phi) and referred to here for simplicity as ``dual-rotators,'' FLEX and starbugs, and (2)~``sequential positioners,'' such as pick-and-place systems. Table~\ref{tab:positioner_systems} shows specifications of some existing fiber positioning systems. 
The following overview briefly summarizes some of these existing fiber technologies, their features and capabilities.

\begin{itemize}
\item \textbf{Tilting Spines:}\\
Tilting spine fiber positioning technology \citep[][and references therein]{2003SPIE.4841..985G} is a flexible and efficient method for precisely positioning optical fibers on the focal plane of an astronomical instrument. Each tilting spine consists of a carbon fiber tube with a steel ball in a magnetic mount, driven by a piezo-actuated stick-slip motor. A saw-tooth voltage pulse overcomes magnetic friction, tilting the spine in controlled micron-scale steps \citep{2015arXiv151100737G, 2003SPIE.4841..985G, 2004SPIE.5492.1228B}. The fiber tip moves in discrete steps, allowing multi-directional adjustments. Many spines form a high-density fiber bed, where the tilt angle determines the focal plane position of the fiber. The compact, scalable design eliminates fiber twisting and allows dense packing, making it ideal for high-multiplexing spectrographs and enabling closely spaced target observations \citep{2018SPIE10702E..1MS}.
Tilting spine fiber positioning technology has been successfully implemented in several large-scale astronomical instruments. One notable example is the ``Echidna'' system of the Australian Astronomical Observatory (AAO). The fiber Multi-Object Spectrograph \citep[FMOS;][]{2010arXiv1006.3102K} of the Subaru telescope employs the Echidna type design and technique as the fiber positioner system. The success of Echidna in FMOS inspired its adaptation for other projects, including the Australian-ESO (European Southern Observatory) Positioner system \citep[AESOP;][]{2022SPIE12184E..6NB,2022SPIE12184E..6MB} for the 4MOST project.
In most traditional fiber positioners, such as dual-rotator systems, the fiber remains telecentric throughout the positioning range, ensuring that the incoming light is coupled with minimal angular misalignment. The angular misalignments between the incoming light and the fiber face predominantly contribute to focal ratio degradation (FRD).
However, in tilting-spine mechanisms, the fiber is tilted relative to the optical axis to extend the patrol radius. This tilt introduces non-telecentricity, which can cause geometric FRD and reduce light throughput due to angular mismatch at the fiber entrance.\\

\item \textbf{Dual-rotators:}\\ 
Dual-rotator fiber positioning technology is a precise and efficient system used in multi-object spectroscopic instruments to position optical fibers on astronomical targets. It consists of two rotational actuators. The theta ($\theta$) motor rotates the entire fiber assembly around a fixed central axis, while the phi ($\phi$) motor moves the fiber around an offset pivot point. By coordinating these two degrees of freedom, the fiber can reach any position within its patrol region, enabling accurate target acquisition while minimizing collisions with neighboring fibers.
Dual-rotator fiber positioners, first implemented by the Large Sky Area Multi-Object fiber Spectroscopy Telescope \citep[LAMOST;][]{2012RAA....12.1197C} with 4,000 fibers enable high-throughput spectroscopy and have collected over 20 million spectra. This design is adopted by the Dark Energy Spectroscopic Instrument \citep[DESI;][]{2016arXiv161100036D, 2016arXiv161100037D}, which uses 5,000 positioners across a 7.5 square-degree field and a 10.4,mm pitch, feeding a spectrograph spanning 360–980,nm. Subaru’s Prime Focus Spectrograph \citep[PFS;][]{2016SPIE.9908E..1MT} uses 2,394 dual-rotator fibers over a 1.3-degree field, linked to four spectrographs covering 380–1260,nm. Designed for a multi-purpose telescope, PFS requires frequent mounting, with 50-meter fiber runs and two connectors, posing added performance challenges.

\item \textbf{Starbugs:}\\ 
Starbugs are miniature piezoelectric robots designed for precise optical fiber positioning on the focal plane of the telescope \citep[][and references therein]{2012SPIE.8450E..1AG, kuehn2014taipan}. Each unit consists of two concentric piezoceramic tubes that function as legs, enabling micro-stepping motion in four directions with sub-$5\,\mu$m accuracy. Controlled by applied voltage, these legs expand and contract, producing stepping and rotational movement. Starbugs adhere to a transparent glass field plate via vacuum pressure and move simultaneously, positioning a field of 150 units in under 10 minutes \citep{2022SPIE12184E..0YB}. They operate at voltages up to ±200V and frequencies between 50 and 150Hz, with step size increasing at higher voltages \citep{2012SPIE.8450E..1AG, 2014SPIE.9151E..1AB}. Each starbug carries an optical fiber or microlens assembly in its central aperture, eliminating the need for retractors. A flexible umbilical routes wires and fibers to a connector plate integrating optical, mechanical, and electrical connections.
Installed in 2017 at the UKST, TAIPAN \citep{kuehn2014taipan, staszak2016taipan} is a 150-fiber starbug-based prototype, extensively commissioned to validate key subsystems \citep{2014SPIE.9151E..1AB, 2022SPIE12182E..36O}. These tests informed the development of starbug systems for future instruments, including FOBOS \citep{2019BAAS...51g.198B, 2020SPIE11447E..1DB}, which will deploy 1800 fibers across Keck’s 0.8-m focal plane significantly surpassing TAIPAN’s capacity.\\

\item \textbf{Pick-and-Place:}\\
Pick-and-place fiber positioning systems have been essential in MOS for decades, using robotic mechanisms to precisely align optical fibers at a telescope's focal plane for spectroscopy. A robotic gantry with a precision gripper \citep{2002MNRAS.333..279L} sequentially picks fibers from a retractable carousel and places them on a focal plate. While ensuring high accuracy, this process results in longer reconfiguration times compared to parallel-actuated systems. Stepper motors and encoders guide the robotic arm in X-Y-Z coordinates, with alignment aided by fiducial marks and metrology cameras. Closed-loop feedback corrects errors, and fibers are secured using magnetic docking or mechanical clamps. Though slower, this method remains reliable and well-suited for large multi-object spectrographs requiring precise positioning across a wide field.
One of the earliest pick-and-place fiber systems was the Autofib robot at the AAT and WHT \citep{1986SPIE..627..118P}. Subsequent systems, such as Hydra \citep{1994SPIE.2198...87B}, improved positioning accuracy and versatility. 
The AAO’s 2dF enabled 400-object observations within a two-degree field on the AAT \citep{2002MNRAS.333..279L}, while 6dF on the UK Schmidt Telescope was designed for wide-area surveys \citep{1998ASPC..152...80P}. More recent systems include OzPoz on the VLT for faint-object spectroscopy \citep{2004SPIE.5492..643G} and WEAVE on the WHT for detailed stellar and galactic mapping \citep{2020SPIE11447E..14D}.\\

\begin{figure*}[ht!]
\centering
\includegraphics[width=0.75\linewidth]{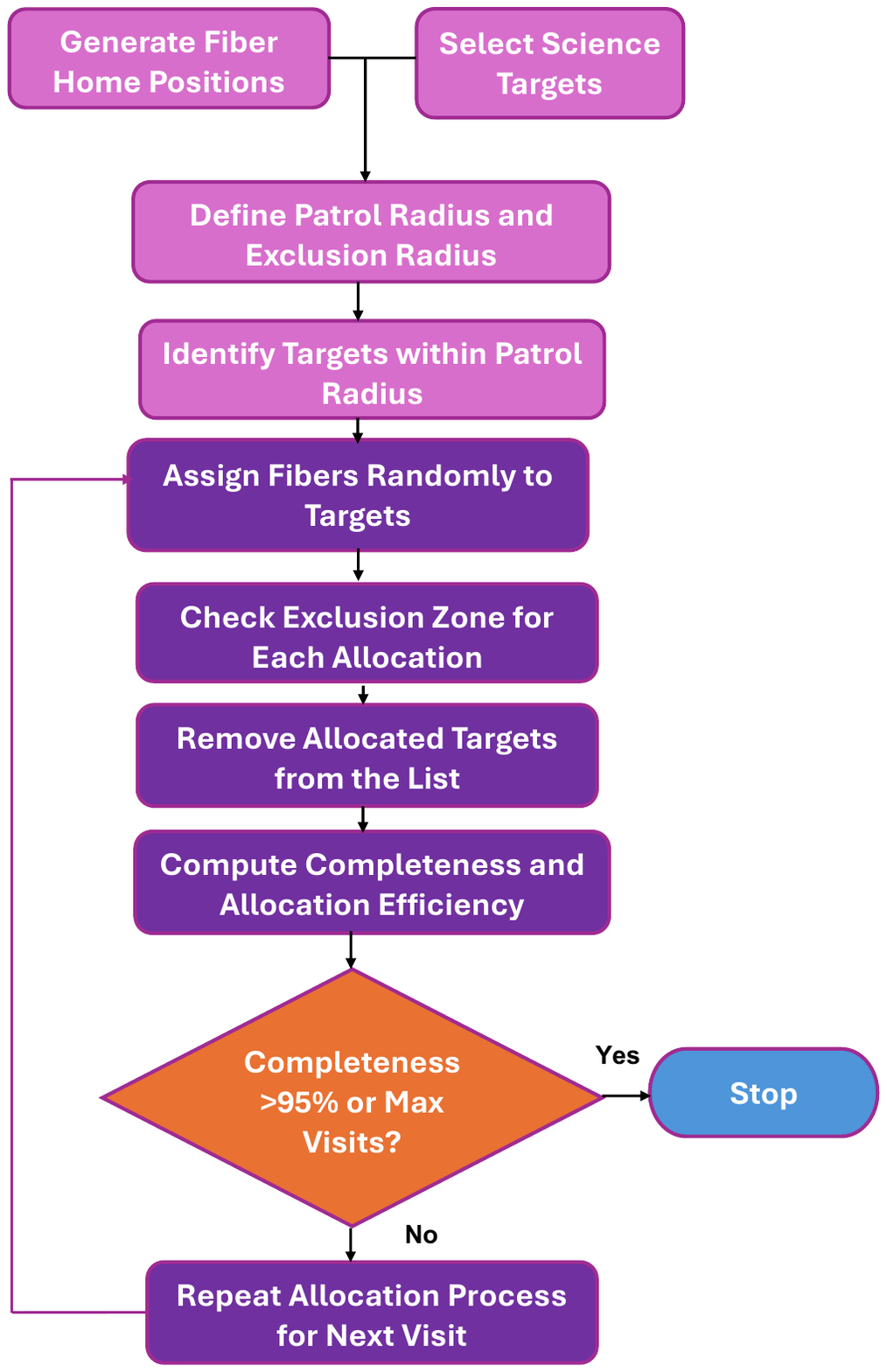}
\caption{The flow chart of the proof-of-concept code illustrates the step-by-step process involved in the fiber allocation procedure. The process begins with target selection, where the input catalog of astronomical objects is filtered based on observational priorities and constraints. At the same time the fiber home positions are generated to uniformly distribute the fibers over a chosen FoV. The algorithm then identifies targets within each patrol radius (PR) (for parallel positioner systems), ensuring that each fiber can only access a defined region to prevent overlaps.
fiber assignment is performed through a random allocation process, where available fibers are assigned to targets while strictly enforcing PR and exclusion constraints (e.g., preventing multiple fibers from selecting the same target or overlapping patrol zones). Once an initial allocation is made, assigned targets are removed from the available pool to mimic the observation process, and these stages are iterated across multiple subsequent visits until the nominated completeness requirement is achieved.
The final output consists of a set of completeness and efficiency metrics as a function of each visit, to establish the performance of any given system that can then be compared against others.}
\label{fig:flowchart}
\end{figure*}

\item \textbf{Fiber Location EXtender (FLEX):}\\
Fiber Location EXtender \citep[FLEX;][]{10.1117/12.3019907} is a novel fiber positioner concept inspired by minimally invasive surgical mechanisms. It is designed to meet the demands of next-generation spectroscopic surveys \citep[e.g.,][]{2016arXiv160600060M, 2022arXiv220904322S, 2024arXiv240305398M}. FLEX combines the advantages of dual-rotator, tilting spine, and grid-based systems while minimizing their drawbacks. It enables high-density placement, a large patrol area per fiber, and effective collision avoidance. The design features a mass-producible parallel mechanism, fabricated via laser cutting, with a tripod actuator for (x,y) positioning and (z) focus maintenance. This ensures precise pupil alignment and accommodates curved focal surfaces. Unlike conventional systems, FLEX allows at least 20 fibers to access any focal plane position. This enhances flexibility and scalability for future spectroscopic facilities.
FLEX is the proposed fiber positioning system for the Wide-field Spectroscopic Telescope \citep[WST;][]{10.1117/12.3018093, 2024arXiv240305398M}, a planned 12-meter telescope designed for high-efficiency spectroscopic surveys.\\

\end{itemize}

\section{Data} 
\label{sec:data}

We use data from multiple regions within GAMA survey fields, selecting targets based on specific magnitude ranges. This provides a representative sample of real galaxies to ensure we correctly account for galaxy clustering in our tests using the proof-of-concept code. These fields were chosen due to their high complete photometric and spectroscopic datasets, which provide a robust foundation for validating and testing the efficiency of fiber to target allocation. The primary aim is to demonstrate the effectiveness of the code in assigning fibers to targets while accommodating realistic observational constraints. In doing so, we demonstrate the potential of this code to support future MOS technology development.

The GAMA survey \citep{2011MNRAS.413..971D, 2022MNRAS.513..439D} is an extensive photometric, spectroscopic, and multi-wavelength investigation of galaxies, carried out using the AAOmega multi-object spectrograph on the Anglo-Australian Telescope (AAT). Covering a total area of 286 deg$^2$ across five sky regions (G02, G09, G12, G15, and G23), it is a wide-field, multi-pass survey designed to gather detailed data on galaxy properties. The survey includes 300\,000 galaxy spectra, achieving a 98.5\% redshift completeness, regardless of target density \citep{2015MNRAS.452.2087L}. Most fields have a limiting magnitude of $r$ $=$ $19.8$ \citep{2015MNRAS.452.2087L}, enabling detailed studies of galaxy evolution and large-scale structure \citep{2010MNRAS.404...86B, 2010PASA...27...76R, 2011MNRAS.413..971D, 2013MNRAS.430.2047H,2015MNRAS.452.2087L, 2018MNRAS.474.3875B, 2020MNRAS.496.3235B}. 
The data used in this analysis come from the GAMA-KiDS-VIKING catalog \citep{2020MNRAS.496.3235B}, which combines the GAMA multi-wavelength dataset with imaging from KiDS \citep{2015A&A...582A..62D} and VIKING \citep{2012sngi.confE..40S}.

\begin{figure*}
\centering
\includegraphics[width=0.45\linewidth]{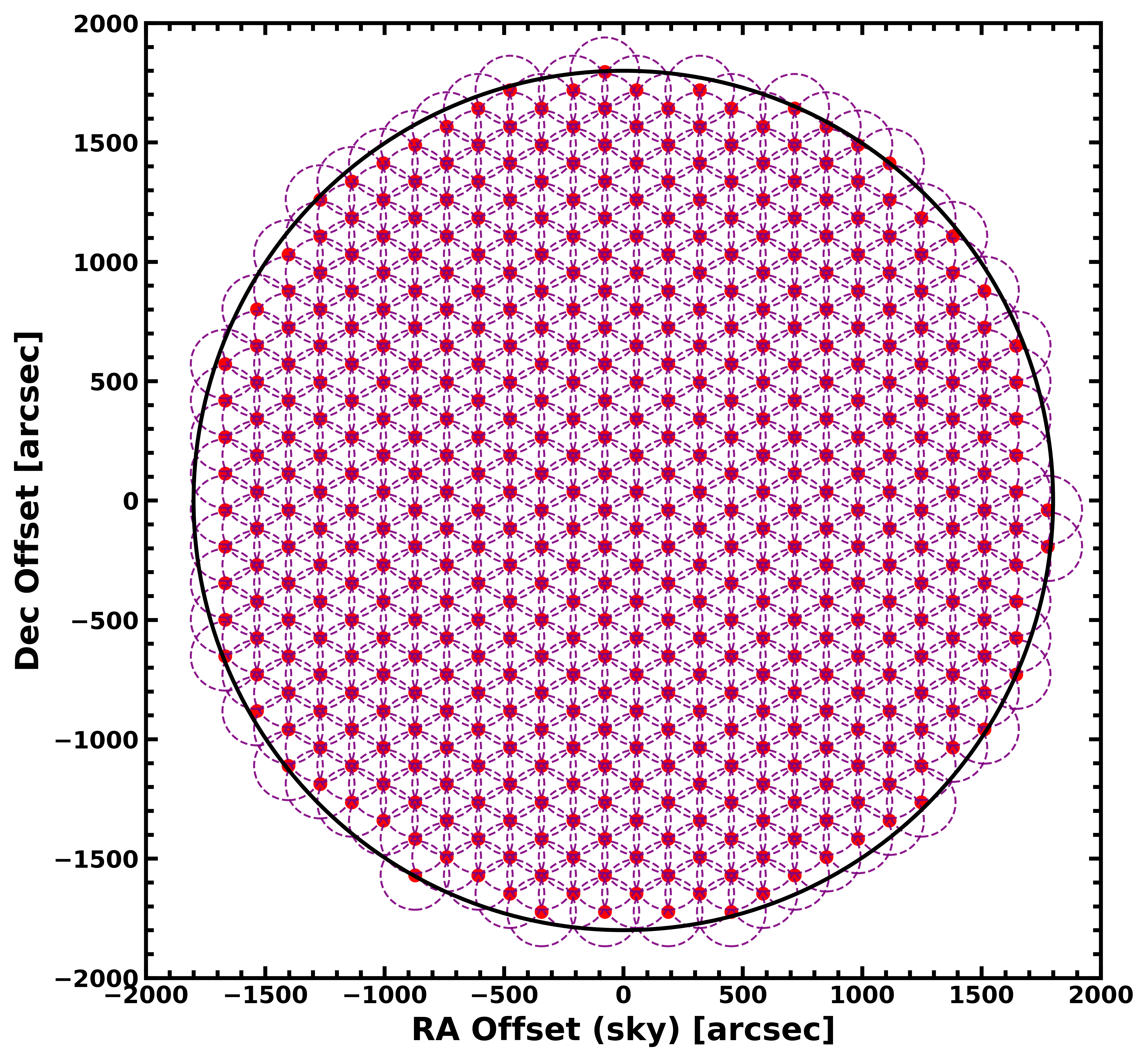}
\includegraphics[width=0.41\linewidth]{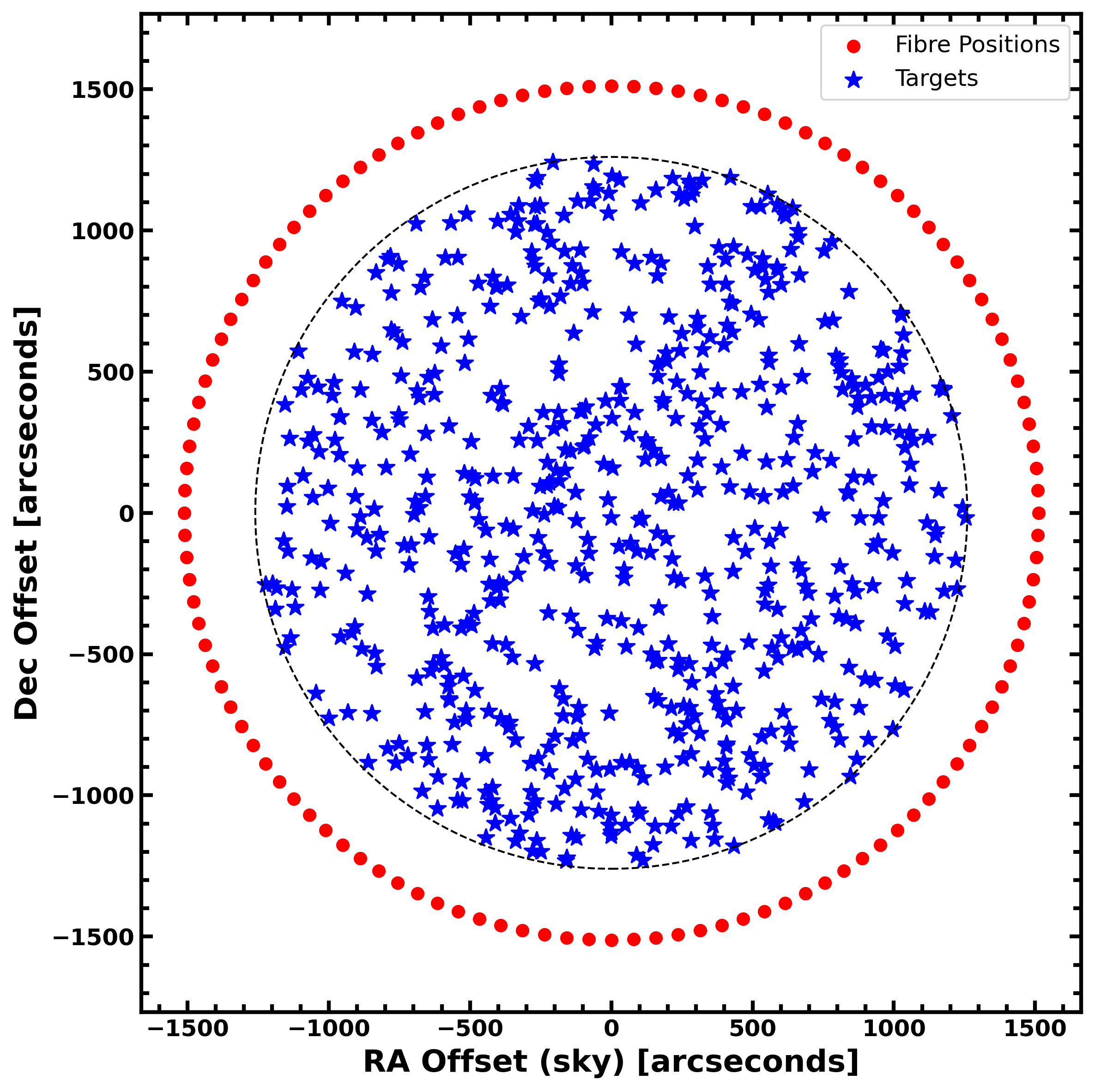}
\caption{Illustrations of the fiber home positions. The left panel illustrates a fixed fiber positioning system (parallel positioner), where hexagonal fiber home positions are overlaid with target positions and corresponding patrol radii. The hexagonal mesh, shown in red, represents the fixed home positions of 500 fibers. Each fiber has a defined PR, depicted as purple dashed circles, indicating the region within which the fiber can be repositioned to acquire a target. This setup ensures a structured and consistent fiber distribution while maintaining a limited range of motion for target acquisition.
In contrast, the right panel demonstrates a pick-and-place fiber positioning system (sequential positioner), where 120 fibers (marked in red) are arranged in their home positions around the edge of a FoV, and in principle can be allocated to any target. Unlike the fixed system, this approach allows greater flexibility in target allocation by enabling fibers to be placed anywhere to support optimal target allocation. This method enhances target accessibility while maintaining necessary constraints on fiber movement to prevent overlap or mechanical interference.}
\label{fig:hexa_grid}
\end{figure*}

\section{Proof-of-Concept Code}
\label{sec:code}

We introduce a proof-of-concept code to model the current generation of fiber positioner instruments. The proposed tool has been tested for existing and planned fiber positioning systems. Our proof-of-concept tool is designed to be simple, as our first goal is to look at the difference between the different fiber positioning specifications, rather than detailed differences between the technologies.
Figure~\ref{fig:flowchart} presents the flow chart outlining the steps taken by the proof-of-concept code. The analysis consists of multiple steps, beginning with target selection and proceeding through fiber assignment over multiple repeated visits to achieve the desired completeness.

\subsection{Nomenclature and Definitions}
\label{subsec:defintions}

{We begin by introducing the fundamental parameters, metrics, and concepts, before describing in more detail the procedures of the proof-of-concept code. Parallel positioners (tilting spine, dual-rotator, FLEX, and starbug systems) can be essentially specified using fiber pitch, patrol radius (PR), and exclusion radius (ER). Sequential positioners (such as pick-and-place systems) are rather different, and we treat them separately.
The fiber pitch is defined as the spacing between individual fibers in the focal plane. Smaller pitch values allow for denser fiber packing but may introduce mechanical constraints. The relationship between pitch and fiber density is $\text{fiber Density} \propto \text{Pitch}^{-2}$. In a hexagonal fiber home position configuration, the fiber density is given by $\text{fiber Density} = \frac {2}{\sqrt{3}} \text{Pitch}^{-2}$.
The PR is the maximum distance each fiber can move to reach a target from its home position. A larger PR increases target allocation flexibility but may lead to an increased likelihood for fiber collisions.
We do not model fiber collisions in this analysis, although each technology has a sophisticated approach to ensuring fibers do not interfere with each other during positioning \citep[e.g.,][]{2014A&A...566A..84M, 2020A&A...635A.101T, 2020MNRAS.497.4626T, 2021AJ....161...92S, 2021MNRAS.500..101Z, 2024AJ....167..276Z, 2026AJ....171...37B}.
The ER sets a minimum separation between fibers to model the prevention of mechanical collisions. A larger ER reduces fiber allocation density but ensures mechanical clearance. We characterize the performance of a system modeled using these parameters by two simple metrics. The completeness measures the fraction of available targets successfully assigned to fibers. The efficiency quantifies the fraction of available fibers that are successfully allocated to targets.

\subsection{Fiber Home Position}
\label{subsec:fiber_home}

As a starting point, we generate a hexagonal mesh within a circular FoV for fiber home positions. The hexagon side length is determined by the given pitch, with horizontal and vertical steps computed to maintain a proper hexagonal alignment. The hexagon side length is defined as:
\begin{equation}
  s = \frac{\text{pitch}}{\sqrt{3}}
\end{equation}
where the horizontal step size is:
\begin{equation}
dx = \frac{3}{2} s
\end{equation}
and the vertical step size is:
\begin{equation}
dy = \sqrt{3} s
\end{equation}

This returns an array of hexagonal centroids, representing the adopted fiber home positions. This ensures an even distribution of fiber home positions over the full FoV, while adhering to the fiber pitch constraint.

The PR is defined by setting the maximum distance a fiber can move from its home position, to be assigned to a target. An ER is also defined, which prevents targets from being placed too close to each other, avoiding conflicts in fiber allocation.
The left panel of the Figure~\ref{fig:hexa_grid} illustrates the arrangement of hexagonal fiber home positions and their corresponding patrol radii, demonstrating the spatial constraints of fiber movement. The red markers indicate fiber positions, while the purple dashed circles represent the patrol areas within which fibers can reach potential targets.

Building on this idea, for a pick-and-place system (Figure~\ref{fig:hexa_grid}), we generate fiber home positions in a circular pattern just outside the FoV before assigning them to targets. As shown in the right panel of the Figure~\ref{fig:hexa_grid}, fibers (red) dynamically reposition to acquire targets (blue), emphasizing the flexibility of target assignment without any PR constraint.
It first calculates evenly spaced angles around a circle and places fibers at a radius 20\% larger than the FoV to ensure they start outside the observation area. In the next step, a fixed number of fibers and the field center is set, and the fiber positions are computed for the simulation.

\begin{figure*}[ht!]
\centering
\gridline{\fig{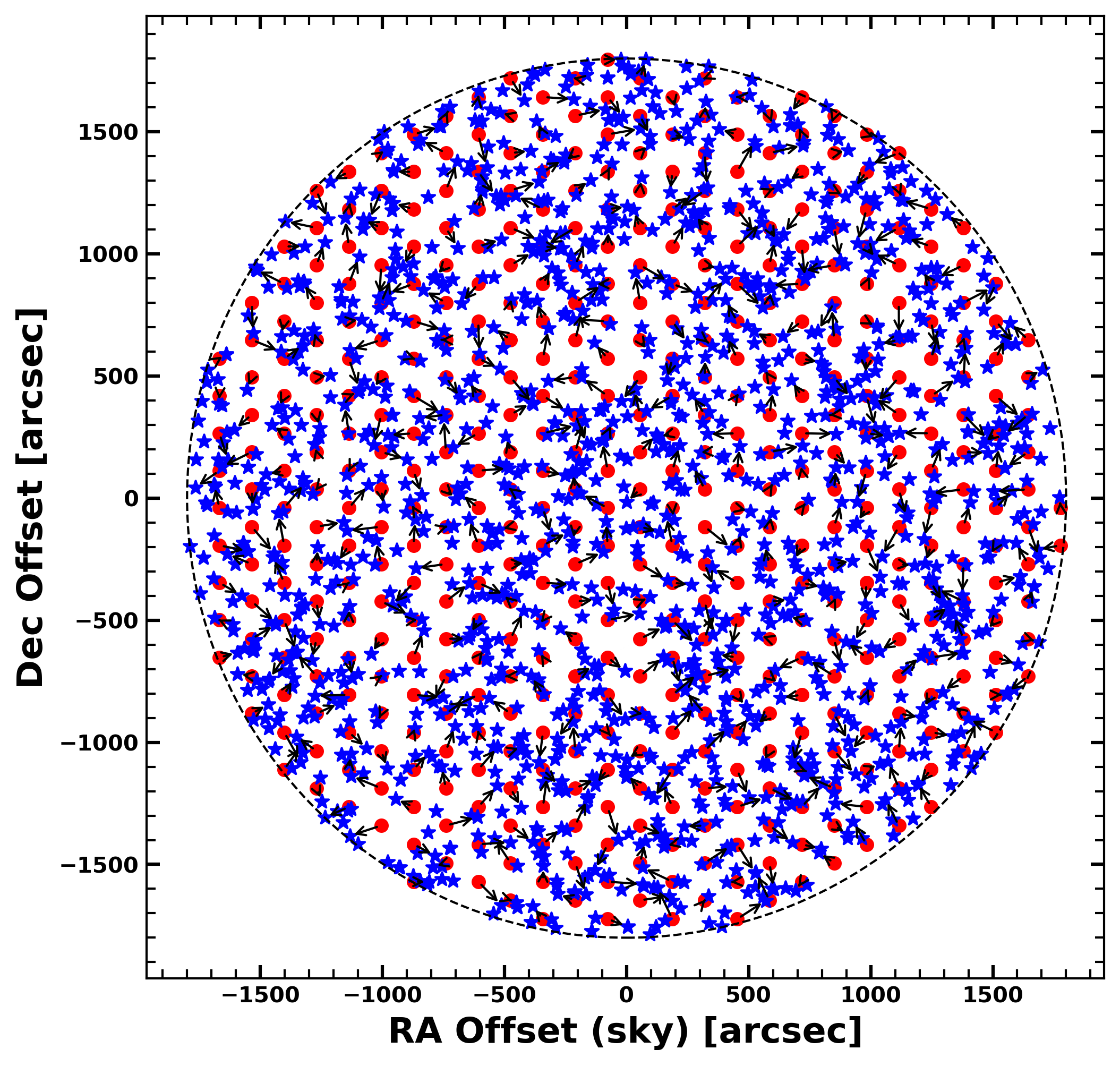}{0.25\textwidth}{(a)}
          \fig{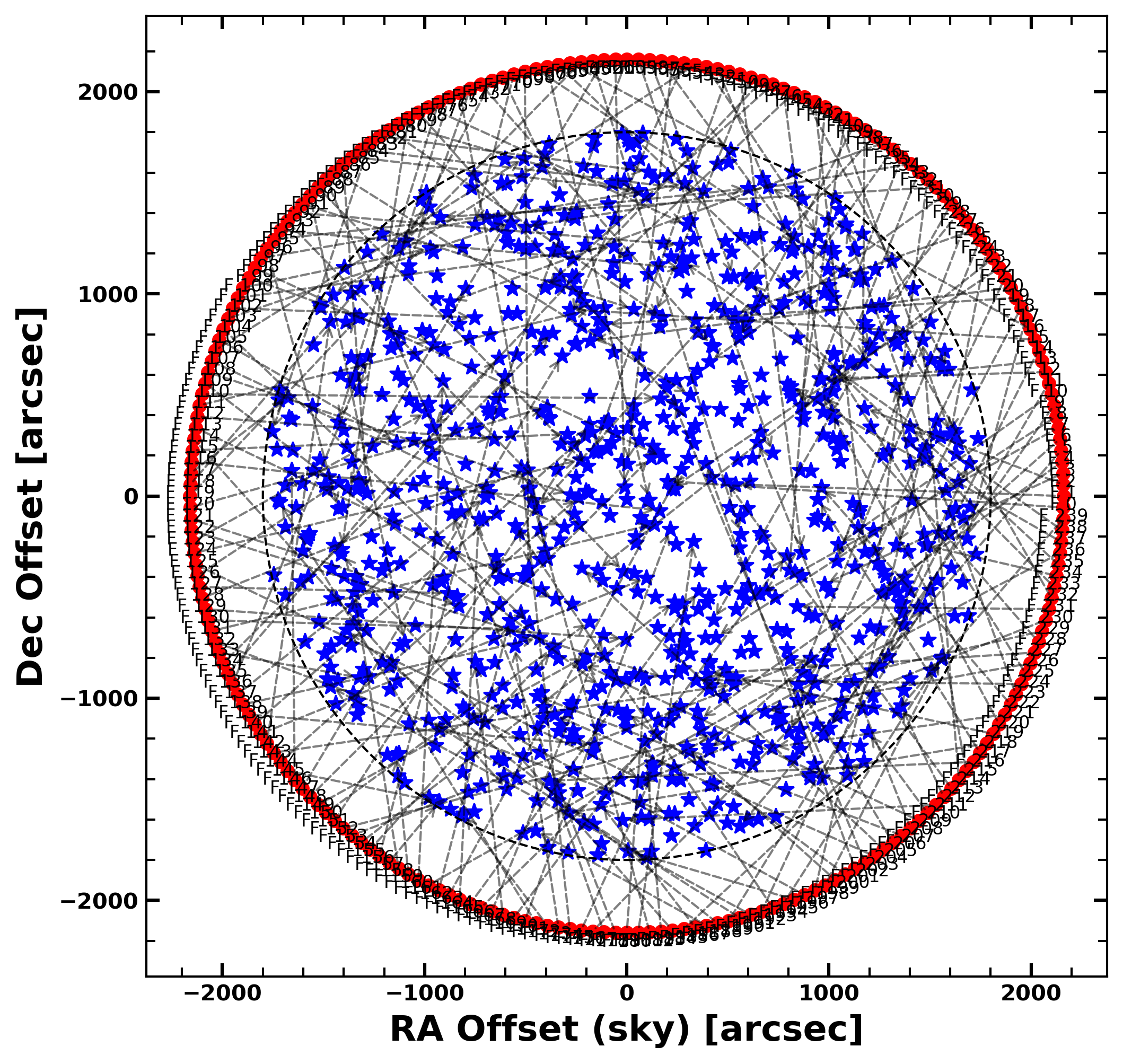}{0.25\textwidth}{(b)}}
\gridline{\fig{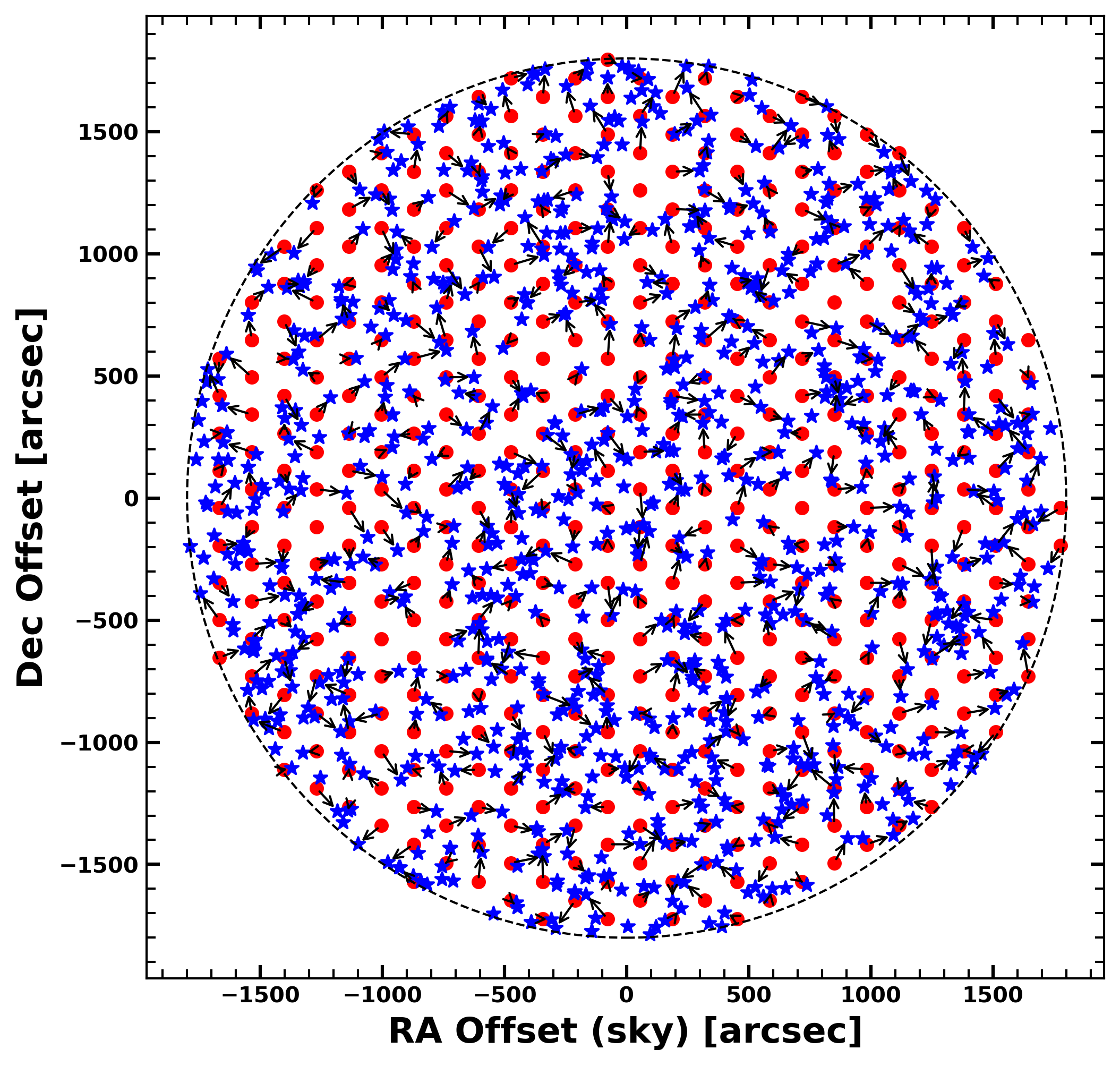}{0.25\textwidth}{(c)}
          \fig{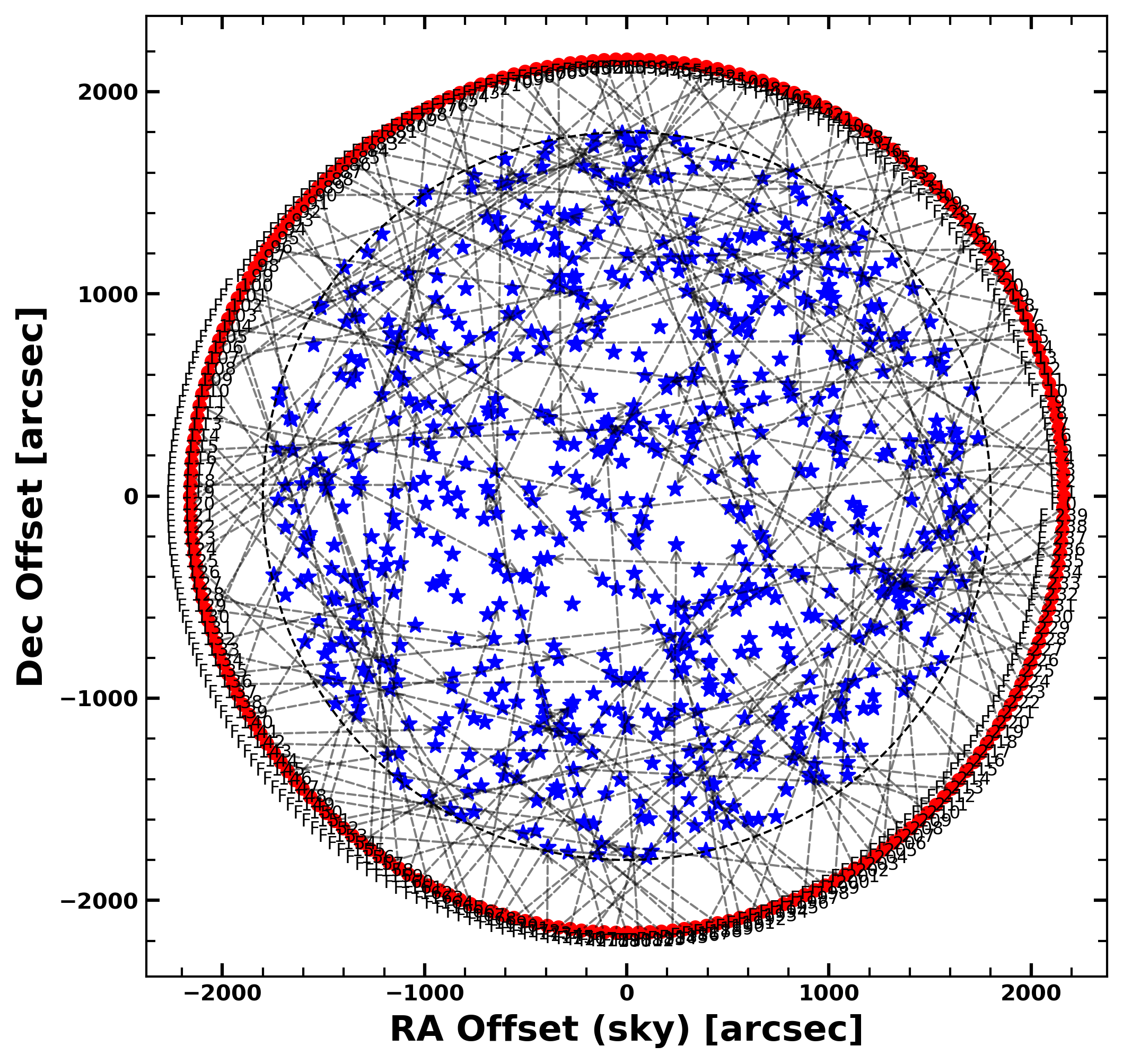}{0.25\textwidth}{(d)}}
\gridline{\fig{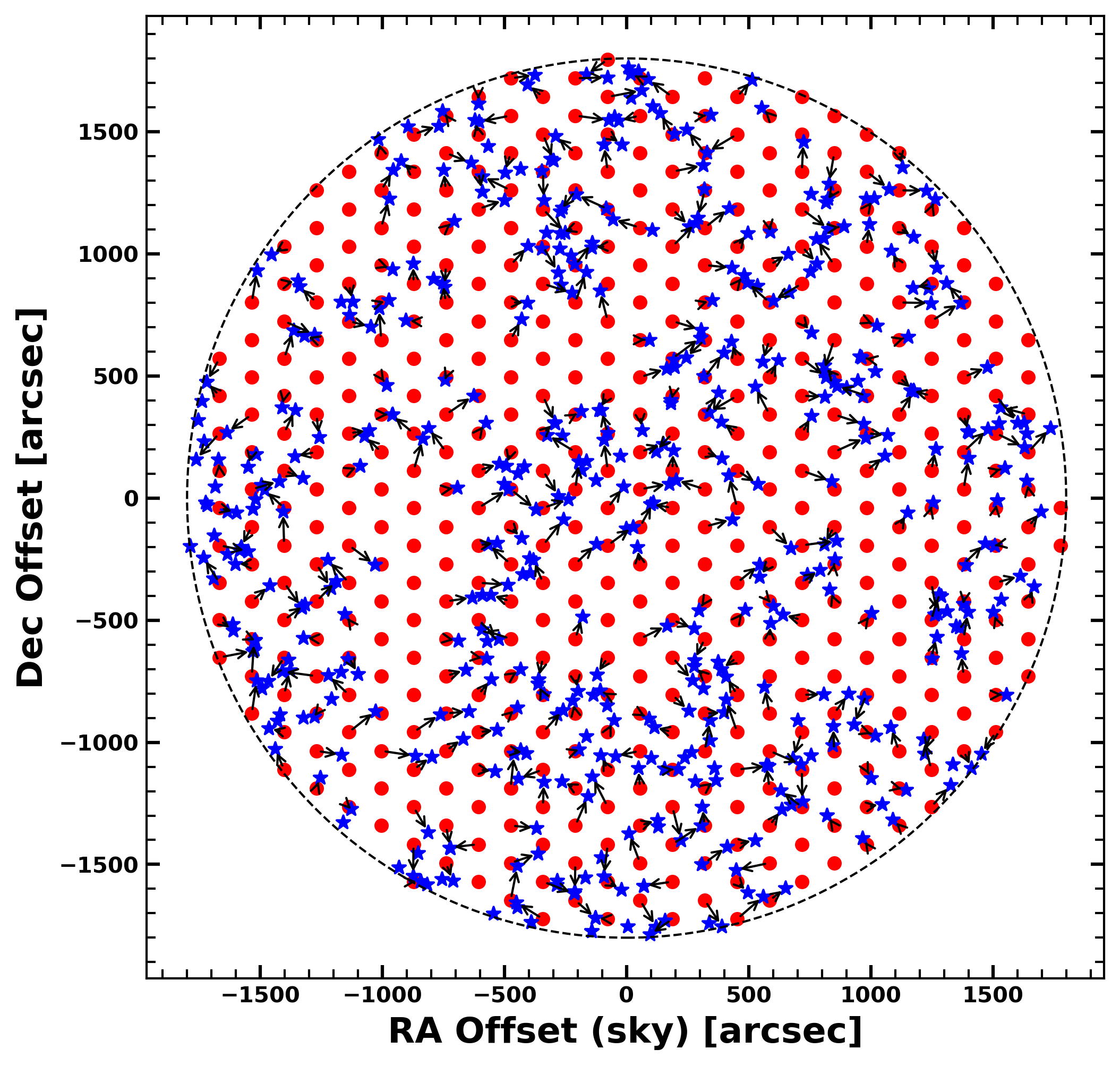}{0.25\textwidth}{(e)}
          \fig{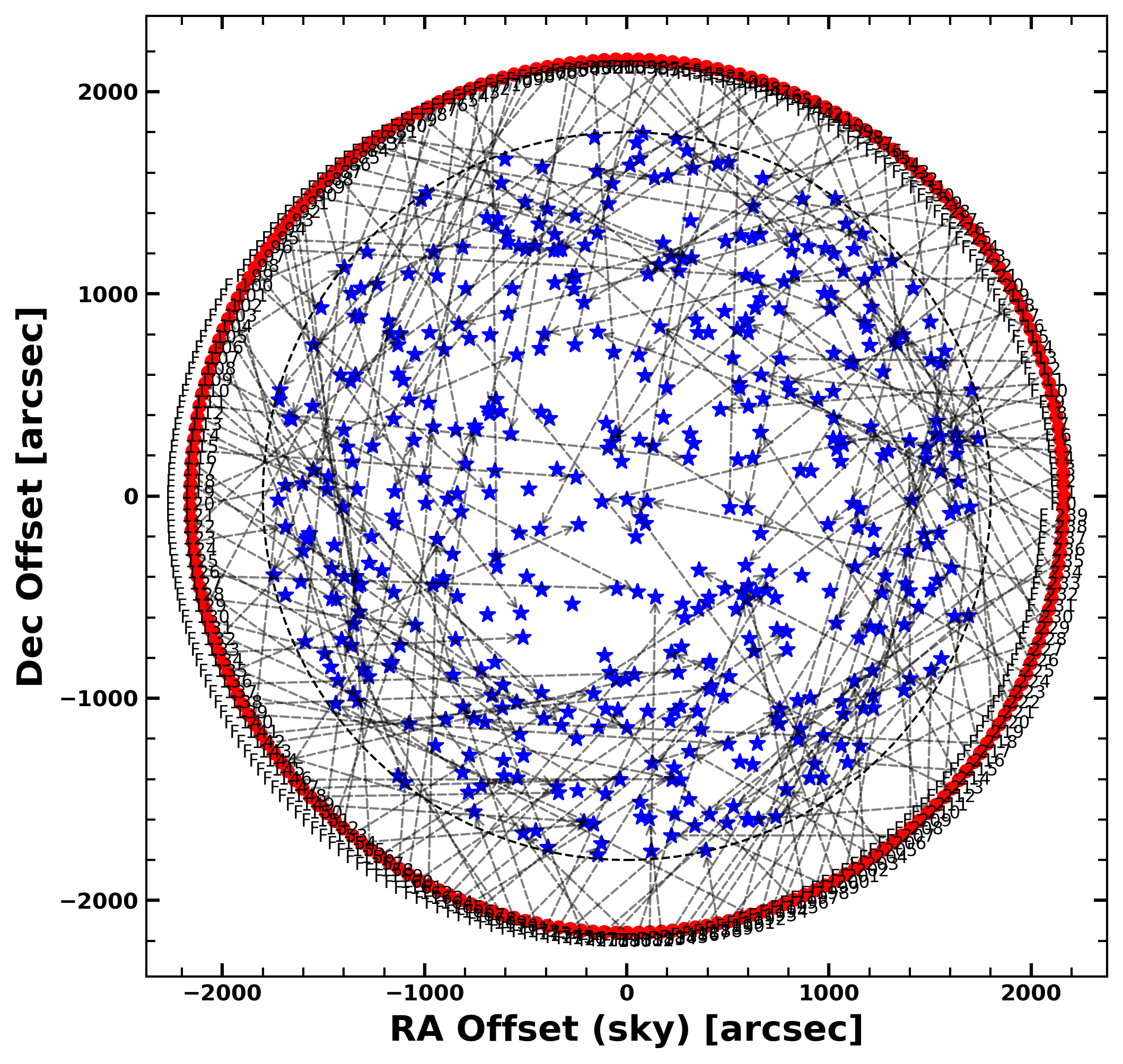}{0.25\textwidth}{(f)}}
\gridline{\fig{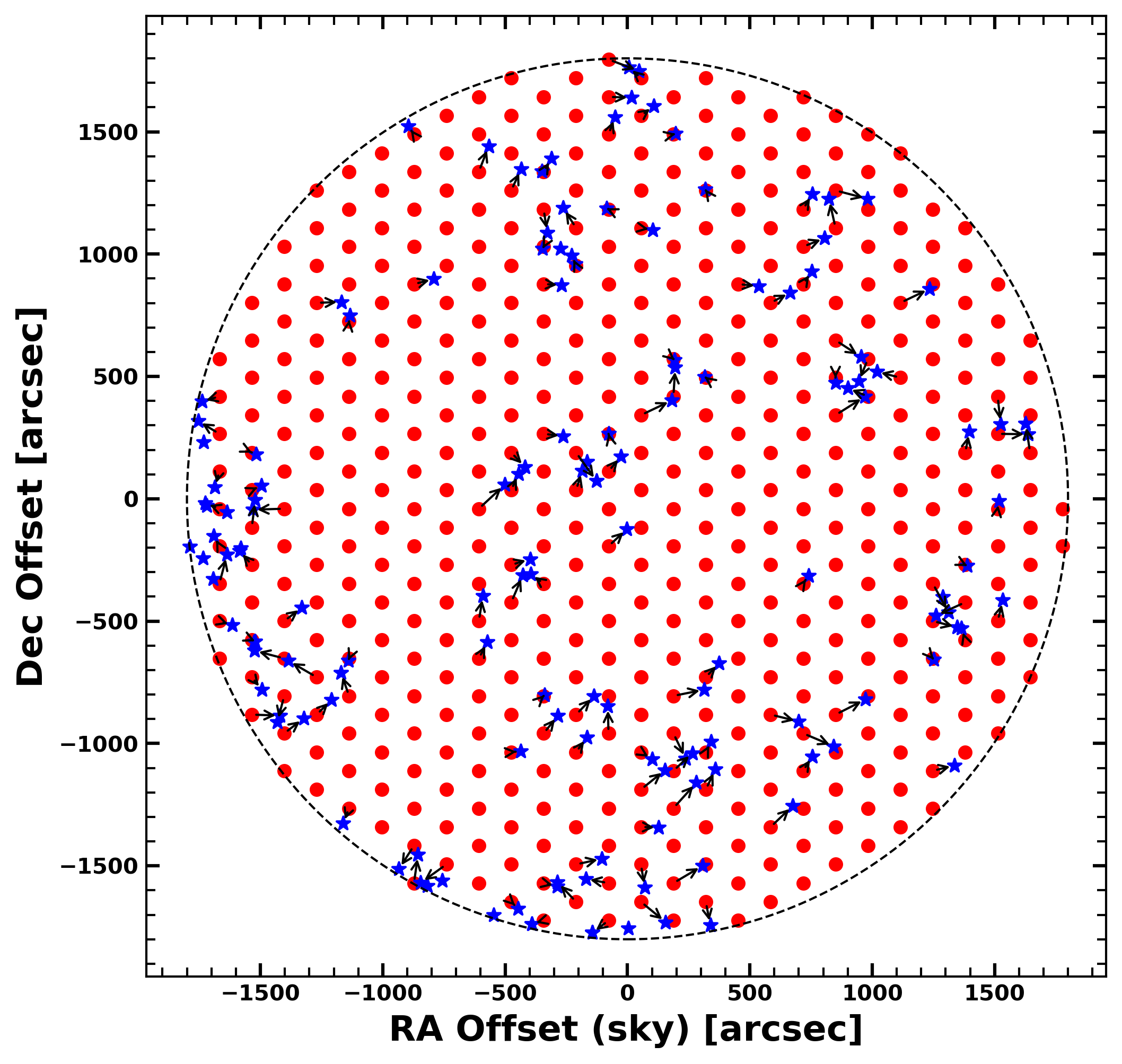}{0.25\textwidth}{(g)}
          \fig{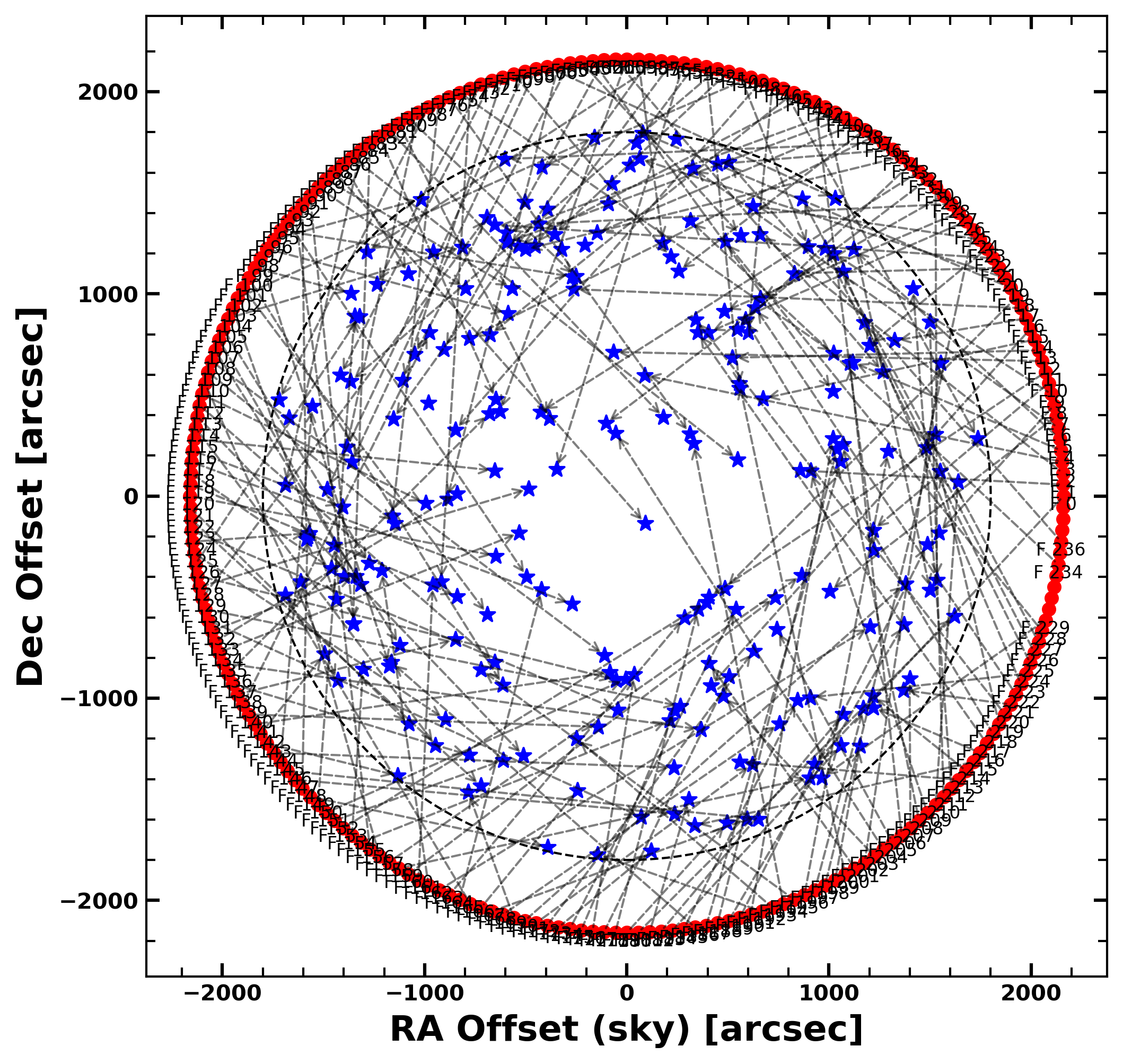}{0.25\textwidth}{(h)}}
\caption{Eight-panel figure showing different stages of fiber-to-target allocation. Each panel is labeled (a) through (h). The left panel illustrates fiber-to-target allocation in a hexagonal fiber positioner system (parallel positioner). Each panel illustrates the allocation of fibers to targets in a fixed configuration. Each panel displays the allocation of 500 fibers to 1500 targets within a $1.0^\circ$ diameter FoV in one of the GAMA (G09) fields. The fiber home positions are represented in red, and target positions are marked in blue. Black arrows show the allocation of fibers to specific targets. The right panel illustrates the fiber-to-target allocation in a pick-and-place fiber positioner system (sequential positioner). Each panel displays the allocation of 240 fibers to 960 targets within a $1.0^\circ$ diameter FoV in one of the GAMA (G09) fields. The fiber home positions are represented in red around the edge of the FoV, and target positions are indicated by blue markers within the FoV. Grey arrows show the allocation of fibers to targets. In both the panels, the first three figures represent the first 3 field visits, while the last figure shows the final visit of fiber allocations.}
\label{fig:fiber_allocation}
\end{figure*}

\subsection{Science Targets}
\label{subsec:target}

To identify suitable targets for fiber-to-target allocation, we apply spatial and magnitude-based selection criteria to our GAMA catalog. The goal is to identify targets within a given circular FoV, that provide realistic galaxy clustering properties. The clustering is important, as highly clustered targets will impact the efficiency of a positioner to efficiently allocate them to fibers, given a specified ER.
A field is chosen,
centered at a given $\alpha,\delta$, which is selected randomly, while ensuring that the FoV lies fully within the given GAMA region. A magnitude range is defined through specifying bright and faint limits, $r_{\text{bright}} < r_{\text{target}} < r_{\text{faint}}$. These limits are chosen to avoid the extreme ends of the GAMA survey range, and in order to provide the desired target density.
This defines the set of targets that are subsequently used throughout a given analysis.

\subsection{Fiber to Target Allocation}
\label{subsec:fiber_target_allocation}

The fiber allocation process involves systematically assigning fibers to targets within their patrol radii while complying with other constraints, such as exclusion radii, to model the avoidance of fiber collisions. This process is repeated over multiple visits of a target field, to achieve some desired eventual completeness. Figure~\ref{fig:fiber_allocation} illustrates the fiber to target allocation process for hexagonal and pick-and-place fiber home positions. Each panel shows the allocation of fibers for different configurations. The main steps of the allocation process are explained as follows:

\subsubsection{\bf Identifying targets within PR}  
Before assigning fibers to targets, we first determine which targets lie within the given PR for each fiber position. This step ensures that each fiber only considers targets it can physically reach. To achieve this, we compute the distance between the position for each fiber position and all potential targets. It iterates through target coordinates, calculates the Euclidean distance to the fiber position, and stores targets within the specified radius. 

\noindent For each fiber \( f_i \in F \), find all targets \( t_j \in T \) that satisfy the condition:
\begin{equation}
d(f_i, t_j) \leq R_{patrol}  
\end{equation}

\noindent where \( R_{\text{patrol}} \) is the maximum distance a fiber can reach. 
So, for a fiber located at \(\mathbf{r}_{\text{fiber}} = (x_{\text{fiber}}, y_{\text{fiber}})\) and a target at \(\mathbf{r}_{\text{target}} = (x_{\text{target}}, y_{\text{target}})\), the distance between them is given by the Euclidean distance:  
\begin{equation}   
d = \sqrt{(x_{\text{fiber}} - x_{\text{target}})^2 + (y_{\text{fiber}} -y_{\text{target}})^2}
\end{equation}

\noindent and the target is considered reachable if this distance satisfies the PR constraint of \(d \leq R_{\text{patrol}}\).
All targets meeting this criterion within the PR are recorded as a possible assignment for that fiber, and the list of available targets for fiber \( f_i \) as \( T_{f_i} \) are stored. This forms the foundation for all subsequent allocation steps.
When an allocation is made, a target is simply chosen randomly from the list of available targets for each fiber.

\subsubsection{\bf Initial fiber allocation} 

After identifying targets fibers are randomly assigned to accessible targets within their patrol radii. Each fiber, 
$f_i \in F$ is considered in turn, and a target is selected randomly from its available options.

\noindent If a fiber has no reachable targets, it remains unassigned:  
\begin{equation}
A_{f_i} = \emptyset \quad \text{if} \quad T_{f_i} = \emptyset
\end{equation}
where \( A_{f_i} \) is the assigned target for fiber \( F \)  and \( T_{f_i} \) is the list of available targets for \( F \).  
Otherwise, if multiple targets are available, a random target is chosen from the available list:

\begin{equation}
A_{f_i} = T_{f_i}[k], \quad k \in \{1, \dots, |T_{f_i}|\}
\end{equation}
where \( k \) is a randomly selected index from the list of available targets. The selected target is then added to the set of allocated targets:  
\begin{equation}
T_{\text{allocated}} = T_{\text{allocated}} \cup \{A_{f_i}\}
\end{equation}

\noindent After assigning
 a target, the list of remaining targets is then updated to exclude assigned targets and those violating the exclusion constraint (see below):
\begin{equation}
T_j' = \{ t \in T_j \mid t \notin T_{\text{allocated}}, \quad d(A_{f_i}, t) > R_{\text{exclusion}} \}
\end{equation}
For all fibers \( F_j \), where \( d(A_{f_i}, t) \) is the Euclidean distance between the newly assigned target \( A_{f_i} \) and any other potential target \( t \), and \( R_{\text{exclusion}} \) is the ER. 
This step ensures that fibers are assigned efficiently while preventing conflicts with previously allocated targets.

\subsubsection{\bf Enforcing exclusion constraints}
Once potential targets have been identified for each fiber, we need to ensure that selected targets do not violate proximity constraints. This step prevents fibers from being assigned to targets that are too close to each other. As each fiber has a newly considered target identified, the distance to all previously assigned targets is computed. If the separation falls within the specified ER, the target is marked as too close and an alternative target is randomly selected from those remaining. If no remaining targets fall beyond the exclusion radii of other assigned fibers, that fiber remains unallocated.

This ensures that no two assigned targets violate the minimum required separation. By enforcing this constraint, the process ensures that fibers are not assigned to overlapping or closely spaced targets, maintaining valid spacings for a given set of fiber allocations.

Therefore, for each assigned target \( t_j \), all targets \( t_k \) are removed that satisfy the following conditions:  
\[
d(t_j, t_k) \leq r_e, \quad t_k \neq t_j
\]  
and the list of available targets \( T \) are updated.

\subsubsection{\bf Removing allocated targets} 
After assigning fibers to targets, it is necessary to update the list of available targets by removing those that have already been allocated. This ensures that future assignment steps, mimicking repeated visits to a target field, do not repeat already assigned targets.

The list of potential targets for each fiber is filtered to exclude allocated targets.
For a given fiber \( F_i \), the updated list of available targets is given by:  
\begin{equation}
T_i' = \{ t \in T_i \mid t \notin T_{\text{allocated}} \}
\end{equation}
where \( T_{f_i} \) is the original list of potential targets for fiber \( F \), \( T_{\text{allocated}} \) is the set of all assigned targets, and \( T_i' \) is the new list of available targets for fiber \( F_i \), excluding assigned targets.  

A global list of remaining unallocated targets is also updated to reflect the removal of those assigned. The overall set of remaining targets, excluding those already allocated, is:  
\begin{equation}
T_{\text{remaining}} = \{ t \in T_{\text{all}} \mid t \notin T_{\text{allocated}} \}
\end{equation}
where \( T_{\text{all}} \) is the complete set of all targets before allocation. This ensures that only unassigned targets remain available for further fiber assignments, preventing duplicate allocations. This list is used in evaluating the completeness over a series of visits, detailed below.

\subsection{\bf Evaluating completeness and efficiency} 
After filtering out allocated targets, the entire fiber allocation process is repeated over multiple visits until some desired completeness goal is achieved, or some maximum number of visits is reached. In our analysis we define the completeness goal as 95\% for all tests. The maximum number of visits is set to either 5 or 10, detailed in the specific results below. This latter requirement is simply imposed to avoid endless loops.

Completeness measures the cumulative total of all allocated targets as a fraction of the total number of input targets. The completeness value is updated after each visit, and the iteration of fiber allocation ceases once the completeness goal is achieved.
The cumulative completeness \( C_v \) after visit \( v \) is given by:
\begin{equation}
C_v = \frac{|T_{\text{allocated}}|}{|T_{\text{all}}|}
\end{equation}
where \( |T_{\text{allocated}}| \) is the cumulative total number of targets assigned after visit \( v \), and \( |T_{\text{all}}| \) is the total number of input targets.
 
Allocation efficiency evaluates how well fibers are assigned to targets in each visit. Efficiency is not a cumulative metric, but a direct measure of the fraction of fibers allocated to targets. Accordingly, the efficiency is expected to decrease over subsequent visits as the available number of targets declines. The efficiency \( E_v \) for visit \( v \) is calculated as the fraction of fibers that have a target assigned:
\begin{equation}
E_v = \frac{|F_{\text{assigned}}|}{|F|}
\end{equation}
where \( |F_{\text{assigned}}| \) is the number of fibers that have a target assigned in visit \( v \), and \( |F| \) is the total number of fibers available.

The process is repeated for multiple visits, with completeness and efficiency updated after each visit. These metrics form the basis of our analysis in evaluating how well different sets of input specification perform compared to each other.

\begin{table*}[ht!]
    \centering
    \caption{Fiber configuration parameters for different fiber pitch values, including fiber density, total number of fibers, PR, ER, target number, and target density, within a $0.5^\circ$ diameter (0.196 deg$^2$) FoV. The physical parameters assume a plate scale of 13.751\,arcsec/mm, representative of a telescope with a 15\,m focal length.}
    \begin{tabular}{ccccccccc}
        \toprule
        \textbf{Pitch} & \textbf{PR} & \textbf{ER} & \textbf{Fiber Density} & \textbf{Fiber Number} \\
        arcmin (mm) & arcmin (mm) & arcmin (mm) & (fibers/deg$^2$) & &  \\
        \midrule \midrule
                               &   2.64 (11.5) & 2.9 (12.8) &    \\
                               &  2.64 (11.5) & 2.64 (11.5) &        \\
                               &   2.64 (11.5) & 2.38 (10.4) &       \\
                               &   2.4 (10.5) & 2.9 (12.8) &       \\
                    1.8 (7.9)  &  2.4 (10.5) & 2.64 (11.5)  &  1282 & 251 \\
                               &   2.4 (10.5)  & 2.38 (10.4) &       \\
                               &   2.16 (9.45) & 2.9 (12.8) &       \\
                               &   2.16 (9.45)  & 2.64 (11.5) &      \\
                               &   2.16 9.45) & 2.38 (10.4) &      \\
           
        \midrule
                               &  2.64 (11.5) & 2.9 (12.8) &  \\
                               &   2.64 (11.5) & 2.64 (11.5) &        \\
                               &   2.64 (11.5) & 2.38 (10.4)  &        \\
                               &   2.4 (10.5) & 2.9 (12.8) &        \\
                2.4 (10.5)     & 2.4 (10.5) & 2.64 (11.5)  & 722 & 142 &   \\
                               &   2.4 (10.5) & 2.38 (10.4)  &        \\
                               &   2.16 (9.45) & 2.9 (12.8) &        \\
                               &   2.16 (9.45) & 2.64 (11.5)  &        \\
                               &   2.16 (9.45) & 2.38 (10.4) &        \\
        \midrule
        
                               & 2.64 (11.5) & 2.9 (12.8) &   \\
                               &  2.64 (11.5) & 2.64 (11.5)  &        \\
                               &  2.64 (11.5) & 2.38 (10.4)  &        \\
                               &  2.4 (10.5) & 2.9 (12.8) &        \\
                       3.0 (13.1) & 2.4 (10.5)  & 2.64 (11.5)  &   465 & 92 &   \\
                               &   2.4 (10.5) & 2.38 (10.4)  &        \\
                               &   2.16 (9.45) & 2.9  (12.8) &        \\
                               &   2.16 (9.45) & 2.64 (11.5)  &        \\
                               &   2.16 (9.45) & 2.38 (10.4)  &  \\                          
        \bottomrule
    \end{tabular}
     \label{tab:parameter_combination}
\end{table*}

\begin{table*}[ht!]
    \centering
    \caption{Target numbers used in testing each combination of system parameters. The target density adopts our nominal $0.5^\circ$ diameter (0.196 deg$^2$) FoV.}
    \begin{tabular}{ccccccccc}
        \toprule
        \textbf{Target Number} & \textbf{Target Density} \\
         &(targets/deg$^2$)\\
        \midrule \midrule
       100 & 510 \\
       150 & 765 \\
       200 & 1020 \\
       250 & 1275 \\
       300 & 1530 \\
       400 & 2040 \\
       450 & 2295 \\
       500 & 2551 \\
       600 & 3061 \\
       750 & 3826 \\
       1000 & 5102 \\
       
        \bottomrule
    \end{tabular}
     \label{tab:target_combination}
\end{table*}

\section{Results and Analysis}
\label{sec:results}

\subsection{Set-up of the problem}
\label{subsec:problem_setup}

The performance of a MOS system depends on the specifications of its fiber positioning system. The efficiency of the system is fundamentally determined by its fiber density, which dictates how many targets can be observed simultaneously. The ability to efficiently allocate fibers to targets in the system also depends on the additional key design parameters of PR (for parallel positioners) and ER. These are important in defining the available movement range and spacing limitations for each fiber. For a given fiber density, a larger PR improves flexibility in target assignment, reducing conflicts and maximizing completeness. Conversely, a larger ER restricts fiber positioning, reducing the efficiency of a system.

The central problem in this analysis is to quantify how these three parameters interact to influence the efficiency of a given system in target allocation. Specifically, we adopt a fixed FoV and a series of nominated target densities, and then, for each set of technical specifications, we quantify the number of visits required to achieve a specific completeness goal, in order to assess the performance of different specification choices.

By systematically exploring these dependencies, our analysis provides insights into the trade-offs between fiber density, PR, and ER, to understand which parameters provide the biggest observational impact. Understanding these relationships will help to establish a set of optimal fiber positioner parameters to maximize the observational efficiency given a desired target density or broad scientific goal. The selection of our parameter ranges was adopted from existing system specifications of several key facilities \citep[e.g.,][]{2012SPIE.8446E..0PD, 2014SPIE.9150E..23S, 2014SPIE.9151E..1XS, kuehn2014taipan, staszak2016taipan, 2022SPIE12184E..6NB,2022SPIE12184E..6MB, 10.1117/12.3019907}.

\begin{figure*}
\centering
\includegraphics[width=1\textwidth]{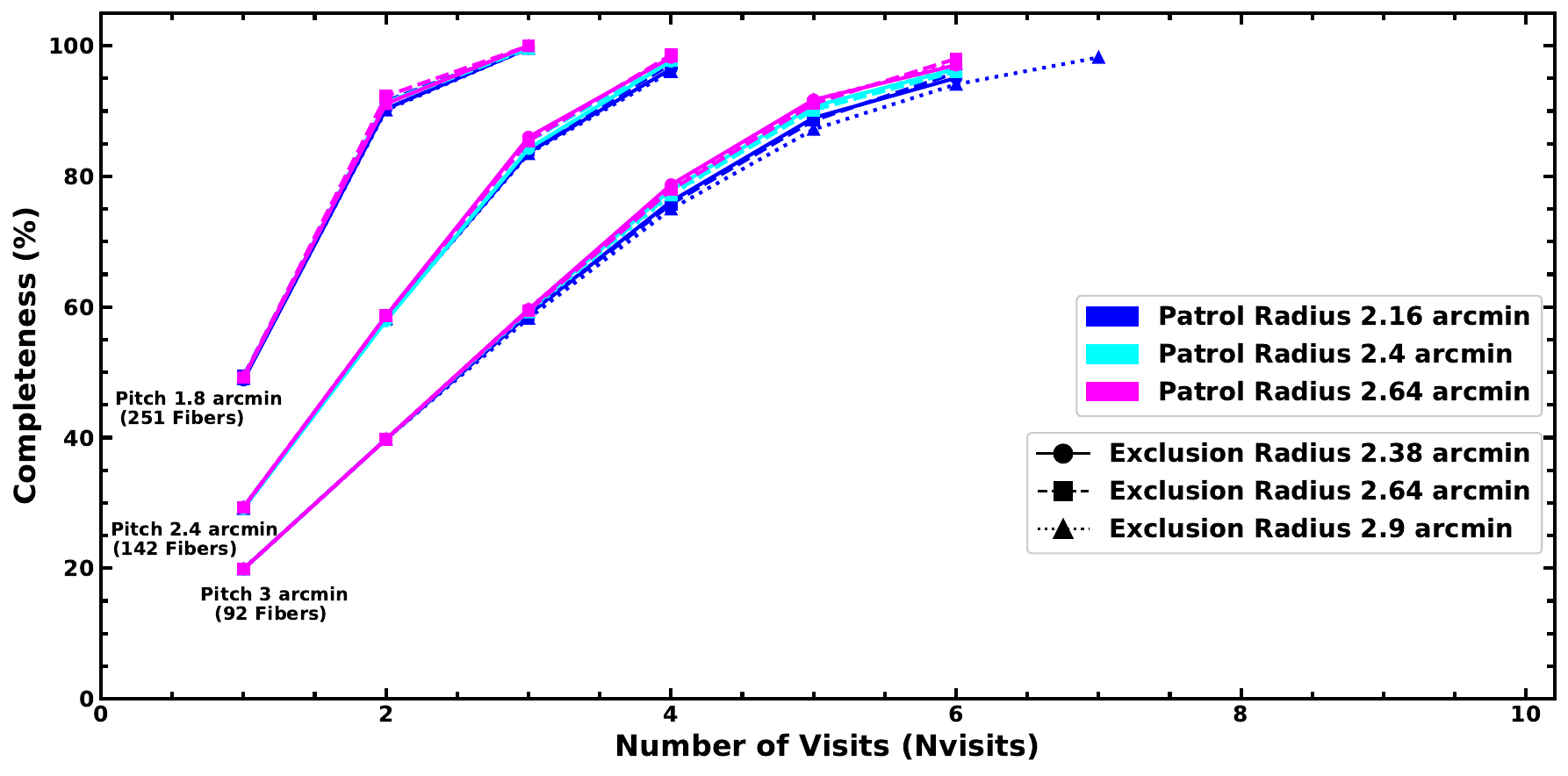}
\includegraphics[width=1\textwidth]{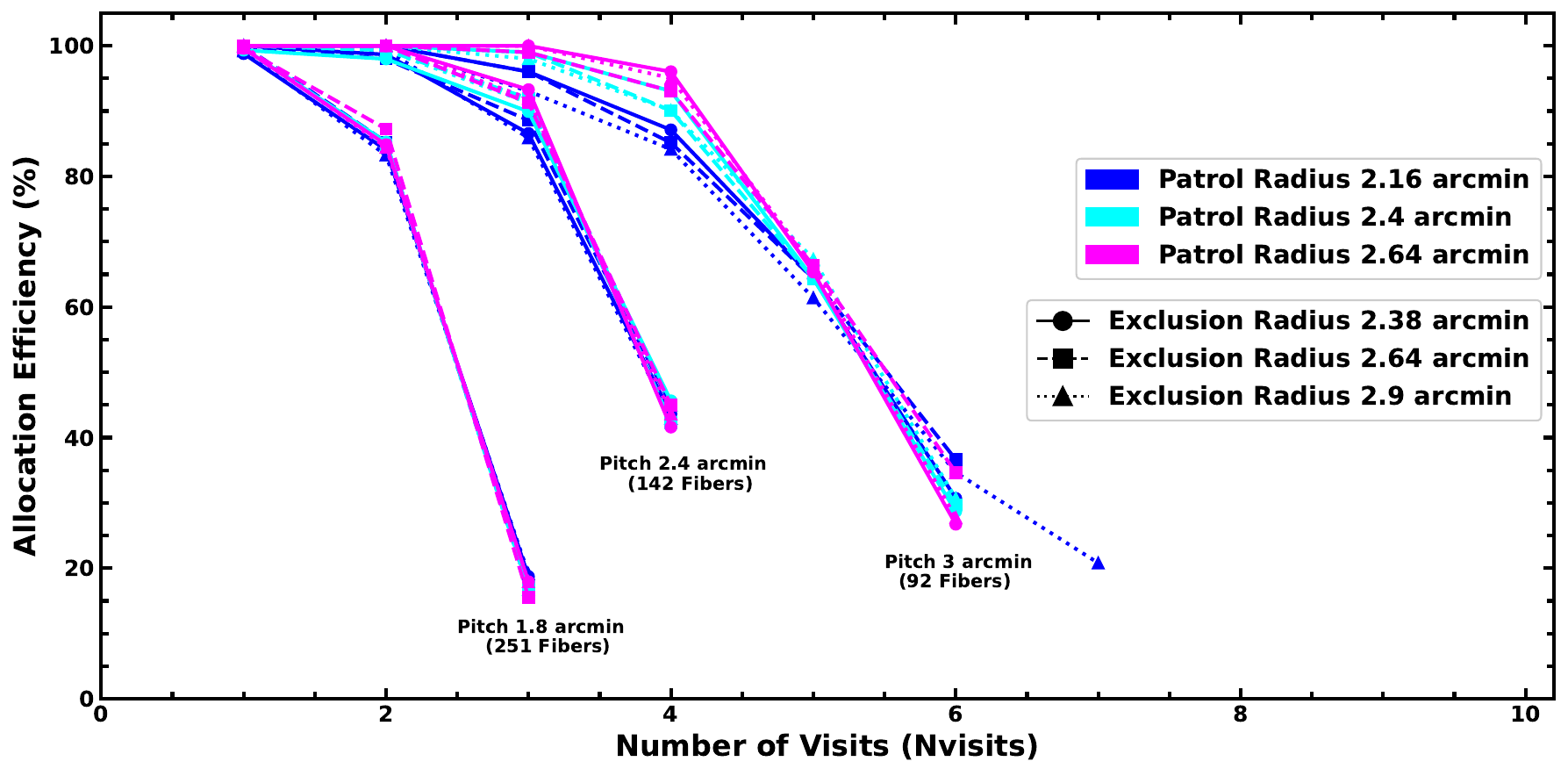}
\caption{The performance of three different fiber pitches with varying patrol and exclusion radii within a $0.5^\circ$ diameter (0.196 deg$^2$) FoV. The top panel displays the completeness as a function of the number of visits (\( N_{\text{visits}} \)) for  2551 targets/deg$^2$ (500 targets). The bottom panel shows the efficiency as a function of \( N_{\text{visits}} \) for the same target density. There are three groupings of results, which correspond to the chosen fiber pitch. The results showing the best performance (highest completeness for the fewest number of visits) come from the simulations with the smallest pitch (largest number of fibers). The colors and the symbols correspond to PR. The choice of line styles is to aid visibility only, without any link to the parameter values. The numbers against each parameter in the key are given in arcsec. It is clear that the primary factor in delivering the best performance is the smallest fiber pitch (largest fiber density or number). The PR and ER have a secondary impact when fiber pitch is fixed.}
\label{fig:comp_eff}
\end{figure*}

\addtocounter{figure}{-1}
\begin{figure*}
\centering
\includegraphics[width=1\textwidth]{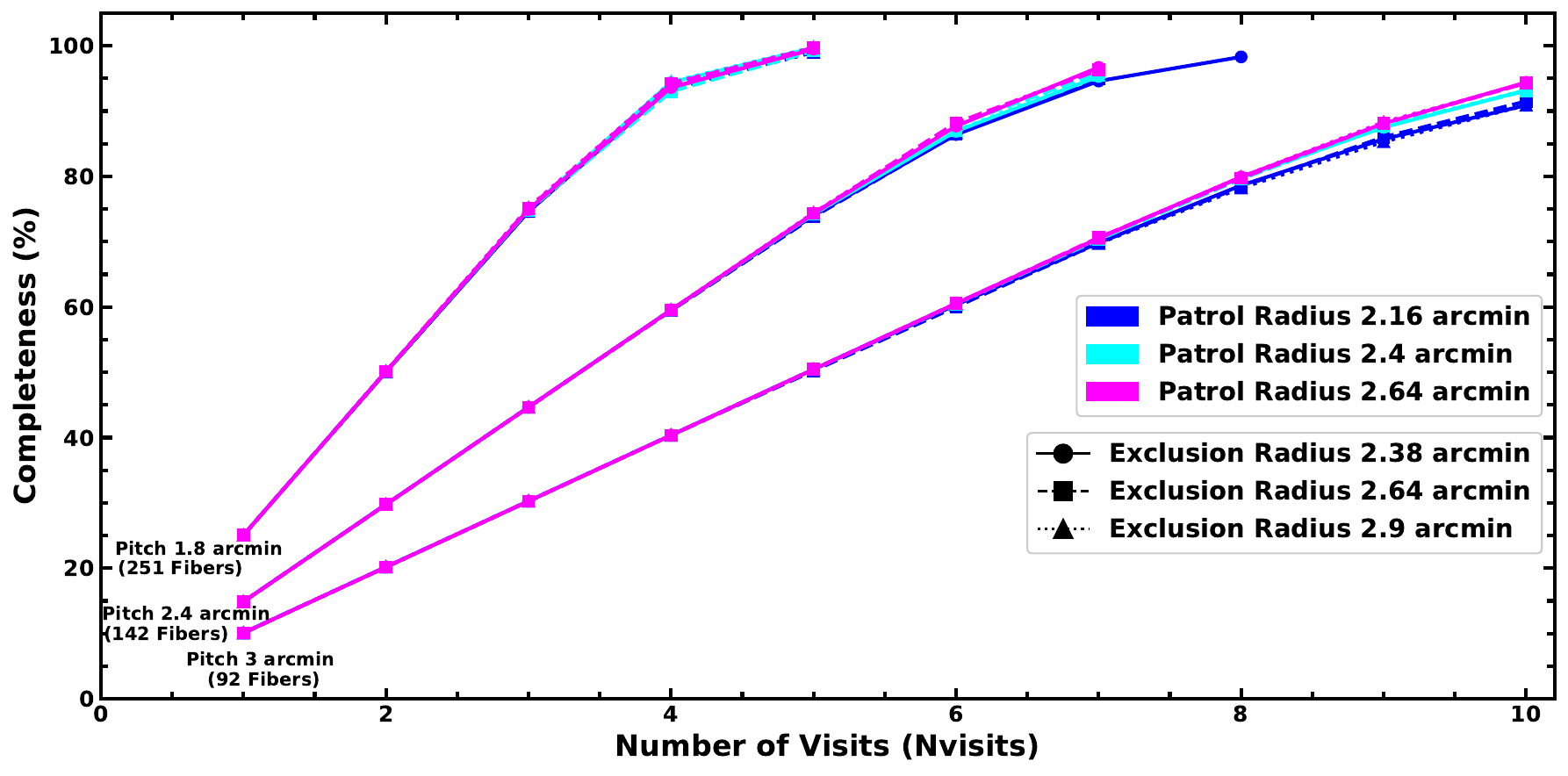}
\includegraphics[width=1\textwidth]{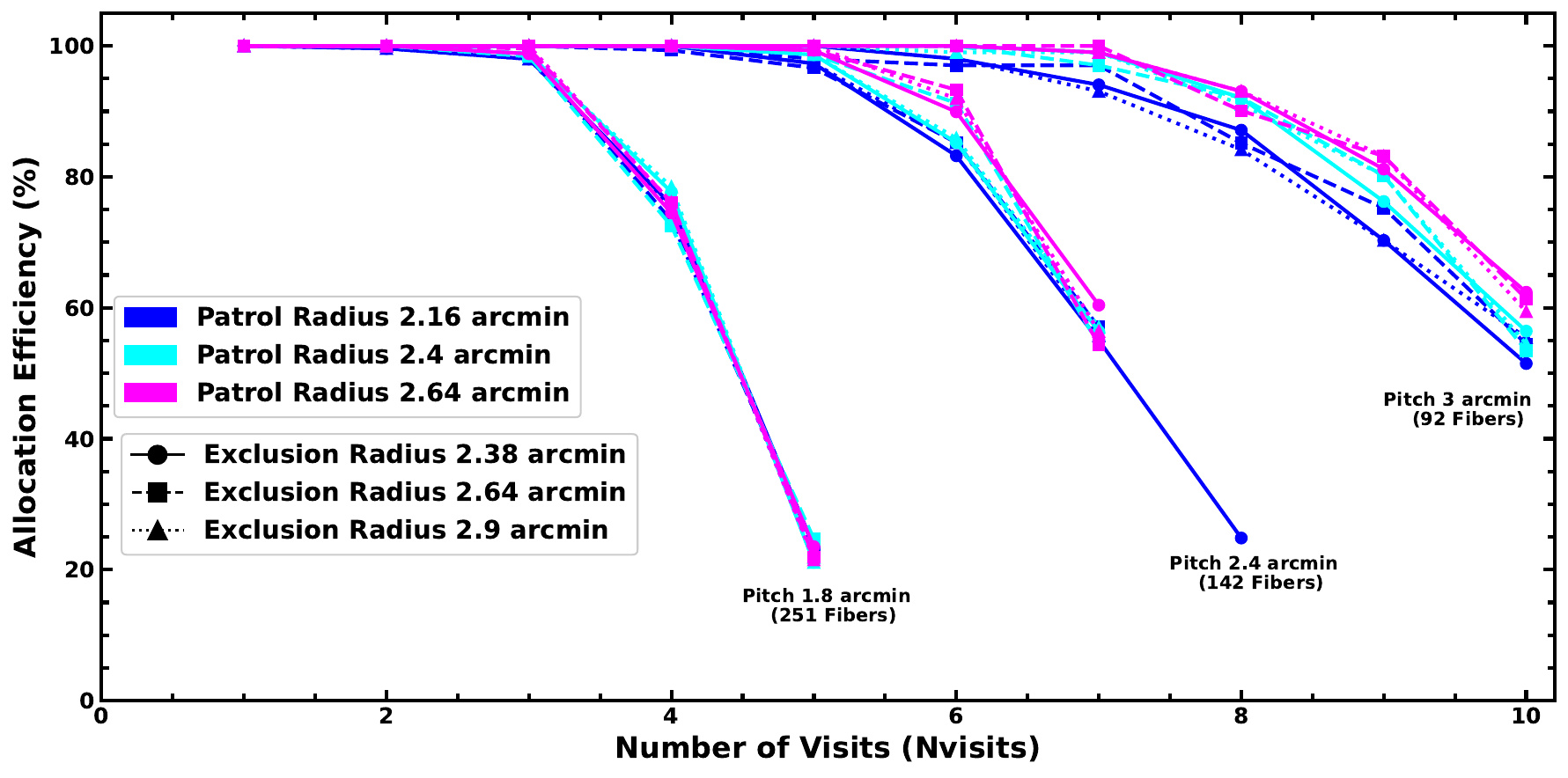}
\caption{Continued, here showing results for 5102 targets/deg$^2$ (1000 targets).}
\label{fig:comp_eff_1000}
\end{figure*}

\subsection{Analyzing the primary simulation}
\label{subsec:main_simulation}
Our model is intended as a general analysis, independent of the specific properties of any given telescope, and accordingly we choose to work in angular (sky) coordinates throughout. The value in this approach is that it links the technical specifications directly to the astronomical target surface densities. It also allows for an analysis that spans a pragmatic range of science-driven parameters without being limited to or driven by the practical considerations of any particular technology or facility. These can of course subsequently be converted to physical size specifications by nominating the plate scale appropriate to a particular telescope.

To systematically evaluate the impact of a set of positioner specifications we iterate through a grid of parameter values, as summarized in Table~\ref{tab:parameter_combination}. To provide some physical context for these angular parameters, we also present their values when adopting 
a plate scale of 13.751 arcsec/mm, typical for a 15 m focal length telescope. Throughout, for computational convenience, we work with a nominal $0.5^\circ$ diameter (0.196 deg$^2$) FoV unless otherwise noted.

In Table~\ref{tab:parameter_combination} the pitch values are grouped together, making it clear that each PR and ER combination is tested under each pitch. The PR and ER iterate over all these values for each pitch value, sampling all combinations of these parameter values. To avoid redundancy, fiber density and fiber numbers are listed only once per pitch group in the table.

We test each combination of system parameters through a mock observation of different numbers of targets (Table~\ref{tab:target_combination}). For each choice of target density (i.e., number) we test each combination of all possible system parameters. Said another way, for each row in Table~\ref{tab:target_combination} we test each combination of parameters from Table~\ref{tab:parameter_combination}. This structured approach ensures a clear assessment of how each parameter influences variations in fiber allocation for a given target density.

Figure~\ref{fig:comp_eff} illustrates the results of this process. We show the completeness (top panels) and efficiency (bottom panels) as a function of the number of visits (\( N_{\text{visits}} \)) for all tested parameter combinations. For clarity, and to avoid needless repetition, we show just two of the choices in target number, from the simulations using 2551 targets/deg$^2$ (500 targets) and 5102 targets/deg$^2$ (1000 targets).

The first obvious trend to see in this Figure is that there are three clear groupings in the set of results. These groupings correspond to the choice of fiber pitch. The combinations with the smallest fiber pitch (largest fiber density, or number) perform the best, regardless of subsequent choice of PR or ER. This means that, unsurprisingly, the most important parameter in any RFP system is simply the fiber density. The more fibers available, the better the performance will be, for any reasonable choice of other parameters.
A clear general result across all combinations is that, unsurprisingly, the completeness increases through each subsequent visit, while the efficiency decreases. Allocation efficiency starts near 100\% for the first visit and gradually decreases due to the reduced availability of unallocated targets, which are often in regions with higher fiber competition, making efficient allocation more difficult.
Smaller pitches (larger fiber numbers) tend to show lower efficiency at later visits compared to larger pitches, but this is simply a consequence of having higher completeness at earlier times. Larger pitches have fewer total fibers, and so are not as efficient in achieving the completeness goal as quickly, and with greater numbers of targets remaining in later visits, the efficiency stays higher longer. The differences between pitch groups underscores the importance of pitch as the primary parameter in determining the performance of an RFP.

To assess the impact of the other parameters, PR and ER, the figure employs different colors, and markers to denote PR. Blue (circles) represents the smallest PR and magenta (triangles) the largest. Focusing first on the completeness (top panels in Figure~\ref{fig:comp_eff}), we can see that the better performance is provided by the largest PR (magenta lines) in contrast to the smallest PR (blue lines) for each fixed value of fiber pitch. This reinforces the intuitive sense that a larger PR allows fibers to reach more targets, enabling improved completeness.

Moving now to the efficiency (bottom panels of Figure~\ref{fig:comp_eff}) we can see the effect of ER. Arrangements with larger ER tend to show poorer fiber allocation efficiency for a given value of pitch and PR. Again, this is consistent with the intuition that larger ER values limit the capacity of an RFP to maximize target allocations, and smaller ER values can perform better. This effect also holds in the completeness results, but it is less apparent visually in the figure than it is for the efficiency results. This reinforces the conclusion that PR is a more important parameter than ER for RFPs. 
We have also tested the effect of adopting ER=0 to assess whether or not our adopted ER parameters have been too conservative. The results are marginal improvements over the smallest ER we otherwise chose. Our chosen values of ER are not substantially biasing our tests away from realistic system specifications.

Overall, this figure illustrates the interplay between fiber pitch, PR, and ER in determining allocation efficiency. The best performance comes from the smallest pitch, independent of PR and ER. With a fixed choice of pitch, increasing the PR is the next most important step to enhance performance, while minimizing ER is a secondary step in improving effectiveness. These results provide the fundamental conclusion that RFP systems maximize observational efficiency best for multi-object spectroscopic surveys by simply having more fibers (smaller fiber pitch). Then, for a given fiber pitch, the next most important parameter is PR (for parallel-positioner systems), with ER being the least important.

\subsection{Evaluating Existing Fiber Positioning Systems}
\label{subsec:evaluation}

We now compare the performance of a selection of existing fiber positioning systems. Each set of parameters is chosen to mimic those of real RFP systems, in order to understand how these constraints affect overall performance, independent of the underlying technology being tested. Our focus is on how specific design specifications influence system performance rather than comparing the technological mechanisms of the positioners themselves. In this part of the analysis we limit the simulations to only five visits, as this was sufficient for the majority of the technologies modeled to achieve a nominal 95\% completeness.

\begin{table*}
\centering
\caption{Comparison of existing fiber positioner system parameters including spines, dual-rotator, pick-and-place and starbugs in terms of fiber pitch, fiber density, fiber number, PR, and ER. This comparison considers a $1.0^\circ$ diameter (0.785 deg$^2$) FoV for all systems, except for the FLEX. Parameters are evaluated with 500 and 1000 targets across all systems. For FLEX, either a $0.3^\circ$ diameter (0.077 deg$^2$) FoV or a $0.5^\circ$ diameter (0.196 deg$^2$) FoV is adopted. The physical parameters assume a plate scale of 13.751\,arcsec/mm, representative of a telescope with a 15\,m focal length.}
\begin{tabular}{@{}lccccccccc@{}}
\toprule
\textbf{Positioner} &  \textbf{Fiber Pitch} & \textbf{PR} & \textbf{ER} & \textbf{Fiber Density} & \textbf{Fiber Number}  \\
\textbf{System} & arcmin (mm)  & arcmin (mm) & arcmin (mm) & (fibers/deg$^2$)  \\

\midrule \midrule
Dual-rotator & 2.4 (10.5) & 1.38 (6) & 0.519 (2.25) & 722  & 500   \\            
       
Tilting Spines & 2.4 (10.5) & 2.4 (10.5) & 0.144 (0.625) & 722 & 500  \\

Starbugs & 4.58 (20) & 4.58 (20)  & 0.97 (4.25) & 198  & 155   \\

Pick-and-place   &   &  & 0.85 (3.7)  & 305 & 240   \\

FLEX                  & 1.6 (7) & 3.44 (15) & 0.344 (1.5) & 6452  & 500, 1000   \\
   
\bottomrule
\end{tabular}
\label{tab:section_5_3}
\end{table*}

Table~\ref{tab:section_5_3} shows the comparison of these system parameters in terms of fiber pitch, PR, ER, fiber density, and fiber number within a $1.0^\circ$ diameter (0.785 deg$^2$) FoV, except for the FLEX system. The reason FLEX is an exception is that its very small pitch means a very large number of fibers, up to a factor of ten to twenty in fiber density compared to the other systems. The current proof-of-concept code becomes very slow if we adopt the same FoV for this system simply due to the sheer number of fibers involved. Accordingly, we adopt a smaller FoV in our tests with FLEX, and interpret those accordingly in the analysis below. We choose a number of 500 and 1000 targets for the test of all the systems. For FLEX, either a $0.3^\circ$ diameter (0.077 deg$^2$) FoV or a $0.5^\circ$ diameter (0.196 deg$^2$) FoV is adopted, as explained further below.

To maintain the simplicity of the model and minimise computational effort, our analysis of dual-rotator configurations is limited to effectively simulating equal-arm or minimally overlapping dual-rotator designs. This is similar to technologies used in systems such as DESI, LAMOST, and PFS \citep[e.g.,][]{2012RAA....12.1197C, 2014SPIE.9151E..1YF}. This constraint comes from our assumption of a simple exclusion radius to model the collision avoidance of such systems, which in practice is much more complex \citep[e.g.,][]{2018MNRAS.481.3070H}. Our approach is not well matched to newer, heavily overlapping unequal arm systems, such as SDSS-V, MOONS, or Via \citep[e.g.,][]{2026AJ....171...52K, 10.1117/12.2629239, 2026arXiv260618332T}. In those dense arrangements, positioners can physically collide along their primary or secondary arm segments even when their fiber tips are far apart, making collision safety significantly more complicated to model \citep{10.1117/12.2629239}. Fully modelling the effects of these technologies would require much more sophisticated collision avoidance algorithms, which are beyond the scope of the current investigation. Given the focus of this work, we leave such modeling to future layout optimization studies.

\subsubsection{Fixed FoV}
\label{subsubsec:fixed_fov}

Figure~\ref{fig:1deg_fov} shows completeness and efficiency for two scenarios with a $1.0^\circ$ diameter (0.785 deg$^2$) FoV, with a choice of 500 targets (upper panels) or 1000 targets (lower panels). FLEX is the exception here, as out of computational necessity we adopt a $0.5^\circ$ diameter (0.196 deg$^2$) FoV. Despite this we retain the same target numbers (500 and 1000) even though this implies a larger target density. Despite this mismatch, as we will see below, the sheer numbers of fibers available to FLEX ensure that its performance remains the best.

The results shown in Figure~\ref{fig:1deg_fov} are an average taken from four simulations, for each technology and target number scenario, using four independent regions selected from within the GAMA (G02, G09, G12, G15, and G23) fields. The uncertainties shown come from the range of these values. The fact that we adopt a fixed FoV in this part of the analysis serves to emphasize the effect of different fiber pitch (or fiber density or number) primarily, between the different technologies. From the results above, we would expect that the technology with the largest fiber number, in this test, would perform the best. We refer below to fiber number only, for brevity, given that we are discussing a fixed FoV.

The starbugs (cyan) show the lowest completeness for any single visit, in either scenario. This is due to its larger pitch (4.58 arcmin), and smallest number of fibers (155). Although starbugs have the largest PR (4.58 arcmin), their large ER (0.97 arcmin) also limits how closely fibers can be packed together.
The pick-and-place system (blue) has the next highest fiber number (240). Its completeness starts off relatively low in consequence, but within a reasonable number of visits (3 or 5 in the scenarios tested here) it reaches nearly 100\%. This is a consequence of not being restricted by a patrol radius, as any available fibers can be allocated to any remaining targets in subsequent visits. This also means that the efficiency of target allocation remains high at essentially 100\% until the final visit. An important aspect of pick-and-place RFPs, though, is that they are not parallel positioning systems. The sequential nature of the fiber allocation may reduce overall system efficiency due to large reconfiguration times, which are not modeled here. We discuss this further in \S\,\ref{sec:discussion} below.

The dual-rotators (red) and tilting spines share the same pitch and consequent fiber number (500) in this model. The dual-rotators show lower completeness at any visit compared to tilting spines (magenta), for either scenario in this test. This is a direct consequence of tilting spines having a larger PR than dual-rotators, and hence being able to reach more targets in any given visit, ensuring higher allocation efficiency, and better completeness.
The FLEX system (black), despite our simulation using a target density which is twice as high as any other modeled system (due to our smaller adopted FoV), achieves the highest completeness the most quickly. It surpasses our nominal 95\% goal with just one to two visits in either tested scenario. This is a direct consequence of having the largest number of fibers (1000).

As seen in the general case above, the efficiency declines more slowly for those systems with fewer fibers, as a consequence of a slower growth in completeness leaving more targets available to allocate in later visits. As noted above, the pick-and-place and starbug systems are highly efficient, although this is a reflection of their slower improvement in completeness compared to other technologies.

\begin{figure*}
\centering
\includegraphics[width=1\textwidth]{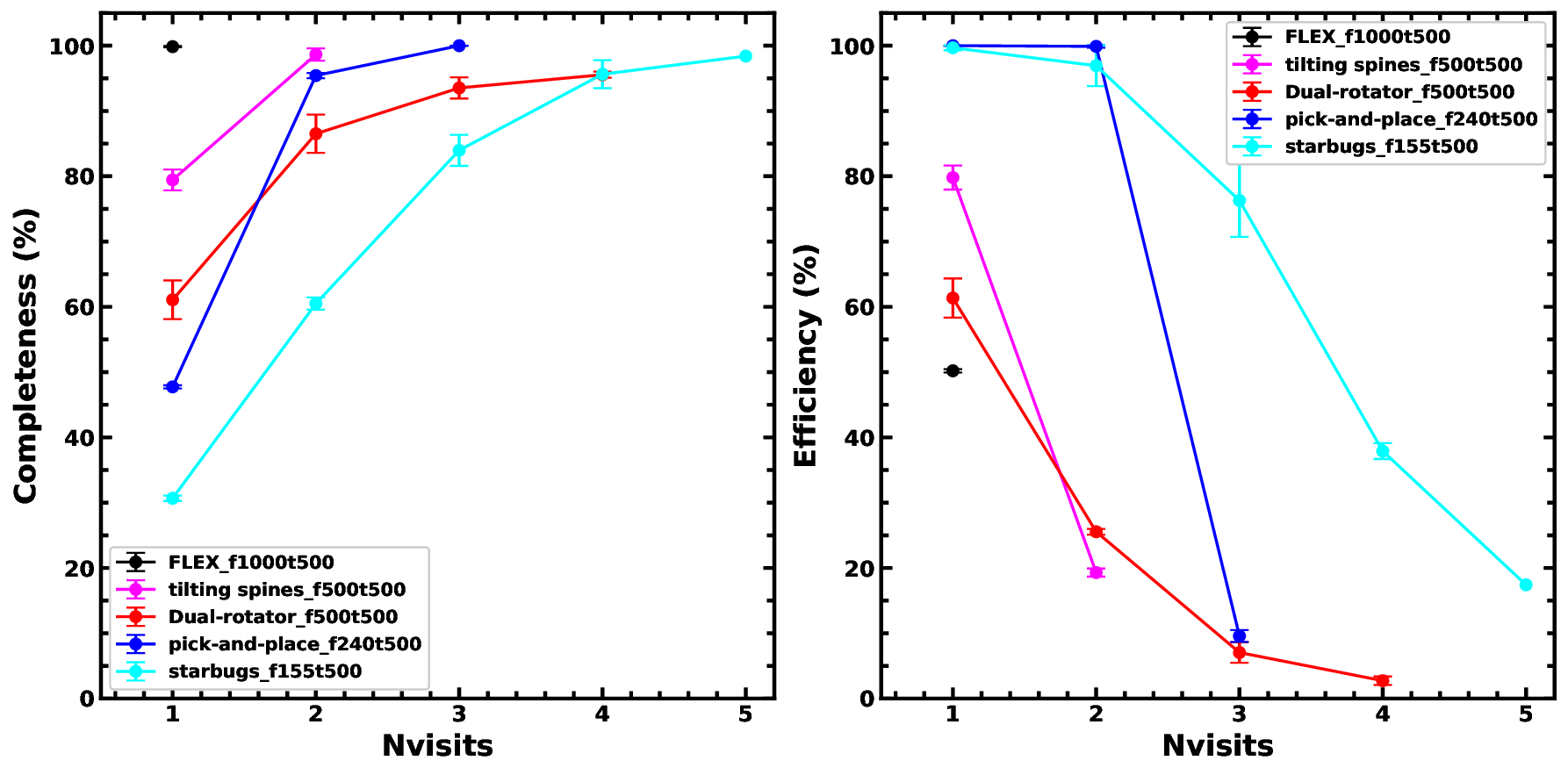}
\includegraphics[width=1\textwidth]{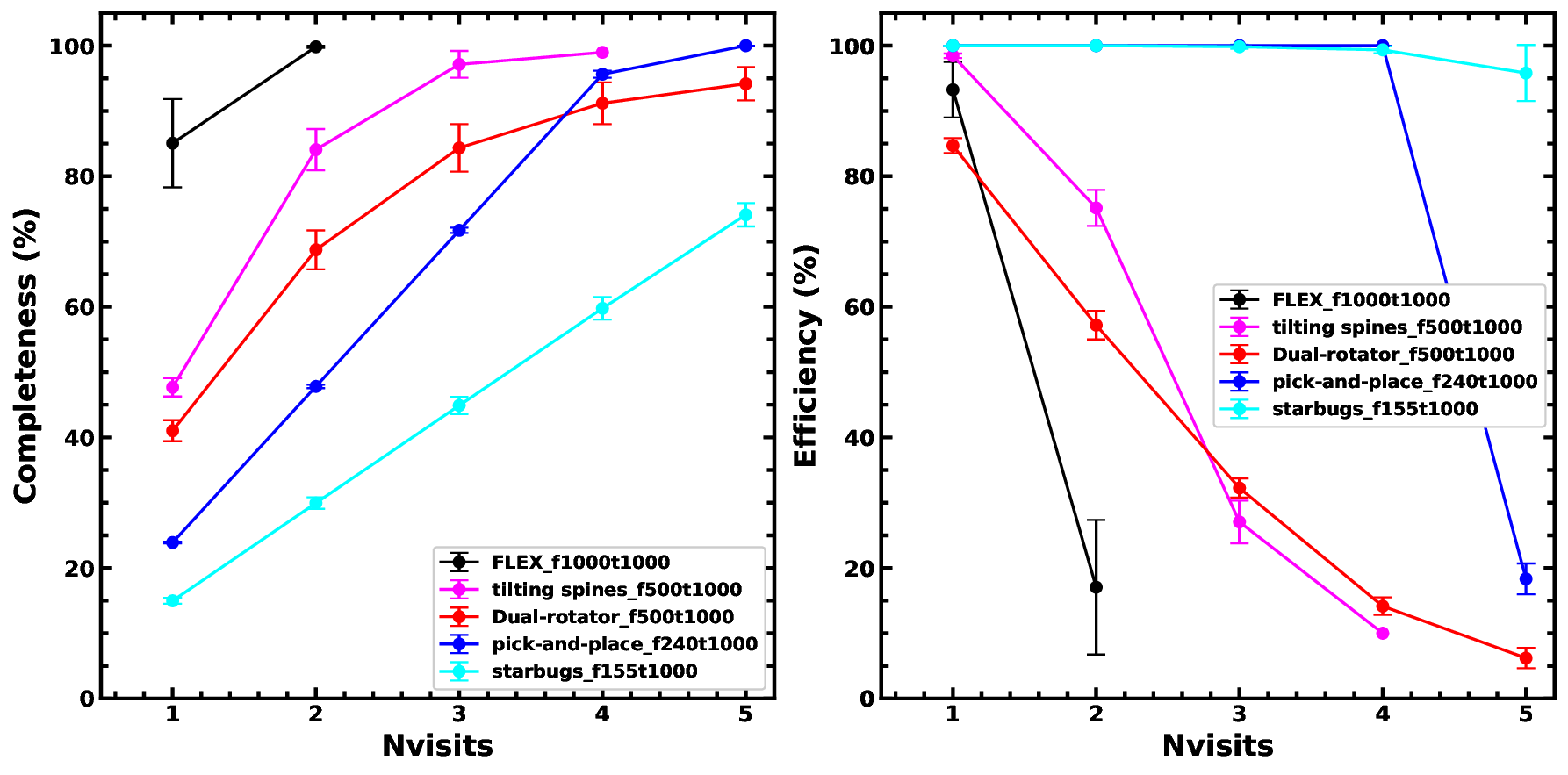}
\caption{Completeness and efficiency as a function of the number of visits (\( N_{\text{visits}} \)) for different fiber positioning systems (tilting spines, dual-rotator, pick-and-place and starbugs) within a $1.0^\circ$ diameter (0.785 deg$^2$) FoV, except for the FLEX system. For FLEX, a $0.5^\circ$ diameter (0.196 deg$^2$) FoV is adopted. 
The left panel in each row shows completeness (\%), while the right panel shows efficiency (\%). Top panel corresponds to 500 targets and bottom panel corresponds to 1000 targets, respectively for different fiber densities. As \( N_{\text{visits}} \) increases, completeness improves, approaching 100\%, while efficiency decreases due to fiber reallocation constraints. The trends vary across different positioning systems, reflecting their relative performance in fiber assignment for different target densities.}
\label{fig:1deg_fov}
\end{figure*}

\begin{figure*}
\centering
\includegraphics[width=1\textwidth]{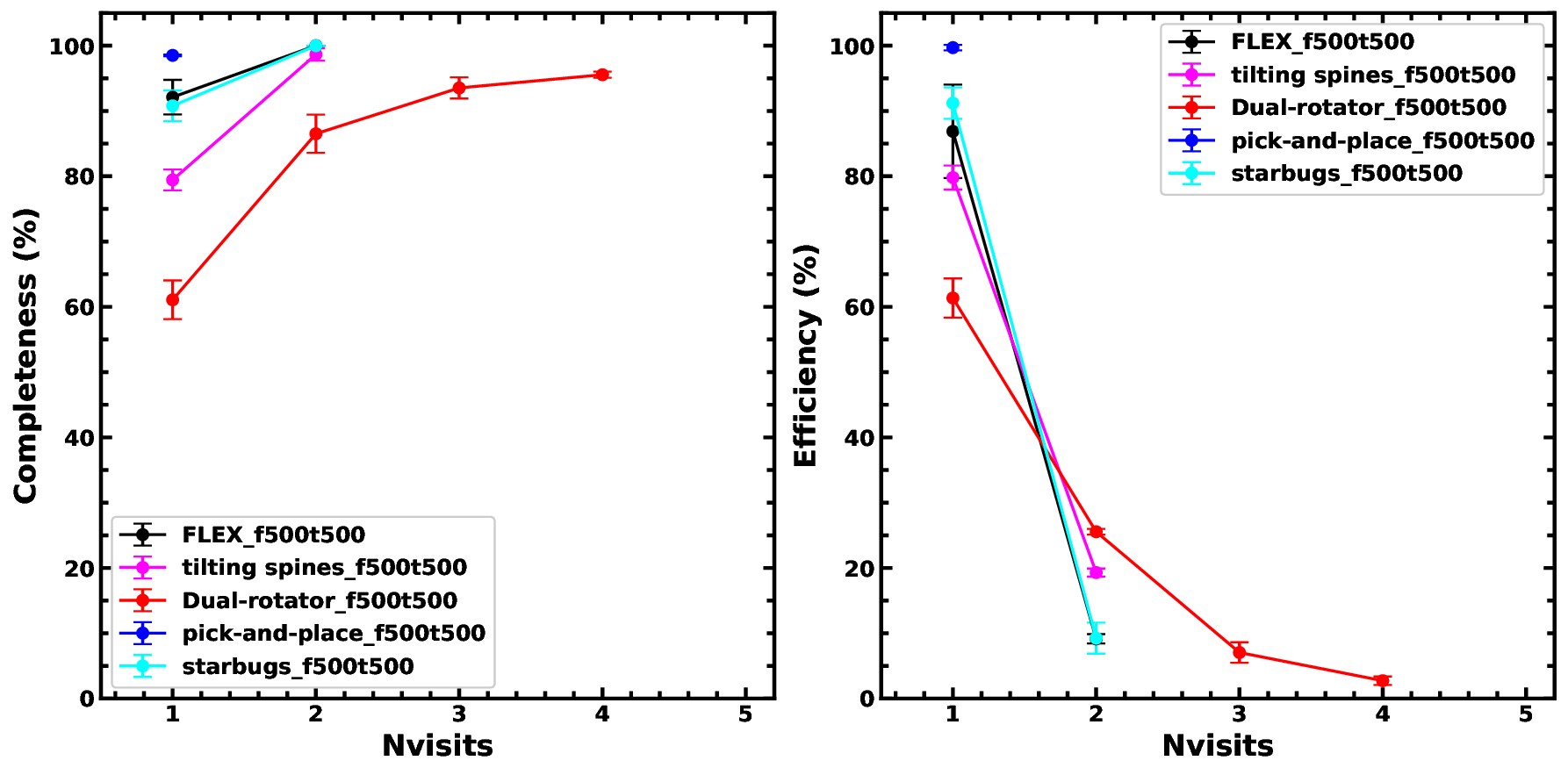}
\includegraphics[width=1\textwidth]{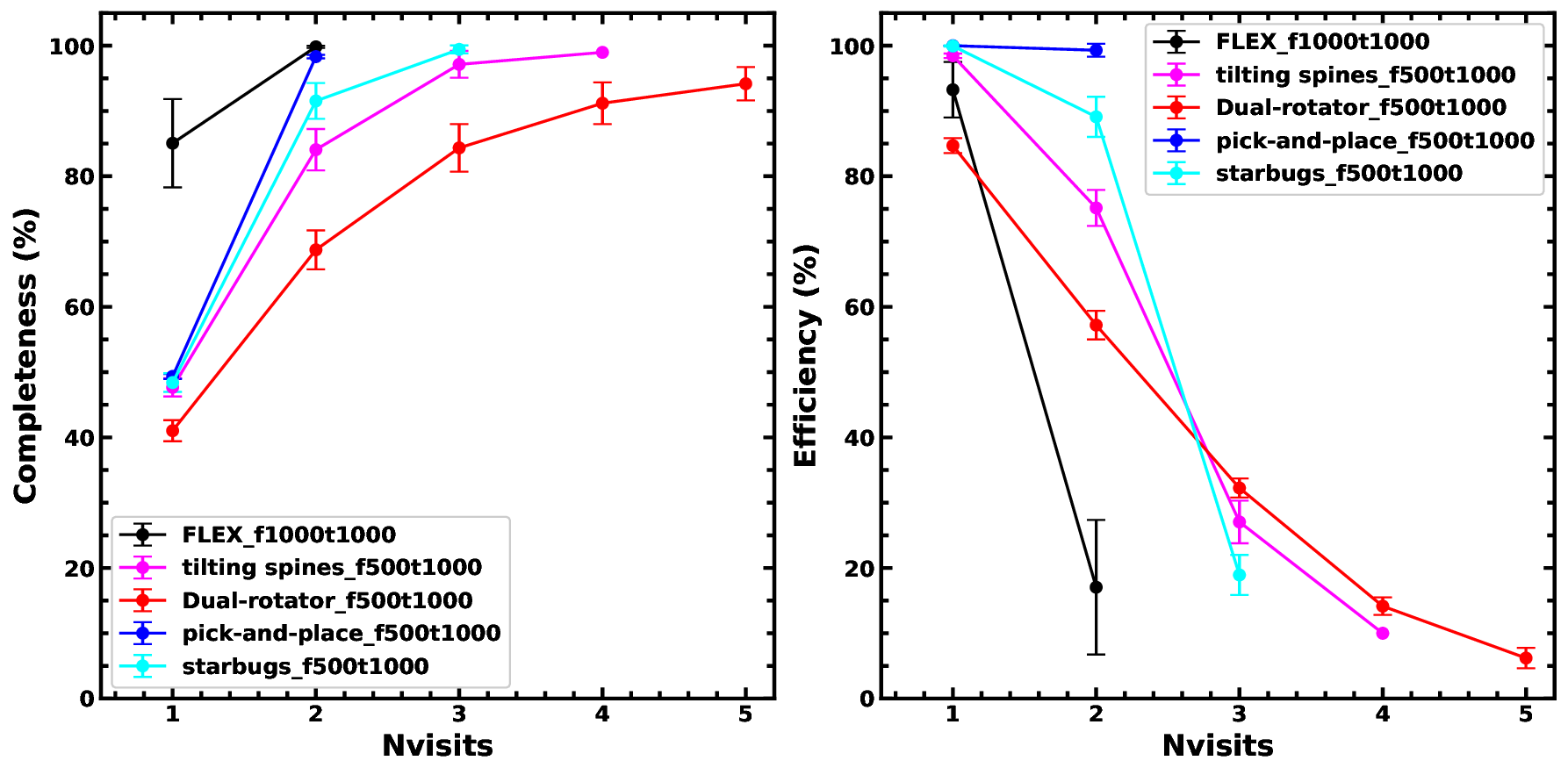}

\caption{Completeness and efficiency as a function of the number of visits (\( N_{\text{visits}} \)) for different fiber positioning systems (FLEX, tilting spines, dual-rotator, pick-and-place and starbugs). Each figure corresponds to 500 fibers, except for the FLEX.
The left panel in each row show completeness (\%), while the right panel show efficiency (\%). Top panel corresponds to 500 targets and bottom panel corresponds to 1000 targets, respectively for different fiber densities.}
\label{fig:500fibers}
\end{figure*}

\subsubsection{Fixed fiber number}
\label{subsubsec:fixed_fiber}
We further extend our analysis by comparing different system specifications using both a fixed number of fibers and targets in Figure~\ref{fig:500fibers}. For the purpose of this part of the investigation, we choose 500 fibers, with a choice of 500 targets (upper panels) and 1000 targets (lower panels). The point here is that by forcing a common fiber number we can emphasize the secondary effects of PR and ER in the technologies explored.
In order to enforce a common fiber number, we make small changes to the FoV. 
As in the case above, FLEX is computationally challenging, and with a restriction of only small changes, for FLEX only we adopted two FoVs for this analysis. For the case where we test 500 targets, we use a $0.3^\circ$ diameter (0.077 deg$^2$) FoV, and this allows for 500 FLEX fibers. For the larger target number, though, we need to use a $0.5^\circ$ diameter (0.196 deg$^2$) FoV, and this leads to 1000 FLEX fibers, rather than the common 500 for all other systems. While we show this latter result for completeness, we do not over interpret it, since its results will be dominated primarily by fiber number.

For the lower number of targets (\( N_{\text{targets}} \) = 500), the fiber number and target number are the same. All systems apart from the dual-rotator achieve our completeness goal (\(\sim 100\%\)) by \( N_{\text{visits}} \) = 2.0, however, their rates of increase vary. Notably, the pick-and-place system only requires a single visit. At \( N_{\text{visits}} \) = 1.0, completeness varies significantly, with one system (pick-and-place) reaching 95\% while others have lower starting points, as low as 60\%. The variation here is almost entirely a consequence of PR, with the systems having larger PR performing better. It is worth commenting on the similar performance in this simulation between FLEX and starbugs, as starbugs have the larger PR, and so might be expected to perform better. Note here that these are directly comparable as the fiber number and target number is the same for both FLEX and starbugs in this simulation. However, starbugs have a larger ER than FLEX, and this is likely to limit its performance, maintaining a similar outcome to the FLEX system.

At the higher target number (\( N_{\text{targets}} \) =1000), completeness and efficiency trends are similar. As expected more visits are required to reach the completeness goal, which helps to emphasize the primary remaining dependence on PR. As noted above, we show the FLEX result for completeness, but neglect it here as it has an unfair fiber number advantage for the purpose of this comparison. The pick-and-place, starbugs, tilting spines and then dual-rotators perform in the order expected given their PR (or lack of PR limitation, for pick-and-place).

\begin{figure*}
\centering
\includegraphics[width=0.5\linewidth]{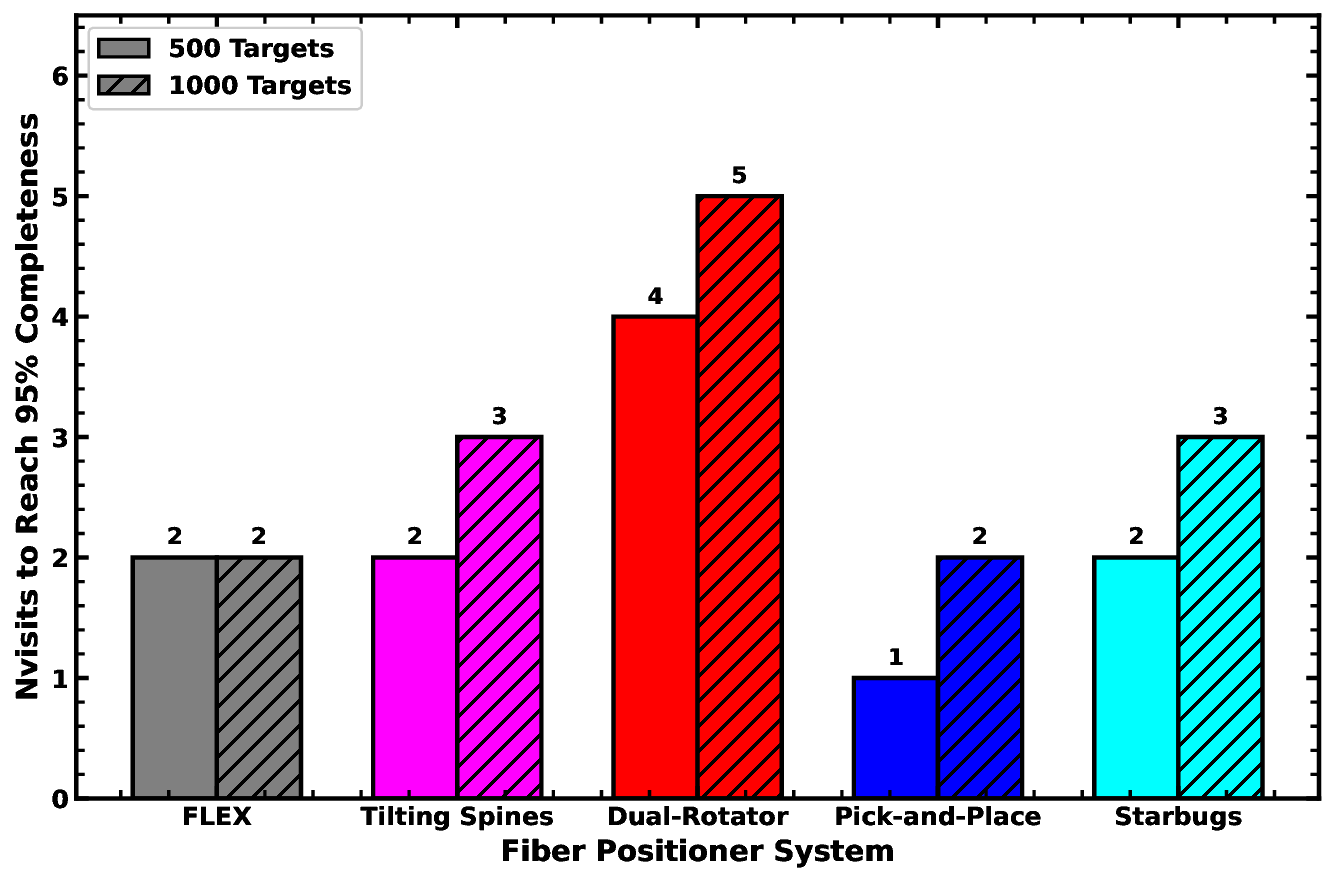}
\caption{The panel shows cost comparison across different fiber positioner configurations for 500 fibers. The bars indicate the minimum number of visits ($N_{\rm visits}$) required for each positioner system to achieve a cumulative target completeness threshold of $95\%$. Solid colored bars denote $500$ targets, while hatched bars indicate performance under $1000$ targets.}
\label{fig:new_figures}
\end{figure*}

To get a clear picture of how these five fiber positioner systems compare, we look at how fast they finish the survey, and how well they handle crowded fields for a fixed number of fibers (500 fibers). Figure~\ref{fig:new_figures} shows the operational cost of using each system by tracking how many visits it takes to observe at least 95\% of the targets. With 500 targets, pick-and-place, FLEX, tilting spines, and starbugs are the fastest, hitting the 95\% completeness in just 1 and 2 visits. Dual-rotators, as we model them here, require more visits compared to the other systems. 
The fact that all other systems need two visits, while pick-and-place needs only one may be a consequence of target clustering in the field modelled here, or alternatively, a coincidental arrangement of targets in such a way that there are more targets in a particular zone than probes able to reach that region, such as around the edges of the field. This would limit the number of probes able to be allocated to such targets by any of the technologies with a grid-like home position layout, and force a second visit. The level of clustering in this model has not restricted the pick-and-place system in the same way.
Doubling the number of targets to 1000 crowds the focal plane. This slows every system down. Pick-and-place, tilting spines, starbugs, and dual-rotators need one extra visit to finish. However, FLEX can still reach 95\% completeness with only 2 visits, likely due to its larger patrol radius.

Together the results from S~\ref{subsubsec:fixed_fov} and S~\ref{subsubsec:fixed_fiber} indicate that the smallest pitch or the highest fiber density gives the best performance. When the fiber number is fixed, increasing the PR is the most effective way to improve fiber allocation. Reducing the ER also helps, but to a smaller extent. The FLEX system demonstrates this clearly. Its compact design and high fiber density allow it to achieve the highest completeness and efficiency with fewer visits. Furthermore, when designing a real system, the fiber number may be fixed due to budget limits, electronics control, or other constraints. In such cases, understanding how the other parameters behave is important. This approach is not just theoretical, but could be of practical value.

\section{DISCUSSION}
\label{sec:discussion}
The analysis above explores in some detail a relatively simple model for comparing the basic specifications of RFPs. The results are not unexpected, aligning with an intuitive sense of how such systems should perform. This is reassuring, but also now establishes a new tool that can be extended or adapted to test a range of potential new RFP specifications when new systems are being designed. There are many directions that could be explored in such developments, as, of necessity, the first stage in establishing this tool has made a range of simplifying assumptions. Here we discuss some of these suggestions for further expansion.

\subsection{Future Direction}

The current implementation of the proof-of-concept code uses random fiber to target allocation without any specific optimization routines. 
A key future direction is the incorporation of optimization routines for fiber to target allocation. Implementing more efficient fiber assignment algorithms would align the outcomes more closely with those of a fully optimized system and significantly enhance overall system performance \citep[e.g.,][]{2012MNRAS.419.1187M, 2020A&A...635A.101T, 2020MNRAS.497.4626T, 2021AJ....161...92S, 2026AJ....171...37B}.

Future development of the code can involve incorporating reconfiguration time or other operational overheads into the model, which are critical to the real world performance of MOS systems.
Parallel positioners have much faster reconfiguration times than sequential positioners (e.g., pick-and-place). Existing pick-and-place systems (e.g., 2dF and WEAVE) mitigate against large reconfiguration times by using a tumbling barrel system that allows for one field plate to be configured while another is being observed. As long as the observation times are longer than the reconfiguration times, this ensures no loss of observing efficiency due to reconfiguration. For parallel positioners, most such systems do not employ such an approach, so reconfiguration time is a measurable overhead, but since these systems position all the fibers simultaneously the reconfiguration times are small. Again, if the observation times are reasonably long compared to this reconfiguration time, the observational efficiency can remain high. It is only in the case of very short exposure times where reconfiguration overheads are likely to dominate a system's observing efficiency.

For tilting spines, there is a position dependent throughput loss due to tilt induced FRD. 
A valuable future extension of the code will be the explicit modeling of fiber tilt induced FRD and the associated position dependent throughput losses in tilting spine systems, to more accurately reflect their impact on overall system performance.
These could be simulated by varying the target offsets, as larger offsets are presumed to induce greater FRD and reduce throughput \citep{2014SPIE.9151E..1XS}. 
\citet{2014SPIE.9150E..23S} compare throughput with different positioner types and demonstrate that tilt-induced losses are generally offset by higher allocation efficiency and lower fiber stress-induced FRD. The losses from spine tilt are balanced by the increased allocation yield from the larger patrol area \citep{2014SPIE.9151E..1XS}. For typical tilts, spines provides better efficiency compared to current dual-rotator systems. Moreover, spine tilt reduces throughput due to FRD and aperture losses caused by defocus \citep{2014SPIE.9150E..23S}. The analysis by \citet{2014SPIE.9150E..23S} suggest that tilt losses cause an average $\approx$ 3.2\% drop in S/N per target for a fixed integration time or $\approx$ 6.5\% loss in survey speed for a fixed S/N. These losses vary slightly between spines and dual-rotator systems and between target types in each survey. These factors do not affect the results presented in this analysis.

Furthermore, a promising direction for future enhancements is to integrate specific technological constraints, such as fiber collision avoidance, into the model to improve its alignment with real-world systems.
Next-generation large MOS surveys plan to use thousands of fiber positioner robots on the focal plane to rapidly move fiber ends between target points in parallel. For RFP systems, one challenge is moving a single fiber end without colliding with others. Many studies have adopted different approaches to solve the collision avoidance problem. 
\citet{2014A&A...566A..84M} proposed a motion planning method for the RFP on decentralized navigation functions based on the artificial potential field.
It can provide a collision free path for all the fiber ends between positioners with equal arms, however, it cannot completely avoid collisions between positioners with unequal arms. \citet{2019JATIS...5d5002M} proposed an improved approach of the artificial potential field taking an account of the convergences of a fiber positioner and its neighboring positioners. 

\citet{2021MNRAS.500..101Z} analyzed the types of collisions and their corresponding probabilities. They proposed a straightforward and effective method based on the safety zone concept, capable of completely preventing collisions for both equal and unequal arms. Recently, \citet{2024AJ....167..276Z} have enhanced their motion planning algorithm with a primary focus on equal arm RFPs. To minimize communication overhead while preventing collisions, they have introduced a new motion planning approach based on a rapidly exploring random tree.
\citet{2021AJ....161...92S} also provides a framework for fast, collision free trajectory planning. Recent work by \citet{2026AJ....171...37B} provides an optimized strategy for the multi-exposure allocation, demonstrating how to maximize target coverage while strictly adhering to real world mechanical collision constraints. By considering these physical limitations across the entire survey timeline, it is possible to significantly improve the efficiency of fiber to target assignments.

There is a growing recognition of the need for a next generation spectroscopic facility. Such a facility would overcome the constraints of existing instruments and better meet the demands of modern astronomy. As imaging surveys continue to deliver unprecedented amounts of data, the scientific community requires a dedicated system. This system must be capable of efficiently following up on discoveries with large scale, high resolution spectroscopic observations.

Building on the success of early instruments, a new generation of multi-object spectrographs has emerged. These systems are designed to tackle increasingly complex scientific objectives. Each represents a significant advancement in both technology and design. They enable more efficient and expansive surveys than ever before. These instruments are tailored to address specific astrophysical challenges, highlighting the ongoing evolution of MOS. Moreover, with the rapid growth of imaging surveys, there is increasing interest in spectroscopic follow-up of the identified targets. This has led to substantial investment in new multi-object spectroscopic facilities. There are many aspects to consider when evaluating the need for a dedicated spectroscopic facility. Key questions include what scientific challenges upcoming facilities will address and what limitations they might encounter. These limitations could stem from restricted survey speeds or the need to share telescopes with other instruments.

Future spectroscopic facilities including the Extremely Large Telescope \citep[ELT;][]{2023ConPh..64...47P}, the proposed Maunakea Spectroscopic Explorer \citep[MSE;][]{2019BAAS...51g.126M, 2023AN....34430108S} and the proposed WST \citep{10.1117/12.3018093, 2024arXiv240305398M} are essential to play a transformative role in advancing our understanding of the Universe. 
MSE \citep{2023AN....34430108S} is a 12.5 meter telescope with a 1.5 deg$^2$ FoV designed for massively multiplexed spectroscopic surveys. Similarly, the planned WST \citep{2024arXiv240305398M} will offer fiber fed MOS across 3.1 deg$^2$ and integral field spectroscopy within a central 3 × 3 arcmin$^2$ region. The ELT \citep{2023ConPh..64...47P}, a 39 meter class telescope under construction by European Southern Observatory (ESO), is set to revolutionize observational astronomy complementing facilities like the WST by providing deeper, high resolution observations. 

Modern multi-object spectrographs are powerful and efficient. However, when targets are spread out beyond the telescope’s FoV, multiple pointings are still often needed. Instruments with configurable, fiber fed spectrographs and a wide FoV are especially useful in these cases. They can observe many clustered targets at once, saving a lot of observing time compared to single object spectrographs. 
Ambitious projects including ELT, MSE, and WST aim to deliver unprecedented survey speed, sensitivity, and multiplexing capabilities, which will place new demands on the technical performance of their fiber positioning systems. In this context, the development and optimization of fiber specifications including positioning accuracy, reconfiguration time, and fiber throughput are critical to ensuring the scientific success of these next generation facilities. 

Our work on MOS instrument specifications provides a timely framework for evaluating current limitations and identifying design priorities. By quantifying performance trade-offs and assessing the operational implications of various fiber technologies, this study lays the groundwork for informed decision making in the planning and implementation of future survey instruments.

\section{CONCLUSION}
\label{sec:concl}

We have developed a proof-of-concept code and evaluated the performance of MOS fiber systems using an arbitrary number of fibers and targets. 
Our approach here is to begin with the general analysis to quantify how three parameters (e.g. fiber density, PR, and ER) work together to affect the efficiency of a fiber positioner system in target allocation. By systematically exploring these dependencies, our analysis reveals the trade-offs between these parameters. This helps us identify which parameters have the greatest impact on observations. We further explore our analysis to a series of realistic scenarios by comparing the specifications of a selection of existing fiber positioner systems. This helps us understand how our chosen specifications for each system affect the overall performance of the fiber positioner system. 
From the overall analysis we discuss trade-offs between different system specifications and the completeness and allocation efficiency in the context of our code. We find that the smallest pitch or the highest fiber density gives the best completeness and efficiency. When the pitch is fixed, increasing the PR is the best way to improve fiber allocation. Reducing the ER also helps but has a smaller effect.


\begin{acknowledgments}
We thank the anonymous referee for important suggestions, which helped improve the quality of this paper.
\end{acknowledgments}

\bibliography{sample701}{}

@ARTICLE{2011MNRAS.413..971D,
       author = {{Driver}, S.~P. and {Hill}, D.~T. and {Kelvin}, L.~S. and {Robotham}, A.~S.~G. and {Liske}, J. and {Norberg}, P. and {Baldry}, I.~K. and {Bamford}, S.~P. and {Hopkins}, A.~M. and {Loveday}, J. and {Peacock}, J.~A. and {Andrae}, E. and {Bland-Hawthorn}, J. and {Brough}, S. and {Brown}, M.~J.~I. and {Cameron}, E. and {Ching}, J.~H.~Y. and {Colless}, M. and {Conselice}, C.~J. and {Croom}, S.~M. and {Cross}, N.~J.~G. and {de Propris}, R. and {Dye}, S. and {Drinkwater}, M.~J. and {Ellis}, S. and {Graham}, Alister W. and {Grootes}, M.~W. and {Gunawardhana}, M. and {Jones}, D.~H. and {van Kampen}, E. and {Maraston}, C. and {Nichol}, R.~C. and {Parkinson}, H.~R. and {Phillipps}, S. and {Pimbblet}, K. and {Popescu}, C.~C. and {Prescott}, M. and {Roseboom}, I.~G. and {Sadler}, E.~M. and {Sansom}, A.~E. and {Sharp}, R.~G. and {Smith}, D.~J.~B. and {Taylor}, E. and {Thomas}, D. and {Tuffs}, R.~J. and {Wijesinghe}, D. and {Dunne}, L. and {Frenk}, C.~S. and {Jarvis}, M.~J. and {Madore}, B.~F. and {Meyer}, M.~J. and {Seibert}, M. and {Staveley-Smith}, L. and {Sutherland}, W.~J. and {Warren}, S.~J.},
        title = "{Galaxy and Mass Assembly (GAMA): survey diagnostics and core data release}",
      journal = {\mnras},
         year = 2011,
        month = may,
       volume = {413},
       number = {2},
        pages = {971-995},
          doi = {10.1111/j.1365-2966.2010.18188.x},
archivePrefix = {arXiv},
       eprint = {1009.0614},
 primaryClass = {astro-ph.CO},
       adsurl = {https://ui.adsabs.harvard.edu/abs/2011MNRAS.413..971D}
}

@ARTICLE{2015MNRAS.452.2087L,
       author = {{Liske}, J. and {Baldry}, I.~K. and {Driver}, S.~P. and {Tuffs}, R.~J. and {Alpaslan}, M. and {Andrae}, E. and {Brough}, S. and {Cluver}, M.~E. and {Grootes}, M.~W. and {Gunawardhana}, M.~L.~P. and {Kelvin}, L.~S. and {Loveday}, J. and {Robotham}, A.~S.~G. and {Taylor}, E.~N. and {Bamford}, S.~P. and {Bland-Hawthorn}, J. and {Brown}, M.~J.~I. and {Drinkwater}, M.~J. and {Hopkins}, A.~M. and {Meyer}, M.~J. and {Norberg}, P. and {Peacock}, J.~A. and {Agius}, N.~K. and {Andrews}, S.~K. and {Bauer}, A.~E. and {Ching}, J.~H.~Y. and {Colless}, M. and {Conselice}, C.~J. and {Croom}, S.~M. and {Davies}, L.~J.~M. and {De Propris}, R. and {Dunne}, L. and {Eardley}, E.~M. and {Ellis}, S. and {Foster}, C. and {Frenk}, C.~S. and {H{\"a}u{\ss}ler}, B. and {Holwerda}, B.~W. and {Howlett}, C. and {Ibarra}, H. and {Jarvis}, M.~J. and {Jones}, D.~H. and {Kafle}, P.~R. and {Lacey}, C.~G. and {Lange}, R. and {Lara-L{\'o}pez}, M.~A. and {L{\'o}pez-S{\'a}nchez}, {\'A}. R. and {Maddox}, S. and {Madore}, B.~F. and {McNaught-Roberts}, T. and {Moffett}, A.~J. and {Nichol}, R.~C. and {Owers}, M.~S. and {Palamara}, D. and {Penny}, S.~J. and {Phillipps}, S. and {Pimbblet}, K.~A. and {Popescu}, C.~C. and {Prescott}, M. and {Proctor}, R. and {Sadler}, E.~M. and {Sansom}, A.~E. and {Seibert}, M. and {Sharp}, R. and {Sutherland}, W. and {V{\'a}zquez-Mata}, J.~A. and {van Kampen}, E. and {Wilkins}, S.~M. and {Williams}, R. and {Wright}, A.~H.},
        title = "{Galaxy And Mass Assembly (GAMA): end of survey report and data release 2}",
      journal = {\mnras},
         year = 2015,
        month = sep,
       volume = {452},
       number = {2},
        pages = {2087-2126},
          doi = {10.1093/mnras/stv1436},
archivePrefix = {arXiv},
       eprint = {1506.08222},
 primaryClass = {astro-ph.GA},
       adsurl = {https://ui.adsabs.harvard.edu/abs/2015MNRAS.452.2087L}
}

@ARTICLE{2013MNRAS.430.2047H,
       author = {{Hopkins}, A.~M. and {Driver}, S.~P. and {Brough}, S. and {Owers}, M.~S. and {Bauer}, A.~E. and {Gunawardhana}, M.~L.~P. and {Cluver}, M.~E. and {Colless}, M. and {Foster}, C. and {Lara-L{\'o}pez}, M.~A. and {Roseboom}, I. and {Sharp}, R. and {Steele}, O. and {Thomas}, D. and {Baldry}, I.~K. and {Brown}, M.~J.~I. and {Liske}, J. and {Norberg}, P. and {Robotham}, A.~S.~G. and {Bamford}, S. and {Bland-Hawthorn}, J. and {Drinkwater}, M.~J. and {Loveday}, J. and {Meyer}, M. and {Peacock}, J.~A. and {Tuffs}, R. and {Agius}, N. and {Alpaslan}, M. and {Andrae}, E. and {Cameron}, E. and {Cole}, S. and {Ching}, J.~H.~Y. and {Christodoulou}, L. and {Conselice}, C. and {Croom}, S. and {Cross}, N.~J.~G. and {De Propris}, R. and {Delhaize}, J. and {Dunne}, L. and {Eales}, S. and {Ellis}, S. and {Frenk}, C.~S. and {Graham}, Alister W. and {Grootes}, M.~W. and {H{\"a}u{\ss}ler}, B. and {Heymans}, C. and {Hill}, D. and {Hoyle}, B. and {Hudson}, M. and {Jarvis}, M. and {Johansson}, J. and {Jones}, D.~H. and {van Kampen}, E. and {Kelvin}, L. and {Kuijken}, K. and {L{\'o}pez-S{\'a}nchez}, {\'A}. and {Maddox}, S. and {Madore}, B. and {Maraston}, C. and {McNaught-Roberts}, T. and {Nichol}, R.~C. and {Oliver}, S. and {Parkinson}, H. and {Penny}, S. and {Phillipps}, S. and {Pimbblet}, K.~A. and {Ponman}, T. and {Popescu}, C.~C. and {Prescott}, M. and {Proctor}, R. and {Sadler}, E.~M. and {Sansom}, A.~E. and {Seibert}, M. and {Staveley-Smith}, L. and {Sutherland}, W. and {Taylor}, E. and {Van Waerbeke}, L. and {V{\'a}zquez-Mata}, J.~A. and {Warren}, S. and {Wijesinghe}, D.~B. and {Wild}, V. and {Wilkins}, S.},
        title = "{Galaxy And Mass Assembly (GAMA): spectroscopic analysis}",
      journal = {\mnras},
         year = 2013,
        month = apr,
       volume = {430},
       number = {3},
        pages = {2047-2066},
          doi = {10.1093/mnras/stt030},
archivePrefix = {arXiv},
       eprint = {1301.7127},
 primaryClass = {astro-ph.CO},
       adsurl = {https://ui.adsabs.harvard.edu/abs/2013MNRAS.430.2047H}
}

@ARTICLE{2015A&A...582A..62D,
       author = {{de Jong}, Jelte T.~A. and {Verdoes Kleijn}, Gijs A. and {Boxhoorn}, Danny R. and {Buddelmeijer}, Hugo and {Capaccioli}, Massimo and {Getman}, Fedor and {Grado}, Aniello and {Helmich}, Ewout and {Huang}, Zhuoyi and {Irisarri}, Nancy and {Kuijken}, Konrad and {La Barbera}, Francesco and {McFarland}, John P. and {Napolitano}, Nicola R. and {Radovich}, Mario and {Sikkema}, Gert and {Valentijn}, Edwin A. and {Begeman}, Kor G. and {Brescia}, Massimo and {Cavuoti}, Stefano and {Choi}, Ami and {Cordes}, Oliver-Mark and {Covone}, Giovanni and {Dall'Ora}, Massimo and {Hildebrandt}, Hendrik and {Longo}, Giuseppe and {Nakajima}, Reiko and {Paolillo}, Maurizio and {Puddu}, Emanuella and {Rifatto}, Agatino and {Tortora}, Crescenzo and {van Uitert}, Edo and {Buddendiek}, Axel and {Harnois-D{\'e}raps}, Joachim and {Erben}, Thomas and {Eriksen}, Martin B. and {Heymans}, Catherine and {Hoekstra}, Henk and {Joachimi}, Benjamin and {Kitching}, Thomas D. and {Klaes}, Dominik and {Koopmans}, L{\'e}on V.~E. and {K{\"o}hlinger}, Fabian and {Roy}, Nivya and {Sif{\'o}n}, Crist{\'o}bal and {Schneider}, Peter and {Sutherland}, Will J. and {Viola}, Massimo and {Vriend}, Willem-Jan},
        title = "{The first and second data releases of the Kilo-Degree Survey}",
      journal = {\aap},
         year = 2015,
        month = oct,
       volume = {582},
          eid = {A62},
        pages = {A62},
          doi = {10.1051/0004-6361/201526601},
archivePrefix = {arXiv},
       eprint = {1507.00742},
 primaryClass = {astro-ph.CO},
       adsurl = {https://ui.adsabs.harvard.edu/abs/2015A&A...582A..62D}
}

@ARTICLE{2010MNRAS.404...86B,
       author = {{Baldry}, I.~K. and {Robotham}, A.~S.~G. and {Hill}, D.~T. and {Driver}, S.~P. and {Liske}, J. and {Norberg}, P. and {Bamford}, S.~P. and {Hopkins}, A.~M. and {Loveday}, J. and {Peacock}, J.~A. and {Cameron}, E. and {Croom}, S.~M. and {Cross}, N.~J.~G. and {Doyle}, I.~F. and {Dye}, S. and {Frenk}, C.~S. and {Jones}, D.~H. and {van Kampen}, E. and {Kelvin}, L.~S. and {Nichol}, R.~C. and {Parkinson}, H.~R. and {Popescu}, C.~C. and {Prescott}, M. and {Sharp}, R.~G. and {Sutherland}, W.~J. and {Thomas}, D. and {Tuffs}, R.~J.},
        title = "{Galaxy And Mass Assembly (GAMA): the input catalogue and star-galaxy separation}",
      journal = {\mnras},
         year = 2010,
        month = may,
       volume = {404},
       number = {1},
        pages = {86-100},
          doi = {10.1111/j.1365-2966.2010.16282.x},
archivePrefix = {arXiv},
       eprint = {0910.5120},
 primaryClass = {astro-ph.CO},
       adsurl = {https://ui.adsabs.harvard.edu/abs/2010MNRAS.404...86B}
}

@ARTICLE{2010PASA...27...76R,
       author = {{Robotham}, A. and {Driver}, S.~P. and {Norberg}, P. and {Baldry}, I.~K. and {Bamford}, S.~P. and {Hopkins}, A.~M. and {Liske}, J. and {Loveday}, J. and {Peacock}, J.~A. and {Cameron}, E. and {Croom}, S.~M. and {Doyle}, I.~F. and {Frenk}, C.~S. and {Hill}, D.~T. and {Jones}, D.~H. and {van Kampen}, E. and {Kelvin}, L.~S. and {Kuijken}, K. and {Nichol}, R.~C. and {Parkinson}, H.~R. and {Popescu}, C.~C. and {Prescott}, M. and {Sharp}, R.~G. and {Sutherland}, W.~J. and {Thomas}, D. and {Tuffs}, R.~J.},
        title = "{Galaxy and Mass Assembly (GAMA): Optimal Tiling of Dense Surveys with a Multi-Object Spectrograph}",
      journal = {\pasa},
         year = 2010,
        month = mar,
       volume = {27},
       number = {1},
        pages = {76-90},
          doi = {10.1071/AS09053},
archivePrefix = {arXiv},
       eprint = {0910.5121},
 primaryClass = {astro-ph.CO},
       adsurl = {https://ui.adsabs.harvard.edu/abs/2010PASA...27...76R}
}

@ARTICLE{2018MNRAS.474.3875B,
       author = {{Baldry}, I.~K. and {Liske}, J. and {Brown}, M.~J.~I. and {Robotham}, A.~S.~G. and {Driver}, S.~P. and {Dunne}, L. and {Alpaslan}, M. and {Brough}, S. and {Cluver}, M.~E. and {Eardley}, E. and {Farrow}, D.~J. and {Heymans}, C. and {Hildebrandt}, H. and {Hopkins}, A.~M. and {Kelvin}, L.~S. and {Loveday}, J. and {Moffett}, A.~J. and {Norberg}, P. and {Owers}, M.~S. and {Taylor}, E.~N. and {Wright}, A.~H. and {Bamford}, S.~P. and {Bland-Hawthorn}, J. and {Bourne}, N. and {Bremer}, M.~N. and {Colless}, M. and {Conselice}, C.~J. and {Croom}, S.~M. and {Davies}, L.~J.~M. and {Foster}, C. and {Grootes}, M.~W. and {Holwerda}, B.~W. and {Jones}, D.~H. and {Kafle}, P.~R. and {Kuijken}, K. and {Lara-Lopez}, M.~A. and {L{\'o}pez-S{\'a}nchez}, {\'A}. R. and {Meyer}, M.~J. and {Phillipps}, S. and {Sutherland}, W.~J. and {van Kampen}, E. and {Wilkins}, S.~M.},
        title = "{Galaxy And Mass Assembly: the G02 field, Herschel-ATLAS target selection and data release 3}",
      journal = {\mnras},
         year = 2018,
        month = mar,
       volume = {474},
       number = {3},
        pages = {3875-3888},
          doi = {10.1093/mnras/stx3042},
archivePrefix = {arXiv},
       eprint = {1711.09139},
 primaryClass = {astro-ph.GA},
       adsurl = {https://ui.adsabs.harvard.edu/abs/2018MNRAS.474.3875B}
}

@ARTICLE{2020MNRAS.496.3235B,
       author = {{Bellstedt}, Sabine and {Driver}, Simon P. and {Robotham}, Aaron S.~G. and {Davies}, Luke J.~M. and {Bogue}, Cameron R.~J. and {Cook}, Robin H.~W. and {Hashemizadeh}, Abdolhosein and {Koushan}, Soheil and {Taylor}, Edward N. and {Thorne}, Jessica E. and {Turner}, Ryan J. and {Wright}, Angus H.},
        title = "{Galaxy And Mass Assembly (GAMA): assimilation of KiDS into the GAMA database}",
      journal = {\mnras},
         year = 2020,
        month = aug,
       volume = {496},
       number = {3},
        pages = {3235-3256},
          doi = {10.1093/mnras/staa1466},
archivePrefix = {arXiv},
       eprint = {2005.11215},
 primaryClass = {astro-ph.GA},
       adsurl = {https://ui.adsabs.harvard.edu/abs/2020MNRAS.496.3235B}
}

@ARTICLE{2022MNRAS.513..439D,
       author = {{Driver}, Simon P. and {Bellstedt}, Sabine and {Robotham}, Aaron S.~G. and {Baldry}, Ivan K. and {Davies}, Luke J. and {Liske}, Jochen and {Obreschkow}, Danail and {Taylor}, Edward N. and {Wright}, Angus H. and {Alpaslan}, Mehmet and {Bamford}, Steven P. and {Bauer}, Amanda E. and {Bland-Hawthorn}, Joss and {Bilicki}, Maciej and {Bravo}, Mat{\'\i}as and {Brough}, Sarah and {Casura}, Sarah and {Cluver}, Michelle E. and {Colless}, Matthew and {Conselice}, Christopher J. and {Croom}, Scott M. and {de Jong}, Jelte and {D'Eugenio}, Franceso and {De Propris}, Roberto and {Dogruel}, Burak and {Drinkwater}, Michael J. and {Dvornik}, Andrej and {Farrow}, Daniel J. and {Frenk}, Carlos S. and {Giblin}, Benjamin and {Graham}, Alister W. and {Grootes}, Meiert W. and {Gunawardhana}, Madusha L.~P. and {Hashemizadeh}, Abdolhosein and {H{\"a}u{\ss}ler}, Boris and {Heymans}, Catherine and {Hildebrandt}, Hendrik and {Holwerda}, Benne W. and {Hopkins}, Andrew M. and {Jarrett}, Tom H. and {Heath Jones}, D. and {Kelvin}, Lee S. and {Koushan}, Soheil and {Kuijken}, Konrad and {Lara-L{\'o}pez}, Maritza A. and {Lange}, Rebecca and {L{\'o}pez-S{\'a}nchez}, {\'A}ngel R. and {Loveday}, Jon and {Mahajan}, Smriti and {Meyer}, Martin and {Moffett}, Amanda J. and {Napolitano}, Nicola R. and {Norberg}, Peder and {Owers}, Matt S. and {Radovich}, Mario and {Raouf}, Mojtaba and {Peacock}, John A. and {Phillipps}, Steven and {Pimbblet}, Kevin A. and {Popescu}, Cristina and {Said}, Khaled and {Sansom}, Anne E. and {Seibert}, Mark and {Sutherland}, Will J. and {Thorne}, Jessica E. and {Tuffs}, Richard J. and {Turner}, Ryan and {van der Wel}, Arjen and {van Kampen}, Eelco and {Wilkins}, Steve M.},
        title = "{Galaxy And Mass Assembly (GAMA): Data Release 4 and the z < 0.1 total and z < 0.08 morphological galaxy stellar mass functions}",
      journal = {\mnras},
         year = 2022,
        month = jun,
       volume = {513},
       number = {1},
        pages = {439-467},
          doi = {10.1093/mnras/stac472},
archivePrefix = {arXiv},
       eprint = {2203.08539},
 primaryClass = {astro-ph.GA},
       adsurl = {https://ui.adsabs.harvard.edu/abs/2022MNRAS.513..439D}
}

@INPROCEEDINGS{2012sngi.confE..40S,
       author = {{Sutherland}, Will},
        title = "{VIKING: the VISTA Kilo-degree INfrared Galaxy survey}",
    booktitle = {Science from the Next Generation Imaging and Spectroscopic Surveys},
         year = 2012,
        month = oct,
          eid = {40},
        pages = {40},
       adsurl = {https://ui.adsabs.harvard.edu/abs/2012sngi.confE..40S}
}

@ARTICLE{2003AJ....126.2081A,
       author = {{Abazajian}, Kevork and {Adelman-McCarthy}, Jennifer K. and {Ag{\"u}eros}, Marcel A. and {Allam}, Sahar S. and {Anderson}, Scott F. and {Annis}, James and {Bahcall}, Neta A. and {Baldry}, Ivan K. and {Bastian}, Steven and {Berlind}, Andreas and {Bernardi}, Mariangela and {Blanton}, Michael R. and {Blythe}, Norman and {Bochanski}, John J., Jr. and {Boroski}, William N. and {Brewington}, Howard and {Briggs}, John W. and {Brinkmann}, J. and {Brunner}, Robert J. and {Budav{\'a}ri}, Tam{\'a}s and {Carey}, Larry N. and {Carr}, Michael A. and {Castander}, Francisco J. and {Chiu}, Kuenley and {Collinge}, Matthew J. and {Connolly}, A.~J. and {Covey}, Kevin R. and {Csabai}, Istv{\'a}n and {Dalcanton}, Julianne J. and {Dodelson}, Scott and {Doi}, Mamoru and {Dong}, Feng and {Eisenstein}, Daniel J. and {Evans}, Michael L. and {Fan}, Xiaohui and {Feldman}, Paul D. and {Finkbeiner}, Douglas P. and {Friedman}, Scott D. and {Frieman}, Joshua A. and {Fukugita}, Masataka and {Gal}, Roy R. and {Gillespie}, Bruce and {Glazebrook}, Karl and {Gonzalez}, Carlos F. and {Gray}, Jim and {Grebel}, Eva K. and {Grodnicki}, Lauren and {Gunn}, James E. and {Gurbani}, Vijay K. and {Hall}, Patrick B. and {Hao}, Lei and {Harbeck}, Daniel and {Harris}, Frederick H. and {Harris}, Hugh C. and {Harvanek}, Michael and {Hawley}, Suzanne L. and {Heckman}, Timothy M. and {Helmboldt}, J.~F. and {Hendry}, John S. and {Hennessy}, Gregory S. and {Hindsley}, Robert B. and {Hogg}, David W. and {Holmgren}, Donald J. and {Holtzman}, Jon A. and {Homer}, Lee and {Hui}, Lam and {Ichikawa}, Shin-ichi and {Ichikawa}, Takashi and {Inkmann}, John P. and {Ivezi{\'c}}, {\v{Z}}eljko and {Jester}, Sebastian and {Johnston}, David E. and {Jordan}, Beatrice and {Jordan}, Wendell P. and {Jorgensen}, Anders M. and {Juri{\'c}}, Mario and {Kauffmann}, Guinevere and {Kent}, Stephen M. and {Kleinman}, S.~J. and {Knapp}, G.~R. and {Kniazev}, Alexei Y. and {Kron}, Richard G. and {Krzesi{\'n}ski}, Jurek and {Kunszt}, Peter Z. and {Kuropatkin}, Nickolai and {Lamb}, Donald Q. and {Lampeitl}, Hubert and {Laubscher}, Bryan E. and {Lee}, Brian C. and {Leger}, R. French and {Li}, Nolan and {Lidz}, Adam and {Lin}, Huan and {Loh}, Yeong-Shang and {Long}, Daniel C. and {Loveday}, Jon and {Lupton}, Robert H. and {Malik}, Tanu and {Margon}, Bruce and {McGehee}, Peregrine M. and {McKay}, Timothy A. and {Meiksin}, Avery and {Miknaitis}, Gajus A. and {Moorthy}, Bhasker K. and {Munn}, Jeffrey A. and {Murphy}, Tara and {Nakajima}, Reiko and {Narayanan}, Vijay K. and {Nash}, Thomas and {Neilsen}, Eric H., Jr. and {Newberg}, Heidi Jo and {Newman}, Peter R. and {Nichol}, Robert C. and {Nicinski}, Tom and {Nieto-Santisteban}, Maria and {Nitta}, Atsuko and {Odenkirchen}, Michael and {Okamura}, Sadanori and {Ostriker}, Jeremiah P. and {Owen}, Russell and {Padmanabhan}, Nikhil and {Peoples}, John and {Pier}, Jeffrey R. and {Pindor}, Bartosz and {Pope}, Adrian C. and {Quinn}, Thomas R. and {Rafikov}, R.~R. and {Raymond}, Sean N. and {Richards}, Gordon T. and {Richmond}, Michael W. and {Rix}, Hans-Walter and {Rockosi}, Constance M. and {Schaye}, Joop and {Schlegel}, David J. and {Schneider}, Donald P. and {Schroeder}, Joshua and {Scranton}, Ryan and {Sekiguchi}, Maki and {Seljak}, Uro{\v{s}} and {Sergey}, Gary and {Sesar}, Branimir and {Sheldon}, Erin and {Shimasaku}, Kazu and {Siegmund}, Walter A. and {Silvestri}, Nicole M. and {Sinisgalli}, Allan J. and {Sirko}, Edwin and {Smith}, J. Allyn and {Smol{\v{c}}i{\'c}}, Vernesa and {Snedden}, Stephanie A. and {Stebbins}, Albert and {Steinhardt}, Charles and {Stinson}, Gregory and {Stoughton}, Chris and {Strateva}, Iskra V. and {Strauss}, Michael A. and {SubbaRao}, Mark and {Szalay}, Alexander S. and {Szapudi}, Istv{\'a}n and {Szkody}, Paula and {Tasca}, Lidia and {Tegmark}, Max and {Thakar}, Aniruddha R. and {Tremonti}, Christy and {Tucker}, Douglas L. and {Uomoto}, Alan and {Vanden Berk}, Daniel E. and {Vandenberg}, Jan and {Vogeley}, Michael S. and {Voges}, Wolfgang and {Vogt}, Nicole P. and {Walkowicz}, Lucianne M. and {Weinberg}, David H. and {West}, Andrew A. and {White}, Simon D.~M. and {Wilhite}, Brian C. and {Willman}, Beth and {Xu}, Yongzhong and {Yanny}, Brian and {Yarger}, Jean and {Yasuda}, Naoki and {Yip}, Ching-Wa and {Yocum}, D.~R. and {York}, Donald G. and {Zakamska}, Nadia L. and {Zehavi}, Idit and {Zheng}, Wei and {Zibetti}, Stefano and {Zucker}, Daniel B.},
        title = "{The First Data Release of the Sloan Digital Sky Survey}",
      journal = {The Astronomical Journal},
         year = 2003,
        month = oct,
       volume = {126},
       number = {4},
        pages = {2081-2086},
          doi = {10.1086/378165},
archivePrefix = {arXiv},
       eprint = {astro-ph/0305492},
 primaryClass = {astro-ph},
       adsurl = {https://ui.adsabs.harvard.edu/abs/2003AJ....126.2081A}
}

@inproceedings{kuehn2014taipan,
  title={TAIPAN: optical spectroscopy with StarBugs},
  author={Kuehn, Kyler and Lawrence, Jon and Brown, David M and Case, Scott and Colless, Matthew and Content, Robert and Gers, Luke and Gilbert, James and Goodwin, Michael and Hopkins, Andrew M and others},
  booktitle={Ground-based and Airborne Instrumentation for Astronomy V},
  volume={9147},
  pages={380--387},
  year={2014},
  organization={SPIE}
}

@article{staszak2016taipan,
  title={TAIPAN instrument fibre positioner and Starbug robots: engineering overview},
  author={Staszak, Nicholas F and Lawrence, Jon and Brown, David M and Brown, Rebecca and Zhelem, Ross and Goodwin, Michael and Kuehn, Kyler and Lorente, Nuria PF and Nichani, Vijay and Waller, Lew and others},
  journal={Advances in Optical and Mechanical Technologies for Telescopes and Instrumentation II},
  volume={9912},
  pages={633--652},
  year={2016},
  publisher={SPIE}
}

@INPROCEEDINGS{1986SPIE..627..118P,
       author = {{Parry}, Ian R. and {Gray}, Peter M.},
        title = "{An automated multiobject fibre optic coupler for the Anglo-AustralianTelescope.}",
    booktitle = {Instrumentation in astronomy VI},
         year = 1986,
       editor = {{Crawford}, David L.},
       series = {Society of Photo-Optical Instrumentation Engineers (SPIE) Conference Series},
       volume = {627},
        month = jan,
        pages = {118-124},
          doi = {10.1117/12.968080},
       adsurl = {https://ui.adsabs.harvard.edu/abs/1986SPIE..627..118P}
}

@INPROCEEDINGS{1994SPIE.2198...87B,
       author = {{Barden}, Samuel C. and {Armandroff}, Taft and {Muller}, Gary and {Rudeen}, Andy C. and {Lewis}, Jeff and {Groves}, Lee},
        title = "{Modifying Hydra for the WIYN telescope: an optimum telescope, fiber MOS combination}",
    booktitle = {Instrumentation in Astronomy VIII},
         year = 1994,
       editor = {{Crawford}, David L. and {Craine}, Eric R.},
       series = {Society of Photo-Optical Instrumentation Engineers (SPIE) Conference Series},
       volume = {2198},
        month = jun,
        pages = {87-97},
          doi = {10.1117/12.176816},
       adsurl = {https://ui.adsabs.harvard.edu/abs/1994SPIE.2198...87B}
}

@ARTICLE{2002MNRAS.333..279L,
       author = {{Lewis}, I.~J. and {Cannon}, R.~D. and {Taylor}, K. and {Glazebrook}, K. and {Bailey}, J.~A. and {Baldry}, I.~K. and {Barton}, J.~R. and {Bridges}, T.~J. and {Dalton}, G.~B. and {Farrell}, T.~J. and {Gray}, P.~M. and {Lankshear}, A. and {McCowage}, C. and {Parry}, I.~R. and {Sharples}, R.~M. and {Shortridge}, K. and {Smith}, G.~A. and {Stevenson}, J. and {Straede}, J.~O. and {Waller}, L.~G. and {Whittard}, J.~D. and {Wilcox}, J.~K. and {Willis}, K.~C.},
        title = "{The Anglo-Australian Observatory 2dF facility}",
      journal = {\mnras},
         year = 2002,
        month = jun,
       volume = {333},
       number = {2},
        pages = {279-299},
          doi = {10.1046/j.1365-8711.2002.05333.x},
archivePrefix = {arXiv},
       eprint = {astro-ph/0202175},
 primaryClass = {astro-ph},
       adsurl = {https://ui.adsabs.harvard.edu/abs/2002MNRAS.333..279L}
}

@INPROCEEDINGS{1998ASPC..152...80P,
       author = {{Parker}, Q.~A. and {Watson}, F.~G. and {Miziarski}, S.},
        title = "{6dF: an Automated Multi-Object Fiber Spectroscopy System for the UKST}",
    booktitle = {Fiber Optics in Astronomy III},
         year = 1998,
       editor = {{Arribas}, S. and {Mediavilla}, E. and {Watson}, F.},
       series = {Astronomical Society of the Pacific Conference Series},
       volume = {152},
        month = jan,
        pages = {80},
       adsurl = {https://ui.adsabs.harvard.edu/abs/1998ASPC..152...80P}
}

@INPROCEEDINGS{2004SPIE.5492..643G,
       author = {{Gillingham}, Peter and {Smoker}, Jonathan and {Colless}, Matthew and {Kaufer}, Andreas},
        title = "{Operational performance of OzPoz: the multifiber positioner on the VLT}",
    booktitle = {Ground-based Instrumentation for Astronomy},
         year = 2004,
       editor = {{Moorwood}, Alan F.~M. and {Iye}, Masanori},
       series = {Society of Photo-Optical Instrumentation Engineers (SPIE) Conference Series},
       volume = {5492},
        month = sep,
        pages = {643-650},
          doi = {10.1117/12.550966},
       adsurl = {https://ui.adsabs.harvard.edu/abs/2004SPIE.5492..643G}
}

@INPROCEEDINGS{2020SPIE11447E..14D,
       author = {{Dalton}, Gavin and {Trager}, Scott and {Abrams}, Don Carlos and {Bonifacio}, Piercarlo and {Aguerri}, J. Alfonso L. and {Vallenari}, Antonella and {Bishop}, Georgia and {Middleton}, Kevin and {Benn}, Chris and {Dee}, Kevin and {Mignot}, Shan and {Lewis}, Ian and {Pragt}, Johannes and {Pico}, Sergio and {Walton}, Nicholas and {Rey}, Juerg and {Allende Prieto}, Carlos and {Lhom{\'e}}, Emilie and {Balcells}, Marc and {Terrett}, David and {Brock}, Matthew and {Ridings}, Andy and {Skvar{\v{c}}}, Jure and {Verheijen}, Marc and {Steele}, Iain and {Stuik}, Remko and {Kroes}, Gabby and {Tromp}, Neils and {Kragt}, Jan and {Lesman}, Dirk and {Mottram}, Chris and {Bates}, Stuart and {Gribbin}, Frank and {Burgal}, Jose Alonso and {Herreros}, Jos{\'e} Miguel and {Delgado}, Jos{\'e} Miguel and {Martin}, Carlos and {Cano}, Diego and {Navarro}, Ramon and {Irwin}, Mike and {Peralta de Arriba}, Luis and {O'Mahoney}, Neil and {Bianco}, Andrea and {Moleinezhad}, Alireza and {ter Horst}, Rik and {Molinari}, Emilio and {Lodi}, Marcello and {Guerra}, Jos{\'e} and {Baruffalo}, Andrea and {Carrasco}, Esperanza and {Farcas}, Szigfrid and {Schallig}, Ellen and {Hughes}, Sarah and {Hill}, Vanessa and {Smith}, Dan and {Drew}, Janet and {Poggianti}, Bianca and {Iovino}, Angela and {Pieri}, Mat and {Jin}, Shoko and {Dominguez Palmero}, Lillian and {Fari{\~n}a}, Cecilia and {Mart{\'\i}n}, Adrian and {Worley}, Clare and {Murphy}, David and {Guest}, Steve and {Morris}, Huw and {Elswijk}, Eddy and {de Haan}, Menno and {Hanenburg}, Hiddo and {Salasnich}, Bernardo and {Mayya}, Divakara and {Izazaga-P{\'e}rez}, Rafael and {Gafton}, Emanuel and {Caffau}, Elisabetta and {Horville}, David and {Paz Chinch{\'o}n}, Francisco and {Falcon-Barosso}, Jesus and {G{\"a}nsicke}, Boris and {San Juan}, Jose and {Hernandez}, Nauzet},
        title = "{Integration and early testing of WEAVE: the next-generation spectroscopy facility for the William Herschel Telescope}",
    booktitle = {Ground-based and Airborne Instrumentation for Astronomy VIII},
         year = 2020,
       editor = {{Evans}, Christopher J. and {Bryant}, Julia J. and {Motohara}, Kentaro},
       series = {Society of Photo-Optical Instrumentation Engineers (SPIE) Conference Series},
       volume = {11447},
        month = dec,
          eid = {1144714},
        pages = {1144714},
          doi = {10.1117/12.2561067},
       adsurl = {https://ui.adsabs.harvard.edu/abs/2020SPIE11447E..14D}
}

@ARTICLE{2016arXiv161100036D,
       author = {{DESI Collaboration} and {Aghamousa}, Amir and {Aguilar}, Jessica and {Ahlen}, Steve and {Alam}, Shadab and {Allen}, Lori E. and {Allende Prieto}, Carlos and {Annis}, James and {Bailey}, Stephen and {Balland}, Christophe and {Ballester}, Otger and {Baltay}, Charles and {Beaufore}, Lucas and {Bebek}, Chris and {Beers}, Timothy C. and {Bell}, Eric F. and {Bernal}, Jos{\'e} Luis and {Besuner}, Robert and {Beutler}, Florian and {Blake}, Chris and {Bleuler}, Hannes and {Blomqvist}, Michael and {Blum}, Robert and {Bolton}, Adam S. and {Briceno}, Cesar and {Brooks}, David and {Brownstein}, Joel R. and {Buckley-Geer}, Elizabeth and {Burden}, Angela and {Burtin}, Etienne and {Busca}, Nicolas G. and {Cahn}, Robert N. and {Cai}, Yan-Chuan and {Cardiel-Sas}, Laia and {Carlberg}, Raymond G. and {Carton}, Pierre-Henri and {Casas}, Ricard and {Castander}, Francisco J. and {Cervantes-Cota}, Jorge L. and {Claybaugh}, Todd M. and {Close}, Madeline and {Coker}, Carl T. and {Cole}, Shaun and {Comparat}, Johan and {Cooper}, Andrew P. and {Cousinou}, M. -C. and {Crocce}, Martin and {Cuby}, Jean-Gabriel and {Cunningham}, Daniel P. and {Davis}, Tamara M. and {Dawson}, Kyle S. and {de la Macorra}, Axel and {De Vicente}, Juan and {Delubac}, Timoth{\'e}e and {Derwent}, Mark and {Dey}, Arjun and {Dhungana}, Govinda and {Ding}, Zhejie and {Doel}, Peter and {Duan}, Yutong T. and {Ealet}, Anne and {Edelstein}, Jerry and {Eftekharzadeh}, Sarah and {Eisenstein}, Daniel J. and {Elliott}, Ann and {Escoffier}, St{\'e}phanie and {Evatt}, Matthew and {Fagrelius}, Parker and {Fan}, Xiaohui and {Fanning}, Kevin and {Farahi}, Arya and {Farihi}, Jay and {Favole}, Ginevra and {Feng}, Yu and {Fernandez}, Enrique and {Findlay}, Joseph R. and {Finkbeiner}, Douglas P. and {Fitzpatrick}, Michael J. and {Flaugher}, Brenna and {Flender}, Samuel and {Font-Ribera}, Andreu and {Forero-Romero}, Jaime E. and {Fosalba}, Pablo and {Frenk}, Carlos S. and {Fumagalli}, Michele and {Gaensicke}, Boris T. and {Gallo}, Giuseppe and {Garcia-Bellido}, Juan and {Gaztanaga}, Enrique and {Pietro Gentile Fusillo}, Nicola and {Gerard}, Terry and {Gershkovich}, Irena and {Giannantonio}, Tommaso and {Gillet}, Denis and {Gonzalez-de-Rivera}, Guillermo and {Gonzalez-Perez}, Violeta and {Gott}, Shelby and {Graur}, Or and {Gutierrez}, Gaston and {Guy}, Julien and {Habib}, Salman and {Heetderks}, Henry and {Heetderks}, Ian and {Heitmann}, Katrin and {Hellwing}, Wojciech A. and {Herrera}, David A. and {Ho}, Shirley and {Holland}, Stephen and {Honscheid}, Klaus and {Huff}, Eric and {Hutchinson}, Timothy A. and {Huterer}, Dragan and {Hwang}, Ho Seong and {Illa Laguna}, Joseph Maria and {Ishikawa}, Yuzo and {Jacobs}, Dianna and {Jeffrey}, Niall and {Jelinsky}, Patrick and {Jennings}, Elise and {Jiang}, Linhua and {Jimenez}, Jorge and {Johnson}, Jennifer and {Joyce}, Richard and {Jullo}, Eric and {Juneau}, St{\'e}phanie and {Kama}, Sami and {Karcher}, Armin and {Karkar}, Sonia and {Kehoe}, Robert and {Kennamer}, Noble and {Kent}, Stephen and {Kilbinger}, Martin and {Kim}, Alex G. and {Kirkby}, David and {Kisner}, Theodore and {Kitanidis}, Ellie and {Kneib}, Jean-Paul and {Koposov}, Sergey and {Kovacs}, Eve and {Koyama}, Kazuya and {Kremin}, Anthony and {Kron}, Richard and {Kronig}, Luzius and {Kueter-Young}, Andrea and {Lacey}, Cedric G. and {Lafever}, Robin and {Lahav}, Ofer and {Lambert}, Andrew and {Lampton}, Michael and {Landriau}, Martin and {Lang}, Dustin and {Lauer}, Tod R. and {Le Goff}, Jean-Marc and {Le Guillou}, Laurent and {Le Van Suu}, Auguste and {Lee}, Jae Hyeon and {Lee}, Su-Jeong and {Leitner}, Daniela and {Lesser}, Michael and {Levi}, Michael E. and {L'Huillier}, Benjamin and {Li}, Baojiu and {Liang}, Ming and {Lin}, Huan and {Linder}, Eric and {Loebman}, Sarah R. and {Luki{\'c}}, Zarija and {Ma}, Jun and {MacCrann}, Niall and {Magneville}, Christophe and {Makarem}, Laleh and {Manera}, Marc and {Manser}, Christopher J. and {Marshall}, Robert and {Martini}, Paul and {Massey}, Richard and {Matheson}, Thomas and {McCauley}, Jeremy and {McDonald}, Patrick and {McGreer}, Ian D. and {Meisner}, Aaron and {Metcalfe}, Nigel and {Miller}, Timothy N. and {Miquel}, Ramon and {Moustakas}, John and {Myers}, Adam and {Naik}, Milind and {Newman}, Jeffrey A. and {Nichol}, Robert C. and {Nicola}, Andrina and {Nicolati da Costa}, Luiz and {Nie}, Jundan and {Niz}, Gustavo and {Norberg}, Peder and {Nord}, Brian and {Norman}, Dara and {Nugent}, Peter and {O'Brien}, Thomas and {Oh}, Minji and {Olsen}, Knut A.~G. and {Padilla}, Cristobal and {Padmanabhan}, Hamsa and {Padmanabhan}, Nikhil and {Palanque-Delabrouille}, Nathalie and {Palmese}, Antonella and {Pappalardo}, Daniel and {P{\^a}ris}, Isabelle and {Park}, Changbom and {Patej}, Anna and {Peacock}, John A. and {Peiris}, Hiranya V. and {Peng}, Xiyan and {Percival}, Will J. and {Perruchot}, Sandrine and {Pieri}, Matthew M. and {Pogge}, Richard and {Pollack}, Jennifer E. and {Poppett}, Claire and {Prada}, Francisco and {Prakash}, Abhishek and {Probst}, Ronald G. and {Rabinowitz}, David and {Raichoor}, Anand and {Ree}, Chang Hee and {Refregier}, Alexandre and {Regal}, Xavier and {Reid}, Beth and {Reil}, Kevin and {Rezaie}, Mehdi and {Rockosi}, Constance M. and {Roe}, Natalie and {Ronayette}, Samuel and {Roodman}, Aaron and {Ross}, Ashley J. and {Ross}, Nicholas P. and {Rossi}, Graziano and {Rozo}, Eduardo and {Ruhlmann-Kleider}, Vanina and {Rykoff}, Eli S. and {Sabiu}, Cristiano and {Samushia}, Lado and {Sanchez}, Eusebio and {Sanchez}, Javier and {Schlegel}, David J. and {Schneider}, Michael and {Schubnell}, Michael and {Secroun}, Aur{\'e}lia and {Seljak}, Uros and {Seo}, Hee-Jong and {Serrano}, Santiago and {Shafieloo}, Arman and {Shan}, Huanyuan and {Sharples}, Ray and {Sholl}, Michael J. and {Shourt}, William V. and {Silber}, Joseph H. and {Silva}, David R. and {Sirk}, Martin M. and {Slosar}, Anze and {Smith}, Alex and {Smoot}, George F. and {Som}, Debopam and {Song}, Yong-Seon and {Sprayberry}, David and {Staten}, Ryan and {Stefanik}, Andy and {Tarle}, Gregory and {Sien Tie}, Suk and {Tinker}, Jeremy L. and {Tojeiro}, Rita and {Valdes}, Francisco and {Valenzuela}, Octavio and {Valluri}, Monica and {Vargas-Magana}, Mariana and {Verde}, Licia and {Walker}, Alistair R. and {Wang}, Jiali and {Wang}, Yuting and {Weaver}, Benjamin A. and {Weaverdyck}, Curtis and {Wechsler}, Risa H. and {Weinberg}, David H. and {White}, Martin and {Yang}, Qian and {Yeche}, Christophe and {Zhang}, Tianmeng and {Zhao}, Gong-Bo and {Zheng}, Yi and {Zhou}, Xu and {Zhou}, Zhimin and {Zhu}, Yaling and {Zou}, Hu and {Zu}, Ying},
        title = "{The DESI Experiment Part I: Science,Targeting, and Survey Design}",
      journal = {arXiv e-prints},
         year = 2016,
        month = oct,
          eid = {arXiv:1611.00036},
        pages = {arXiv:1611.00036},
          doi = {10.48550/arXiv.1611.00036},
archivePrefix = {arXiv},
       eprint = {1611.00036},
 primaryClass = {astro-ph.IM},
       adsurl = {https://ui.adsabs.harvard.edu/abs/2016arXiv161100036D}
}

@INPROCEEDINGS{2020SPIE11447E..1DB,
       author = {{Bundy}, Kevin and {Westfall}, Kyle B. and {MacDonald}, Nick and {Kupke}, Renate and {Poppett}, Claire and {Miller}, Timothy N. and {Lawrence}, Jon and {Saavedra Lacombea}, Celestina and {Yan}, Renbin and {Goodwin}, Michael and {Kassis}, Marc and {O'Meara}, John M. and {Masters}, Daniel C. and {Burchett}, Joseph N. and {Williams}, Benjamin F. and {Rich}, Robert M. and {Villar}, V.~A. and {Sandford}, Nathan and {Ting}, Yuan-Sen and {Hinz}, Phil and {Schafer}, Chad and {Mandelbaum}, Rachel and {Huang}, Marina and {Prochaska}, J.~X. and {Guhathakurta}, Puragra},
        title = "{The Keck-FOBOS spectroscopic facility: conceptual design}",
    booktitle = {Ground-based and Airborne Instrumentation for Astronomy VIII},
         year = 2020,
       editor = {{Evans}, Christopher J. and {Bryant}, Julia J. and {Motohara}, Kentaro},
       series = {Society of Photo-Optical Instrumentation Engineers (SPIE) Conference Series},
       volume = {11447},
        month = dec,
          eid = {114471D},
        pages = {114471D},
          doi = {10.1117/12.2562914},
       adsurl = {https://ui.adsabs.harvard.edu/abs/2020SPIE11447E..1DB}
}

@ARTICLE{2001MNRAS.328.1039C,
       author = {{Colless}, Matthew and {Dalton}, Gavin and {Maddox}, Steve and {Sutherland}, Will and {Norberg}, Peder and {Cole}, Shaun and {Bland-Hawthorn}, Joss and {Bridges}, Terry and {Cannon}, Russell and {Collins}, Chris and {Couch}, Warrick and {Cross}, Nicholas and {Deeley}, Kathryn and {De Propris}, Roberto and {Driver}, Simon P. and {Efstathiou}, George and {Ellis}, Richard S. and {Frenk}, Carlos S. and {Glazebrook}, Karl and {Jackson}, Carole and {Lahav}, Ofer and {Lewis}, Ian and {Lumsden}, Stuart and {Madgwick}, Darren and {Peacock}, John A. and {Peterson}, Bruce A. and {Price}, Ian and {Seaborne}, Mark and {Taylor}, Keith},
        title = "{The 2dF Galaxy Redshift Survey: spectra and redshifts}",
      journal = {\mnras},
         year = 2001,
        month = dec,
       volume = {328},
       number = {4},
        pages = {1039-1063},
          doi = {10.1046/j.1365-8711.2001.04902.x},
archivePrefix = {arXiv},
       eprint = {astro-ph/0106498},
 primaryClass = {astro-ph},
       adsurl = {https://ui.adsabs.harvard.edu/abs/2001MNRAS.328.1039C}
}

@inproceedings{10.1117/12.3019907,
    author       = {Roelof S. de Jong and Thomas Liebner and Frank Dionies},
    title        = {{FLEX: a new grid-based fibre positioner concept with large patrol area, small pitch, and very good clustering capabilities}},
    volume       = {13100},
    series       = {Proceedings of SPIE},
    booktitle    = {Advances in Optical and Mechanical Technologies for Telescopes and Instrumentation VI},
    editor       = {Ram{\'o}n Navarro and Ralf Jedamzik},
    organization = {International Society for Optics and Photonics},
    publisher    = {SPIE},
    pages        = {131002A},
    year         = {2024},
    doi          = {10.1117/12.3019907},
    URL          = {https://doi.org/10.1117/12.3019907}
}

@INPROCEEDINGS{2012SPIE.8446E..0PD,
       author = {{Dalton}, Gavin and {Trager}, Scott C. and {Abrams}, Don Carlos and {Carter}, David and {Bonifacio}, Piercarlo and {Aguerri}, J. Alfonso L. and {MacIntosh}, Mike and {Evans}, Chris and {Lewis}, Ian and {Navarro}, Ramon and {Agocs}, Tibor and {Dee}, Kevin and {Rousset}, Sophie and {Tosh}, Ian and {Middleton}, Kevin and {Pragt}, Johannes and {Terrett}, David and {Brock}, Matthew and {Benn}, Chris and {Verheijen}, Marc and {Cano Infantes}, Diego and {Bevil}, Craige and {Steele}, Iain and {Mottram}, Chris and {Bates}, Stuart and {Gribbin}, Francis J. and {Rey}, J{\"u}rg and {Rodriguez}, Luis Fernando and {Delgado}, Jose Miguel and {Guinouard}, Isabelle and {Walton}, Nic and {Irwin}, Michael J. and {Jagourel}, Pascal and {Stuik}, Remko and {Gerlofsma}, Gerrit and {Roelfsma}, Ronald and {Skillen}, Ian and {Ridings}, Andy and {Balcells}, Marc and {Daban}, Jean-Baptiste and {Gouvret}, Carole and {Venema}, Lars and {Girard}, Paul},
        title = "{WEAVE: the next generation wide-field spectroscopy facility for the William Herschel Telescope}",
    booktitle = {Ground-based and Airborne Instrumentation for Astronomy IV},
         year = 2012,
       editor = {{McLean}, Ian S. and {Ramsay}, Suzanne K. and {Takami}, Hideki},
       series = {Society of Photo-Optical Instrumentation Engineers (SPIE) Conference Series},
       volume = {8446},
        month = sep,
          eid = {84460P},
        pages = {84460P},
          doi = {10.1117/12.925950},
       adsurl = {https://ui.adsabs.harvard.edu/abs/2012SPIE.8446E..0PD}
}

@ARTICLE{2024MNRAS.530.2688J,
       author = {{Jin}, Shoko and {Trager}, Scott C. and {Dalton}, Gavin B. and {Aguerri}, J. Alfonso L. and {Drew}, J.~E. and {Falc{\'o}n-Barroso}, Jes{\'u}s and {G{\"a}nsicke}, Boris T. and {Hill}, Vanessa and {Iovino}, Angela and {Pieri}, Matthew M. and {Poggianti}, Bianca M. and {Smith}, D.~J.~B. and {Vallenari}, Antonella and {Abrams}, Don Carlos and {Aguado}, David S. and {Antoja}, Teresa and {Arag{\'o}n-Salamanca}, Alfonso and {Ascasibar}, Yago and {Babusiaux}, Carine and {Balcells}, Marc and {Barrena}, R. and {Battaglia}, Giuseppina and {Belokurov}, Vasily and {Bensby}, Thomas and {Bonifacio}, Piercarlo and {Bragaglia}, Angela and {Carrasco}, Esperanza and {Carrera}, Ricardo and {Cornwell}, Daniel J. and {Dom{\'\i}nguez-Palmero}, Lilian and {Duncan}, Kenneth J. and {Famaey}, Benoit and {Fari{\~n}a}, Cecilia and {Gonzalez}, Oscar A. and {Guest}, Steve and {Hatch}, Nina A. and {Hess}, Kelley M. and {Hoskin}, Matthew J. and {Irwin}, Mike and {Knapen}, Johan H. and {Koposov}, Sergey E. and {Kuchner}, Ulrike and {Laigle}, Clotilde and {Lewis}, Jim and {Longhetti}, Marcella and {Lucatello}, Sara and {M{\'e}ndez-Abreu}, Jairo and {Mercurio}, Amata and {Molaeinezhad}, Alireza and {Mongui{\'o}}, Maria and {Morrison}, Sean and {Murphy}, David N.~A. and {Peralta de Arriba}, Luis and {P{\'e}rez}, Isabel and {P{\'e}rez-R{\`a}fols}, Ignasi and {Pic{\'o}}, Sergio and {Raddi}, Roberto and {Romero-G{\'o}mez}, Merc{\`e} and {Royer}, Fr{\'e}d{\'e}ric and {Siebert}, Arnaud and {Seabroke}, George M. and {Som}, Debopam and {Terrett}, David and {Thomas}, Guillaume and {Wesson}, Roger and {Worley}, C. Clare and {Alfaro}, Emilio J. and {Allende Prieto}, Carlos and {Alonso-Santiago}, Javier and {Amos}, Nicholas J. and {Ashley}, Richard P. and {Balaguer-N{\'u}{\~n}ez}, Lola and {Balbinot}, Eduardo and {Bellazzini}, Michele and {Benn}, Chris R. and {Berlanas}, Sara R. and {Bernard}, Edouard J. and {Best}, Philip and {Bettoni}, Daniela and {Bianco}, Andrea and {Bishop}, Georgia and {Blomqvist}, Michael and {Boeche}, Corrado and {Bolzonella}, Micol and {Bonoli}, Silvia and {Bosma}, Albert and {Britavskiy}, Nikolay and {Busarello}, Gianni and {Caffau}, Elisabetta and {Cantat-Gaudin}, Tristan and {Castro-Ginard}, Alfred and {Couto}, Guilherme and {Carbajo-Hijarrubia}, Juan and {Carter}, David and {Casamiquela}, Laia and {Conrado}, Ana M. and {Corcho-Caballero}, Pablo and {Costantin}, Luca and {Deason}, Alis and {de Burgos}, Abel and {De Grandi}, Sabrina and {Di Matteo}, Paola and {Dom{\'\i}nguez-G{\'o}mez}, Jes{\'u}s and {Dorda}, Ricardo and {Drake}, Alyssa and {Dutta}, Rajeshwari and {Erkal}, Denis and {Feltzing}, Sofia and {Ferr{\'e}-Mateu}, Anna and {Feuillet}, Diane and {Figueras}, Francesca and {Fossati}, Matteo and {Franciosini}, Elena and {Frasca}, Antonio and {Fumagalli}, Michele and {Gallazzi}, Anna and {Garc{\'\i}a-Benito}, Rub{\'e}n and {Gentile Fusillo}, Nicola and {Gebran}, Marwan and {Gilbert}, James and {Gledhill}, T.~M. and {Gonz{\'a}lez Delgado}, Rosa M. and {Greimel}, Robert and {Guarcello}, Mario Giuseppe and {Guerra}, Jose and {Gullieuszik}, Marco and {Haines}, Christopher P. and {Hardcastle}, Martin J. and {Harris}, Amy and {Haywood}, Misha and {Helmi}, Amina and {Hernandez}, Nauzet and {Herrero}, Artemio and {Hughes}, Sarah and {Ir{\v{s}}i{\v{c}}}, Vid and {Jablonka}, Pascale and {Jarvis}, Matt J. and {Jordi}, Carme and {Kondapally}, Rohit and {Kordopatis}, Georges and {Krogager}, Jens-Kristian and {La Barbera}, Francesco and {Lam}, Man I. and {Larsen}, S{\o}ren S. and {Lemasle}, Bertrand and {Lewis}, Ian J. and {Lhom{\'e}}, Emilie and {Lind}, Karin and {Lodi}, Marcello and {Longobardi}, Alessia and {Lonoce}, Ilaria and {Magrini}, Laura and {Ma{\'\i}z Apell{\'a}niz}, Jes{\'u}s and {Marchal}, Olivier and {Marco}, Amparo and {Martin}, Nicolas F. and {Matsuno}, Tadafumi and {Maurogordato}, Sophie and {Merluzzi}, Paola and {Miralda-Escud{\'e}}, Jordi and {Molinari}, Emilio and {Monari}, Giacomo and {Morelli}, Lorenzo and {Mottram}, Christopher J. and {Naylor}, Tim and {Negueruela}, Ignacio and {O{\~n}orbe}, Jose and {Pancino}, Elena and {Peirani}, S{\'e}bastien and {Peletier}, Reynier F. and {Pozzetti}, Lucia and {Rainer}, Monica and {Ramos}, Pau and {Read}, Shaun C. and {Rossi}, Elena Maria and {R{\"o}ttgering}, Huub J.~A. and {Rubi{\~n}o-Mart{\'\i}n}, Jose Alberto and {Sabater}, Jose and {San Juan}, Jos{\'e} and {Sanna}, Nicoletta and {Schallig}, Ellen and {Schiavon}, Ricardo P. and {Schultheis}, Mathias and {Serra}, Paolo and {Shimwell}, Timothy W. and {Sim{\'o}n-D{\'\i}az}, Sergio and {Smith}, Russell J. and {Sordo}, Rosanna and {Sorini}, Daniele and {Soubiran}, Caroline and {Starkenburg}, Else and {Steele}, Iain A. and {Stott}, John and {Stuik}, Remko and {Tolstoy}, Eline and {Tortora}, Crescenzo and {Tsantaki}, Maria and {Van der Swaelmen}, Mathieu and {van Weeren}, Reinout J. and {Vergani}, Daniela},
        title = "{The wide-field, multiplexed, spectroscopic facility WEAVE: Survey design, overview, and simulated implementation}",
      journal = {\mnras},
         year = 2024,
        month = may,
       volume = {530},
       number = {3},
        pages = {2688-2730},
          doi = {10.1093/mnras/stad557},
archivePrefix = {arXiv},
       eprint = {2212.03981},
 primaryClass = {astro-ph.IM},
       adsurl = {https://ui.adsabs.harvard.edu/abs/2024MNRAS.530.2688J}
}

@ARTICLE{2010arXiv1006.3102K,
       author = {{Kimura}, Masahiko and {Maihara}, Toshinori and {Iwamuro}, Fumihide and {Akiyama}, Masayuki and {Tamura}, Naoyuki and {Dalton}, Gavin B. and {Takato}, Naruhisa and {Tait}, Philip and {Ohta}, Kouji and {Eto}, Shigeru and {Mochida}, Daisaku and {Elms}, Brian and {Kawate}, Kaori and {Kurakami}, Tomio and {Moritani}, Yuuki and {Noumaru}, Junichi and {Ohshima}, Norio and {Sumiyoshi}, Masanao and {Yabe}, Kiyoto and {Brzeski}, Jurek and {Farrell}, Tony and {Frost}, Gabriella and {Gillingham}, Peter R. and {Haynes}, Roger and {Moore}, Anna M. and {Muller}, Rolf and {Smedley}, Scott and {Smith}, Greg and {Bonfield}, David G. and {Brooks}, Charles B. and {Holmes}, Alan R. and {Lake}, Emma Curtis and {Lee}, Hanshin and {Lewis}, Ian J. and {Froud}, Tim R. and {Tosh}, Ian A. and {Woodhouse}, Guy F. and {Blackburn}, Colin and {Dipper}, Nigel and {Murray}, Graham and {Sharples}, Ray and {Robertson}, David J.},
        title = "{The Fibre Multi-Object Spectrograph (FMOS) for Subaru Telescope}",
      journal = {arXiv e-prints},
         year = 2010,
        month = jun,
          eid = {arXiv:1006.3102},
        pages = {arXiv:1006.3102},
          doi = {10.48550/arXiv.1006.3102},
archivePrefix = {arXiv},
       eprint = {1006.3102},
 primaryClass = {astro-ph.IM},
       adsurl = {https://ui.adsabs.harvard.edu/abs/2010arXiv1006.3102K}
}

@INPROCEEDINGS{2018SPIE10702E..1MS,
       author = {{Smedley}, Scott and {Baker}, Gabriella and {Brown}, Rebecca and {Gilbert}, James and {Gillingham}, Peter and {Saunders}, Will and {Sheinis}, Andrew and {Venkatesan}, Sudharshan and {Waller}, Lew},
        title = "{Sphinx: a massively multiplexed fiber positioner for MSE}",
    booktitle = {Ground-based and Airborne Instrumentation for Astronomy VII},
         year = 2018,
       editor = {{Evans}, Christopher J. and {Simard}, Luc and {Takami}, Hideki},
       series = {Society of Photo-Optical Instrumentation Engineers (SPIE) Conference Series},
       volume = {10702},
        month = jul,
          eid = {107021M},
        pages = {107021M},
          doi = {10.1117/12.2310021},
archivePrefix = {arXiv},
       eprint = {1807.09181},
 primaryClass = {astro-ph.IM},
       adsurl = {https://ui.adsabs.harvard.edu/abs/2018SPIE10702E..1MS}
}

@INPROCEEDINGS{2022SPIE12184E..0YB,
       author = {{Brown}, Rebecca and {Farrell}, Tony and {Fernando}, Nuwanthika and {Goodwin}, Michael and {Horton}, Anthony and {Lacombe}, Celestina and {Lawrence}, Jon and {Lorente}, Nuria and {McGregor}, Helen and {O'Brien}, Ellie and {Waller}, Lewis and {Zafar}, Tayyaba and {Zheng}, Jessica},
        title = "{TAIPAN starbugs: commissioning and the start of science observations}",
    booktitle = {Ground-based and Airborne Instrumentation for Astronomy IX},
         year = 2022,
       editor = {{Evans}, Christopher J. and {Bryant}, Julia J. and {Motohara}, Kentaro},
       series = {Society of Photo-Optical Instrumentation Engineers (SPIE) Conference Series},
       volume = {12184},
        month = aug,
          eid = {121840Y},
        pages = {121840Y},
          doi = {10.1117/12.2629109},
       adsurl = {https://ui.adsabs.harvard.edu/abs/2022SPIE12184E..0YB}
}

@INPROCEEDINGS{2014SPIE.9151E..1YF,
       author = {{Fisher}, Charles and {Morantz}, Chaz and {Braun}, David and {Seiffert}, Michael and {Aghazarian}, Hrand and {Partos}, Eamon and {King}, Matthew and {Hovland}, Larry E. and {Schwochert}, Mark and {Kaluzny}, Joel and {Capocasale}, Christopher and {Houck}, Andrew and {Gross}, Johannes and {Reiley}, Daniel and {Mao}, Peter and {Riddle}, Reed and {Bui}, Khanh and {Henderson}, David and {Haran}, Todd and {Culhane}, Robert and {Piazza}, Daniele and {Walkama}, Eric},
        title = "{Developing engineering model Cobra fiber positioners for the Subaru Telescope's prime focus spectrometer}",
    booktitle = {Advances in Optical and Mechanical Technologies for Telescopes and Instrumentation},
         year = 2014,
       editor = {{Navarro}, Ram{\'o}n and {Cunningham}, Colin R. and {Barto}, Allison A.},
       series = {Society of Photo-Optical Instrumentation Engineers (SPIE) Conference Series},
       volume = {9151},
        month = jul,
          eid = {91511Y},
        pages = {91511Y},
          doi = {10.1117/12.2054700},
archivePrefix = {arXiv},
       eprint = {1408.2833},
 primaryClass = {astro-ph.IM},
       adsurl = {https://ui.adsabs.harvard.edu/abs/2014SPIE.9151E..1YF}
}

@ARTICLE{2024AJ....167...62D,
       author = {{DESI Collaboration} and {Adame}, A.~G. and {Aguilar}, J. and {Ahlen}, S. and {Alam}, S. and {Aldering}, G. and {Alexander}, D.~M. and {Alfarsy}, R. and {Allende Prieto}, C. and {Alvarez}, M. and {Alves}, O. and {Anand}, A. and {Andrade-Oliveira}, F. and {Armengaud}, E. and {Asorey}, J. and {Avila}, S. and {Aviles}, A. and {Bailey}, S. and {Balaguera-Antol{\'\i}nez}, A. and {Ballester}, O. and {Baltay}, C. and {Bault}, A. and {Bautista}, J. and {Behera}, J. and {Beltran}, S.~F. and {BenZvi}, S. and {Beraldo e Silva}, L. and {Bermejo-Climent}, J.~R. and {Berti}, A. and {Besuner}, R. and {Beutler}, F. and {Bianchi}, D. and {Blake}, C. and {Blum}, R. and {Bolton}, A.~S. and {Brieden}, S. and {Brodzeller}, A. and {Brooks}, D. and {Brown}, Z. and {Buckley-Geer}, E. and {Burtin}, E. and {Cabayol-Garcia}, L. and {Cai}, Z. and {Canning}, R. and {Cardiel-Sas}, L. and {Carnero Rosell}, A. and {Castander}, F.~J. and {Cervantes-Cota}, J.~L. and {Chabanier}, S. and {Chaussidon}, E. and {Chaves-Montero}, J. and {Chen}, S. and {Chen}, X. and {Chuang}, C. and {Claybaugh}, T. and {Cole}, S. and {Cooper}, A.~P. and {Cuceu}, A. and {Davis}, T.~M. and {Dawson}, K. and {de Belsunce}, R. and {de la Cruz}, R. and {de la Macorra}, A. and {de Mattia}, A. and {Demina}, R. and {Demirbozan}, U. and {DeRose}, J. and {Dey}, A. and {Dey}, B. and {Dhungana}, G. and {Ding}, J. and {Ding}, Z. and {Doel}, P. and {Doshi}, R. and {Douglass}, K. and {Edge}, A. and {Eftekharzadeh}, S. and {Eisenstein}, D.~J. and {Elliott}, A. and {Escoffier}, S. and {Fagrelius}, P. and {Fan}, X. and {Fanning}, K. and {Fawcett}, V.~A. and {Ferraro}, S. and {Ereza}, J. and {Flaugher}, B. and {Font-Ribera}, A. and {Forero-S{\'a}nchez}, D. and {Forero-Romero}, J.~E. and {Frenk}, C.~S. and {G{\"a}nsicke}, B.~T. and {Garc{\'\i}a}, L. {\'A}. and {Garc{\'\i}a-Bellido}, J. and {Garcia-Quintero}, C. and {Garrison}, L.~H. and {Gil-Mar{\'\i}n}, H. and {Golden-Marx}, J. and {Gontcho A Gontcho}, S. and {Gonzalez-Morales}, A.~X. and {Gonzalez-Perez}, V. and {Gordon}, C. and {Graur}, O. and {Green}, D. and {Gruen}, D. and {Guy}, J. and {Hadzhiyska}, B. and {Hahn}, C. and {Han}, J.~J. and {Hanif}, M.~M.~S. and {Herrera-Alcantar}, H.~K. and {Honscheid}, K. and {Hou}, J. and {Howlett}, C. and {Huterer}, D. and {Ir{\v{s}}i{\v{c}}}, V. and {Ishak}, M. and {Jana}, A. and {Jiang}, L. and {Jimenez}, J. and {Jing}, Y.~P. and {Joudaki}, S. and {Jullo}, E. and {Joyce}, R. and {Juneau}, S. and {Kizhuprakkat}, N. and {Kara{\c{c}}ayl{\i}}, N.~G. and {Karim}, T. and {Kehoe}, R. and {Kent}, S. and {Khederlarian}, A. and {Kim}, S. and {Kirkby}, D. and {Kisner}, T. and {Kitaura}, F. and {Kneib}, J. and {Koposov}, S.~E. and {Kov{\'a}cs}, A. and {Kremin}, A. and {Krolewski}, A. and {L'Huillier}, B. and {Lahav}, O. and {Lambert}, A. and {Lamman}, C. and {Lan}, T. -W. and {Landriau}, M. and {Lang}, D. and {Lange}, J.~U. and {Lasker}, J. and {Le Guillou}, L. and {Leauthaud}, A. and {Levi}, M.~E. and {Li}, T.~S. and {Linder}, E. and {Lyons}, A. and {Magneville}, C. and {Manera}, M. and {Manser}, C.~J. and {Margala}, D. and {Martini}, P. and {McDonald}, P. and {Medina}, G.~E. and {Medina-Varela}, L. and {Meisner}, A. and {Mena-Fern{\'a}ndez}, J. and {Meneses-Rizo}, J. and {Mezcua}, M. and {Miquel}, R. and {Montero-Camacho}, P. and {Moon}, J. and {Moore}, S. and {Moustakas}, J. and {Mueller}, E. and {Mundet}, J. and {Mu{\~n}oz-Guti{\'e}rrez}, A. and {Myers}, A.~D. and {Nadathur}, S. and {Napolitano}, L. and {Neveux}, R. and {Newman}, J.~A. and {Nie}, J. and {Niz}, G. and {Norberg}, P. and {Noriega}, H.~E. and {Paillas}, E. and {Palanque-Delabrouille}, N. and {Palmese}, A. and {Zhiwei}, P. and {Parkinson}, D. and {Penmetsa}, S. and {Percival}, W.~J. and {P{\'e}rez-Fern{\'a}ndez}, A. and {P{\'e}rez-R{\`a}fols}, I. and {Pieri}, M. and {Poppett}, C. and {Porredon}, A. and {Prada}, F. and {Pucha}, R. and {Raichoor}, A. and {Ram{\'\i}rez-P{\'e}rez}, C.},
        title = "{Validation of the Scientific Program for the Dark Energy Spectroscopic Instrument}",
      journal = {\aj},
         year = 2024,
        month = feb,
       volume = {167},
       number = {2},
          eid = {62},
        pages = {62},
          doi = {10.3847/1538-3881/ad0b08},
archivePrefix = {arXiv},
       eprint = {2306.06307},
 primaryClass = {astro-ph.CO},
       adsurl = {https://ui.adsabs.harvard.edu/abs/2024AJ....167...62D}
}

@ARTICLE{2012RAA....12.1197C,
       author = {{Cui}, Xiang-Qun and {Zhao}, Yong-Heng and {Chu}, Yao-Quan and {Li}, Guo-Ping and {Li}, Qi and {Zhang}, Li-Ping and {Su}, Hong-Jun and {Yao}, Zheng-Qiu and {Wang}, Ya-Nan and {Xing}, Xiao-Zheng and {Li}, Xin-Nan and {Zhu}, Yong-Tian and {Wang}, Gang and {Gu}, Bo-Zhong and {Luo}, A. -Li and {Xu}, Xin-Qi and {Zhang}, Zhen-Chao and {Liu}, Gen-Rong and {Zhang}, Hao-Tong and {Yang}, De-Hua and {Cao}, Shu-Yun and {Chen}, Hai-Yuan and {Chen}, Jian-Jun and {Chen}, Kun-Xin and {Chen}, Ying and {Chu}, Jia-Ru and {Feng}, Lei and {Gong}, Xue-Fei and {Hou}, Yong-Hui and {Hu}, Hong-Zhuan and {Hu}, Ning-Sheng and {Hu}, Zhong-Wen and {Jia}, Lei and {Jiang}, Fang-Hua and {Jiang}, Xiang and {Jiang}, Zi-Bo and {Jin}, Ge and {Li}, Ai-Hua and {Li}, Yan and {Li}, Ye-Ping and {Liu}, Guan-Qun and {Liu}, Zhi-Gang and {Lu}, Wen-Zhi and {Mao}, Yin-Dun and {Men}, Li and {Qi}, Yong-Jun and {Qi}, Zhao-Xiang and {Shi}, Huo-Ming and {Tang}, Zheng-Hong and {Tao}, Qing-Sheng and {Wang}, Da-Qi and {Wang}, Dan and {Wang}, Guo-Min and {Wang}, Hai and {Wang}, Jia-Ning and {Wang}, Jian and {Wang}, Jian-Ling and {Wang}, Jian-Ping and {Wang}, Lei and {Wang}, Shu-Qing and {Wang}, You and {Wang}, Yue-Fei and {Xu}, Ling-Zhe and {Xu}, Yan and {Yang}, Shi-Hai and {Yu}, Yong and {Yuan}, Hui and {Yuan}, Xiang-Yan and {Zhai}, Chao and {Zhang}, Jing and {Zhang}, Yan-Xia and {Zhang}, Yong and {Zhao}, Ming and {Zhou}, Fang and {Zhou}, Guo-Hua and {Zhu}, Jie and {Zou}, Si-Cheng},
        title = "{The Large Sky Area Multi-Object Fiber Spectroscopic Telescope (LAMOST)}",
      journal = {Research in Astronomy and Astrophysics},
         year = 2012,
        month = sep,
       volume = {12},
       number = {9},
        pages = {1197-1242},
          doi = {10.1088/1674-4527/12/9/003},
       adsurl = {https://ui.adsabs.harvard.edu/abs/2012RAA....12.1197C}
}

@INPROCEEDINGS{2022SPIE12189E..0VB,
       author = {{Beard}, Steven and {Willemse}, Bart and {Watson}, Stephen and {Atkinson}, David and {Gutierrez Cheetham}, Pablo and {Franzetti}, Paolo and {Nix}, Johannes},
        title = "{MOONS fibre positioner control and path planning software}",
    booktitle = {Software and Cyberinfrastructure for Astronomy VII},
         year = 2022,
       series = {Society of Photo-Optical Instrumentation Engineers (SPIE) Conference Series},
       volume = {12189},
        month = aug,
          eid = {121890V},
        pages = {121890V},
          doi = {10.1117/12.2629239},
       adsurl = {https://ui.adsabs.harvard.edu/abs/2022SPIE12189E..0VB}
}

@INPROCEEDINGS{2019BAAS...51g.198B,
       author = {{Bundy}, Kevin and {Westfall}, K. and {MacDonald}, N. and {Kupke}, R. and {Savage}, M. and {Poppett}, C. and {Alabi}, A. and {Becker}, G. and {Burchett}, J. and {Capak}, P. and {Coil}, A. and {Cooper}, M. and {Cowley}, D. and {Deich}, W. and {Dillon}, D. and {Edelstein}, J. and {Guhathakurta}, P. and {Hennawi}, J. and {Kassis}, M. and {Lee}, K. -G. and {Masters}, D. and {Miller}, T. and {Newman}, J. and {O'Meara}, J. and {Prochaska}, J.~X. and {Rau}, M. and {Rhodes}, J. and {Rich}, R.~M. and {Rockosi}, C. and {Romanowsky}, A. and {Schafer}, C. and {Schlegel}, D. and {Shapley}, A. and {Siana}, B. and {Ting}, Y. -S. and {Weisz}, D. and {White}, M. and {Williams}, B. and {Wilson}, G. and {Wilson}, M. and {Yan}, R.},
        title = "{FOBOS: A Next-Generation Spectroscopic Facility}",
    booktitle = {Bulletin of the American Astronomical Society},
         year = 2019,
       volume = {51},
        month = sep,
          eid = {198},
        pages = {198},
          doi = {10.48550/arXiv.1907.07195},
archivePrefix = {arXiv},
       eprint = {1907.07195},
 primaryClass = {astro-ph.IM},
       adsurl = {https://ui.adsabs.harvard.edu/abs/2019BAAS...51g.198B}
}

@INPROCEEDINGS{2012SPIE.8450E..1AG,
       author = {{Gilbert}, James and {Goodwin}, Michael and {Heijmans}, Jeroen and {Muller}, Rolf and {Miziarski}, Stan and {Brzeski}, Jurek and {Waller}, Lew and {Saunders}, Will and {Bennet}, Alex and {Tims}, Julia},
        title = "{Starbugs: all-singing, all-dancing fibre positioning robots}",
    booktitle = {Modern Technologies in Space- and Ground-based Telescopes and Instrumentation II},
         year = 2012,
       editor = {{Navarro}, Ram{\'o}n and {Cunningham}, Colin R. and {Prieto}, Eric},
       series = {Society of Photo-Optical Instrumentation Engineers (SPIE) Conference Series},
       volume = {8450},
        month = sep,
          eid = {84501A},
        pages = {84501A},
          doi = {10.1117/12.924502},
archivePrefix = {arXiv},
       eprint = {1311.7371},
 primaryClass = {astro-ph.IM},
       adsurl = {https://ui.adsabs.harvard.edu/abs/2012SPIE.8450E..1AG}
}

@INPROCEEDINGS{2014SPIE.9151E..1AB,
       author = {{Brown}, David M. and {Case}, Scott and {Gilbert}, James and {Goodwin}, Michael and {Jacobs}, Daniel and {Kuehn}, Kyler and {Lawrence}, Jon and {Lorente}, Nuria P.~F. and {Nichani}, Vijay and {Saunders}, Will and {Staszac}, Nick and {Tims}, Julia},
        title = "{Starbug fibre positioning robots: performance and reliability enhancements}",
    booktitle = {Advances in Optical and Mechanical Technologies for Telescopes and Instrumentation},
         year = 2014,
       editor = {{Navarro}, Ram{\'o}n and {Cunningham}, Colin R. and {Barto}, Allison A.},
       series = {Society of Photo-Optical Instrumentation Engineers (SPIE) Conference Series},
       volume = {9151},
        month = jul,
          eid = {91511A},
        pages = {91511A},
          doi = {10.1117/12.2055594},
archivePrefix = {arXiv},
       eprint = {1805.10761},
 primaryClass = {astro-ph.IM},
       adsurl = {https://ui.adsabs.harvard.edu/abs/2014SPIE.9151E..1AB}
}

@INPROCEEDINGS{2022SPIE12182E..36O,
       author = {{O'Brien}, Ellie G. and {Lawrence}, Jon and {Lacombe}, Celestina S. and {Thomakos}, Michael and {Goodwin}, Michael and {Gilbert}, James and {Mali}, Slavko and {Muller}, Rolf and {Kunwar}, Nirmala and {Zahoor}, Jahanzeb and {Waller}, Lew and {Farrell}, Tony},
        title = "{A Starbug's life: a material trade study using fatigue life criteria for high-altitude robotic fibre positioning instruments}",
    booktitle = {Ground-based and Airborne Telescopes IX},
         year = 2022,
       editor = {{Marshall}, Heather K. and {Spyromilio}, Jason and {Usuda}, Tomonori},
       series = {Society of Photo-Optical Instrumentation Engineers (SPIE) Conference Series},
       volume = {12182},
        month = aug,
          eid = {1218236},
        pages = {1218236},
          doi = {10.1117/12.2629152},
       adsurl = {https://ui.adsabs.harvard.edu/abs/2022SPIE12182E..36O}
}

@ARTICLE{2016arXiv160600060M,
       author = {{McConnachie}, Alan W. and {Babusiaux}, Carine and {Balogh}, Michael and {Caffau}, Elisabetta and {C{\^o}t{\'e}}, Pat and {Driver}, Simon and {Robotham}, Aaron and {Starkenburg}, Else and {Venn}, Kim and {Walker}, Matthew and {Bauman}, Steven E. and {Flagey}, Nicolas and {Ho}, Kevin and {Isani}, Sidik and {Laychak}, Mary Beth and {Mignot}, Shan and {Murowinski}, Rick and {Salmon}, Derrick and {Simons}, Doug and {Szeto}, Kei and {Vermeulen}, Tom and {Withington}, Kanoa},
        title = "{A concise overview of the Maunakea Spectroscopic Explorer}",
      journal = {arXiv e-prints},
         year = 2016,
        month = may,
          eid = {arXiv:1606.00060},
        pages = {arXiv:1606.00060},
          doi = {10.48550/arXiv.1606.00060},
archivePrefix = {arXiv},
       eprint = {1606.00060},
 primaryClass = {astro-ph.IM},
       adsurl = {https://ui.adsabs.harvard.edu/abs/2016arXiv160600060M}
}

@ARTICLE{2022arXiv220904322S,
       author = {{Schlegel}, David J. and {Kollmeier}, Juna A. and {Aldering}, Greg and {Bailey}, Stephen and {Baltay}, Charles and {Bebek}, Christopher and {BenZvi}, Segev and {Besuner}, Robert and {Blanc}, Guillermo and {Bolton}, Adam S. and {Bonaca}, Ana and {Bouri}, Mohamed and {Brooks}, David and {Buckley-Geer}, Elizabeth and {Cai}, Zheng and {Crane}, Jeffrey and {Demina}, Regina and {DeRose}, Joseph and {Dey}, Arjun and {Doel}, Peter and {Fan}, Xiaohui and {Ferraro}, Simone and {Finkbeiner}, Douglas and {Font-Ribera}, Andreu and {Gontcho}, Satya Gontcho A and {Green}, Daniel and {Gutierrez}, Gaston and {Guy}, Julien and {Heetderks}, Henry and {Huterer}, Dragan and {Infante}, Leopoldo and {Jelinsky}, Patrick and {Karagiannis}, Dionysios and {Kent}, Stephen M. and {Kim}, Alex G. and {Kneib}, Jean-Paul and {Kremin}, Anthony and {Kronig}, Luzius and {Konidaris}, Nick and {Lahav}, Ofer and {Lampton}, Michael L. and {Landriau}, Martin and {Lang}, Dustin and {Leauthaud}, Alexie and {Levi}, Michael E. and {Liguori}, Michele and {Linder}, Eric V. and {Magneville}, Christophe and {Martini}, Paul and {Mateo}, Mario and {McDonald}, Patrick and {Miller}, Christopher J. and {Moustakas}, John and {Myers}, Adam D. and {Mulchaey}, John and {Newman}, Jeffrey A. and {Nugent}, Peter E. and {Padmanabhan}, Nikhil and {Palanque-Delabrouille}, Nathalie and {Piro}, Antonella Palmese Anthony L. and {Poppett}, Claire and {Prochaska}, Jason X. and {Pullen}, Anthony R. and {Rabinowitz}, David and {Raichoor}, Anand and {Ramirez}, Solange and {Rix}, Hans-Walter and {Ross}, Ashley J. and {Samushia}, Lado and {Schaan}, Emmanuel and {Schubnell}, Michael and {Seljak}, Uros and {Seo}, Hee-Jong and {Shectman}, Stephen A. and {Schlafly}, Edward F. and {Silber}, Joseph and {Simon}, Joshua D. and {Slepian}, Zachary and {Slosar}, An{\v{z}}e and {Soares-Santos}, Marcelle and {Tarl{\'e}}, Greg and {Thompson}, Ian and {Valluri}, Monica and {Wechsler}, Risa H. and {White}, Martin and {Wilson}, Michael J. and {Y{\`e}che}, Christophe and {Zaritsky}, Dennis and {Zhou}, Rongpu},
        title = "{The MegaMapper: A Stage-5 Spectroscopic Instrument Concept for the Study of Inflation and Dark Energy}",
      journal = {arXiv e-prints},
         year = 2022,
        month = sep,
          eid = {arXiv:2209.04322},
        pages = {arXiv:2209.04322},
          doi = {10.48550/arXiv.2209.04322},
archivePrefix = {arXiv},
       eprint = {2209.04322},
 primaryClass = {astro-ph.IM},
       adsurl = {https://ui.adsabs.harvard.edu/abs/2022arXiv220904322S}
}

@inproceedings{10.1117/12.3018093,
author = {Roland Bacon and Vincenzo Maineiri and Sofia Randich and Andrea Cimatti and Jean-Paul Kneib and Jarle Brinchmann and Richard Ellis and Eline Tolstoi and Rodolfo Smiljanic and Vanessa Hill and Richard I. Anderson and Paula Sanchez Saez and Cyrielle Opitom and Ian Bryson and Philippe Dierickx and Bianca Garilli and Oscar Gonzalez and Roelof de Jong and David Lee and Steffen Mieske and Angel Otarola and Pietro Schipani and Tony Travouillon and Joel Vernet and Julia Bryant and Marc Casali and Matthew Colless and Warrick Couch and Simon Driver and Adriano Fontana and Matthew Lehnert and Laura Magrini and Ben Montet and Luca Pasquini and Martin Roth and Ruben Sanchez-Janssen and Matthias Steinmetz and Laurence Tresse and Christophe Yeche and Bodo Ziegler},
title = {{WST - Widefield Spectroscopic Telescope: motivation, science drivers and top level requirements for a new dedicated facility}},
volume = {13094},
booktitle = {Ground-based and Airborne Telescopes X},
editor = {Heather K. Marshall and Jason Spyromilio and Tomonori Usuda},
series = {International Society for Optics and Photonics},
publisher = {SPIE},
pages = {130941O},
year = {2024},
doi = {10.1117/12.3018093},
URL = {https://doi.org/10.1117/12.3018093}
}

@ARTICLE{2024arXiv240305398M,
       author = {{Mainieri}, Vincenzo and {Anderson}, Richard I. and {Brinchmann}, Jarle and {Cimatti}, Andrea and {Ellis}, Richard S. and {Hill}, Vanessa and {Kneib}, Jean-Paul and {McLeod}, Anna F. and {Opitom}, Cyrielle and {Roth}, Martin M. and {Sanchez-Saez}, Paula and {Smiljanic}, Rodolfo and {Tolstoy}, Eline and {Bacon}, Roland and {Randich}, Sofia and {Adamo}, Angela and {Annibali}, Francesca and {Arevalo}, Patricia and {Audard}, Marc and {Barsanti}, Stefania and {Battaglia}, Giuseppina and {Bayo Aran}, Amelia M. and {Belfiore}, Francesco and {Bellazzini}, Michele and {Bellini}, Emilio and {Beltran}, Maria Teresa and {Berni}, Leda and {Bianchi}, Simone and {Biazzo}, Katia and {Bisero}, Sofia and {Bisogni}, Susanna and {Bland-Hawthorn}, Joss and {Blondin}, Stephane and {Bodensteiner}, Julia and {Boffin}, Henri M.~J. and {Bonito}, Rosaria and {Bono}, Giuseppe and {Bouche}, Nicolas F. and {Bowman}, Dominic and {Braga}, Vittorio F. and {Bragaglia}, Angela and {Branchesi}, Marica and {Brucalassi}, Anna and {Bryant}, Julia J. and {Bryson}, Ian and {Busa}, Innocenza and {Camera}, Stefano and {Carbone}, Carmelita and {Casali}, Giada and {Casali}, Mark and {Casasola}, Viviana and {Castro}, Norberto and {Catelan}, Marcio and {Cavallo}, Lorenzo and {Chiappini}, Cristina and {Cioni}, Maria-Rosa and {Colless}, Matthew and {Colzi}, Laura and {Contarini}, Sofia and {Couch}, Warrick and {D'Ammando}, Filippo and {d'Assignies D.}, William and {D'Orazi}, Valentina and {da Silva}, Ronaldo and {Dainotti}, Maria Giovanna and {Damiani}, Francesco and {Danielski}, Camilla and {De Cia}, Annalisa and {de Jong}, Roelof S. and {Dhawan}, Suhail and {Dierickx}, Philippe and {Driver}, Simon P. and {Dupletsa}, Ulyana and {Escoffier}, Stephanie and {Escorza}, Ana and {Fabrizio}, Michele and {Fiorentino}, Giuliana and {Fontana}, Adriano and {Fontani}, Francesco and {Forero Sanchez}, Daniel and {Franois}, Patrick and {Galindo-Guil}, Francisco Jose and {Gallazzi}, Anna Rita and {Galli}, Daniele and {Garcia}, Miriam and {Garcia-Rojas}, Jorge and {Garilli}, Bianca and {Grand}, Robert and {Guarcello}, Mario Giuseppe and {Hazra}, Nandini and {Helmi}, Amina and {Herrero}, Artemio and {Iglesias}, Daniela and {Ilic}, Dragana and {Irsic}, Vid and {Ivanov}, Valentin D. and {Izzo}, Luca and {Jablonka}, Pascale and {Joachimi}, Benjamin and {Kakkad}, Darshan and {Kamann}, Sebastian and {Koposov}, Sergey and {Kordopatis}, Georges and {Kovacevic}, Andjelka B. and {Kraljic}, Katarina and {Kuncarayakti}, Hanindyo and {Kwon}, Yuna and {La Forgia}, Fiorangela and {Lahav}, Ofer and {Laigle}, Clotilde and {Lazzarin}, Monica and {Leaman}, Ryan and {Leclercq}, Floriane and {Lee}, Khee-Gan and {Lee}, David and {Lehnert}, Matt D. and {Lira}, Paulina and {Loffredo}, Eleonora and {Lucatello}, Sara and {Magrini}, Laura and {Maguire}, Kate and {Mahler}, Guillaume and {Zahra Majidi}, Fatemeh and {Malavasi}, Nicola and {Mannucci}, Filippo and {Marconi}, Marcella and {Martin}, Nicolas and {Marulli}, Federico and {Massari}, Davide and {Matsuno}, Tadafumi and {Mattheee}, Jorryt and {McGee}, Sean and {Merc}, Jaroslav and {Merle}, Thibault and {Miglio}, Andrea and {Migliorini}, Alessandra and {Minchev}, Ivan and {Minniti}, Dante and {Miret-Roig}, Nuria and {Monreal Ibero}, Ana and {Montano}, Federico and {Montet}, Ben T. and {Moresco}, Michele and {Moretti}, Chiara and {Moscardini}, Lauro and {Moya}, Andres and {Mueller}, Oliver and {Nanayakkara}, Themiya and {Nicholl}, Matt and {Nordlander}, Thomas and {Onori}, Francesca and {Padovani}, Marco and {Pala}, Anna Francesca and {Panda}, Swayamtrupta and {Pandey-Pommier}, Mamta and {Pasquini}, Luca and {Pawlak}, Michal and {Pessi}, Priscila J. and {Pisani}, Alice and {Popovic}, Lukav C. and {Prisinzano}, Loredana and {Raddi}, Roberto and {Rainer}, Monica and {Rebassa-Mansergas}, Alberto and {Richard}, Johan and {Rigault}, Mickael and {Rocher}, Antoine and {Romano}, Donatella and {Rosati}, Piero and {Sacco}, Germano and {Sanchez-Janssen}, Ruben and {Sander}, Andreas A.~C. and {Sanders}, Jason L. and {Sargent}, Mark and {Sarpa}, Elena and {Schimd}, Carlo and {Schipani}, Pietro and {Sefusatti}, Emiliano and {Smith}, Graham P. and {Spina}, Lorenzo and {Steinmetz}, Matthias and {Tacchella}, Sandro and {Tautvaisiene}, Grazina and {Theissen}, Christopher and {Thomas}, Guillaume and {Ting}, Yuan-Sen and {Travouillon}, Tony and {Tresse}, Laurence and {Trivedi}, Oem and {Tsantaki}, Maria and {Tsedrik}, Maria and {Urrutia}, Tanya and {Valenti}, Elena and {Van der Swaelmen}, Mathieu and {Van Eck}, Sophie and {Verdiani}, Francesco and {Verdier}, Aurelien and {Vergani}, Susanna Diana and {Verhamme}, Anne and {Vernet}, Joel and {Verza}, Giovanni and {Viel}, Matteo and {Vielzeuf}, Pauline and {Vietri}, Giustina and {Vink}, Jorick S. and {Viscasillas Vazquez}, Carlos and {Wang}, Hai-Feng and {Weilbacher}, Peter M. and {Wendt}, Martin and {Wright}, Nicholas and {Ye}, Quanzhi and {Yeche}, Christophe and {Yu}, Jiaxi and {Zafar}, Tayyaba and {Zibetti}, Stefano and {Ziegler}, Bodo and {Zinchenko}, Igor},
        title = "{The Wide-field Spectroscopic Telescope (WST) Science White Paper}",
      journal = {arXiv e-prints},
         year = 2024,
        month = mar,
          eid = {arXiv:2403.05398},
        pages = {arXiv:2403.05398},
          doi = {10.48550/arXiv.2403.05398},
archivePrefix = {arXiv},
       eprint = {2403.05398},
 primaryClass = {astro-ph.IM},
       adsurl = {https://ui.adsabs.harvard.edu/abs/2024arXiv240305398M}
}

@inproceedings{hill1988history,
  title={The history of multiobject fiber spectroscopy},
  author={Hill, John M},
  booktitle={Fiber Optics in Astronomy},
  volume={3},
  pages={77},
  year={1988}
}

@ARTICLE{2013MNRAS.428..447S,
       author = {{Sharp}, R. and {Brough}, S. and {Cannon}, R.~D.},
        title = "{Long-term stability of fibre-optic transmission for multi-object spectroscopy}",
      journal = {\mnras},
         year = 2013,
        month = jan,
       volume = {428},
       number = {1},
        pages = {447-458},
          doi = {10.1093/mnras/sts044},
archivePrefix = {arXiv},
       eprint = {1209.5131},
 primaryClass = {astro-ph.IM},
       adsurl = {https://ui.adsabs.harvard.edu/abs/2013MNRAS.428..447S}
}

@INPROCEEDINGS{2016SPIE.9908E..1SC,
       author = {{Colless}, Matthew},
        title = "{Cosmological surveys with multi-object spectrographs}",
    booktitle = {Ground-based and Airborne Instrumentation for Astronomy VI},
         year = 2016,
       editor = {{Evans}, Christopher J. and {Simard}, Luc and {Takami}, Hideki},
       series = {Society of Photo-Optical Instrumentation Engineers (SPIE) Conference Series},
       volume = {9908},
        month = aug,
          eid = {99081S},
        pages = {99081S},
          doi = {10.1117/12.2231829},
archivePrefix = {arXiv},
       eprint = {1608.04454},
 primaryClass = {astro-ph.IM},
       adsurl = {https://ui.adsabs.harvard.edu/abs/2016SPIE.9908E..1SC}
}

@ARTICLE{2017arXiv170101976E,
       author = {{Ellis}, Richard S. and {Bland-Hawthorn}, Joss and {Bremer}, Malcolm and {Brinchmann}, Jarle and {Guzzo}, Luigi and {Richard}, Johan and {Rix}, Hans-Walter and {Tolstoy}, Eline and {Watson}, Darach},
        title = "{The Future of Multi-Object Spectroscopy: a ESO Working Group Report}",
      journal = {arXiv e-prints},
         year = 2017,
        month = jan,
          eid = {arXiv:1701.01976},
        pages = {arXiv:1701.01976},
          doi = {10.48550/arXiv.1701.01976},
archivePrefix = {arXiv},
       eprint = {1701.01976},
 primaryClass = {astro-ph.IM},
       adsurl = {https://ui.adsabs.harvard.edu/abs/2017arXiv170101976E}
}

@article{massey2010astronomical,
  title={Astronomical spectroscopy},
  author={Massey, Philip and Hanson, Margaret M},
  journal={arXiv preprint arXiv:1010.5270},
  year={2010}
}

@INPROCEEDINGS{1988ASPC....3..125W,
       author = {{Watson}, F.~G.},
        title = "{Multi-object spectroscopy with FLAIR.}",
    booktitle = {Fiber Optics in Astronomy},
         year = 1988,
       editor = {{Barden}, Samuel C.},
       series = {Astronomical Society of the Pacific Conference Series},
       volume = {3},
        month = jan,
        pages = {125-132},
       adsurl = {https://ui.adsabs.harvard.edu/abs/1988ASPC....3..125W}
}

@article{lewis2002anglo,
  title={The Anglo-Australian Observatory 2dF facility},
  author={Lewis, Ian J and Cannon, Russell D and Taylor, Keith and Glazebrook, K and Bailey, JA and Baldry, IK and Barton, JR and Bridges, TJ and Dalton, GB and Farrell, TJ and others},
  journal={Monthly Notices of the Royal Astronomical Society},
  volume={333},
  number={2},
  pages={279--298},
  year={2002},
  publisher={The Royal Astronomical Society}
}

@inproceedings{saunders2004aaomega,
  title={AAOmega: a scientific and optical overview},
  author={Saunders, Will and Bridges, Terry and Gillingham, Peter and Haynes, Roger and Smith, Greg A and Whittard, John D and Churilov, Vladimir and Lankshear, Allan and Croom, Scott and Jones, Damien and others},
  booktitle={Ground-based Instrumentation for Astronomy},
  volume={5492},
  pages={389--400},
  year={2004},
  organization={SPIE}
}

@inproceedings{smith2004aaomega,
  title={AAOmega: a multipurpose fiber-fed spectrograph for the AAT},
  author={Smith, Greg A and Saunders, Will and Bridges, Terry and Churilov, Vladimir and Lankshear, Allan and Dawson, John and Correll, David and Waller, Lew and Haynes, Roger and Frost, Gabriella},
  booktitle={Ground-based Instrumentation for Astronomy},
  volume={5492},
  pages={410--420},
  year={2004},
  organization={SPIE}
}

@inproceedings{sharp2006performance,
  title={Performance of AAOmega: the AAT multi-purpose fiber-fed spectrograph},
  author={Sharp, Robert and Saunders, Will and Smith, Greg and Churilov, Vladimir and Correll, David and Dawson, John and Farrel, Tony and Frost, Gabriella and Haynes, Roger and Heald, Ron and others},
  booktitle={Ground-based and Airborne Instrumentation for Astronomy},
  volume={6269},
  pages={152--164},
  year={2006},
  organization={SPIE}
}

@article{pasquini2002installation,
  title={Installation and commissioning of FLAMES, the VLT Multifibre Facility},
  author={Pasquini, L and Avila, G and Blecha, A and Cacciari, C and Cayatte, V and Colless, M and Damiani, F and De Propris, R and Dekker, H and Di Marcantonio, P and others},
  journal={The Messenger (ISSN 0722-6691), No. 110, p. 1-9 (December 2002)},
  volume={110},
  pages={1--9},
  year={2002}
}

@article{kimura2010fibre,
  title={Fibre multi-object spectrograph (FMOS) for the Subaru telescope},
  author={Kimura, Masahiko and Maihara, Toshinori and Iwamuro, Fumihide and Akiyama, Masayuki and Tamura, Naoyuki and Dalton, Gavin B and Takato, Naruhisa and Tait, Philip and Ohta, Kouji and Eto, Shigeru and others},
  journal={Publications of the Astronomical Society of Japan},
  volume={62},
  number={5},
  pages={1135--1147},
  year={2010},
  publisher={Oxford University Press}
}

@article{szentgyorgyi2011hectochelle,
  title={Hectochelle: a multiobject optical echelle spectrograph for the MMT},
  author={Szentgyorgyi, Andrew and Furesz, Gabor and Cheimets, Peter and Conroy, Maureen and Eng, Roger and Fabricant, Daniel and Fata, Robert and Gauron, Thomas and Geary, John and McLeod, Brian and others},
  journal={Publications of the Astronomical Society of the Pacific},
  volume={123},
  number={908},
  pages={1188},
  year={2011},
  publisher={IOP Publishing}
}

@article{fabricant2005hectospec,
  title={Hectospec, the MMT’s 300 Optical Fiber-Fed Spectrograph},
  author={Fabricant, Daniel and Fata, Robert and Roll, John and Hertz, Edward and Caldwell, Nelson and Gauron, Thomas and Geary, John and McLeod, Brian and Szentgyorgyi, Andrew and Zajac, Joseph and others},
  journal={Publications of the Astronomical Society of the Pacific},
  volume={117},
  number={838},
  pages={1411},
  year={2005},
  publisher={IOP Publishing}
}

@INPROCEEDINGS{1995SPIE.2476....2B,
       author = {{Barden}, Samuel C.},
        title = "{Review of fiber optic properties for astronomical spectroscopy}",
    booktitle = {Fiber Optics in Astronomical Applications},
         year = 1995,
       editor = {{Barden}, Samuel C.},
       series = {Society of Photo-Optical Instrumentation Engineers (SPIE) Conference Series},
       volume = {2476},
        month = jun,
        pages = {2-9},
          doi = {10.1117/12.211837},
       adsurl = {https://ui.adsabs.harvard.edu/abs/1995SPIE.2476....2B}
}

@INPROCEEDINGS{2022SPIE12184E..6NB,
       author = {{Brzeski}, Jurek and {Adams}, Dave and {Baker}, Gabriella and {Baker}, Sufyan and {Brown}, Rebecca and {Case}, Scott and {Chin}, Timothy and {Coyne}, Jack and {Farrell}, Tony and {Gibson}, James and {Gillingham}, Peter and {Houston}, Ellen and {Klauser}, Urs and {Kripak}, Yevgen and {Kunwar}, Nirmala and {Lawrence}, Jon and {Mali}, Slavko and {Maslak}, Wojtek and {McGregor}, Helen and {Muller}, Rolf and {Nichani}, Vijay and {Pai}, Naveen and {O'brien}, Ellie and {Saunders}, Will and {Smedley}, Scott and {Venkatesan}, Sudharshan and {Waller}, Lew and {Zahoor}, Jahanzeb and {Zheng}, Jessica},
        title = "{AESOP, the 4MOST fibre positioner: engineering principles}",
    booktitle = {Ground-based and Airborne Instrumentation for Astronomy IX},
         year = 2022,
       editor = {{Evans}, Christopher J. and {Bryant}, Julia J. and {Motohara}, Kentaro},
       series = {Society of Photo-Optical Instrumentation Engineers (SPIE) Conference Series},
       volume = {12184},
        month = aug,
          eid = {121846N},
        pages = {121846N},
          doi = {10.1117/12.2627903},
       adsurl = {https://ui.adsabs.harvard.edu/abs/2022SPIE12184E..6NB}
}

@ARTICLE{2023ConPh..64...47P,
       author = {{Padovani}, Paolo and {Cirasuolo}, Michele},
        title = "{The Extremely Large Telescope}",
      journal = {Contemporary Physics},
         year = 2023,
        month = jan,
       volume = {64},
       number = {1},
        pages = {47-64},
          doi = {10.1080/00107514.2023.2266921},
archivePrefix = {arXiv},
       eprint = {2312.04299},
 primaryClass = {astro-ph.IM},
       adsurl = {https://ui.adsabs.harvard.edu/abs/2023ConPh..64...47P}
}

@INPROCEEDINGS{2003SPIE.4841..985G,
       author = {{Gillingham}, Peter R. and {Moore}, Anna M. and {Akiyama}, Masayuki and {Brzeski}, Jurek and {Correll}, David and {Dawson}, John and {Farrell}, Tony J. and {Frost}, Gabriella and {Griesbach}, Jason S. and {Haynes}, Roger and {Jones}, Damien and {Miziarski}, Stan and {Muller}, Rolf and {Smedley}, Scott and {Smith}, Greg and {Waller}, Lew G. and {Noakes}, Katie and {Arridge}, Chris},
        title = "{The Fiber Multi-object Spectrograph (FMOS) Project: the Anglo-Australian Observatory role}",
    booktitle = {Instrument Design and Performance for Optical/Infrared Ground-based Telescopes},
         year = 2003,
       editor = {{Iye}, Masanori and {Moorwood}, Alan F.~M.},
       series = {Society of Photo-Optical Instrumentation Engineers (SPIE) Conference Series},
       volume = {4841},
        month = mar,
        pages = {985-996},
          doi = {10.1117/12.462002},
       adsurl = {https://ui.adsabs.harvard.edu/abs/2003SPIE.4841..985G}
}

@INPROCEEDINGS{2004SPIE.5492.1228B,
       author = {{Brzeski}, Jurek K. and {Gillingham}, Peter and {Correll}, David and {Dawson}, John and {Moore}, Anna M. and {Muller}, Rolf and {Smedley}, Scott and {Smith}, Greg A.},
        title = "{Echidna: the engineering challenges}",
    booktitle = {Ground-based Instrumentation for Astronomy},
         year = 2004,
       editor = {{Moorwood}, Alan F.~M. and {Iye}, Masanori},
       series = {Society of Photo-Optical Instrumentation Engineers (SPIE) Conference Series},
       volume = {5492},
        month = sep,
        pages = {1228-1242},
          doi = {10.1117/12.550963},
       adsurl = {https://ui.adsabs.harvard.edu/abs/2004SPIE.5492.1228B}
}

@INPROCEEDINGS{2022SPIE12184E..6MB,
       author = {{Brzeski}, Jurek and {Adams}, David and {Baker}, Gabriella and {Baker}, Sufyan and {Brown}, Rebecca and {Case}, Scott and {Chin}, Timothy and {Coyne}, Jack and {Farrell}, Tony and {Gillingham}, Peter and {Houston}, Ellen and {Klauser}, Urs and {Kripak}, Yevgen and {Kunwar}, Nirmala and {Lawrence}, Jon and {Mali}, Slavko and {Maslak}, Wojtek and {McGregor}, Helen and {Muller}, Rolf and {Nichani}, Vijay and {Pai}, Naveen and {O'brien}, Ellie and {Saunders}, Will and {Smedley}, Scott and {Venkatesan}, Sudharshan and {Waller}, Lew and {Zahoor}, Jahanzeb and {Zheng}, Jessica},
        title = "{Overall performance of AESOP: the 4MOST fibre positioner}",
    booktitle = {Ground-based and Airborne Instrumentation for Astronomy IX},
         year = 2022,
       editor = {{Evans}, Christopher J. and {Bryant}, Julia J. and {Motohara}, Kentaro},
       series = {Society of Photo-Optical Instrumentation Engineers (SPIE) Conference Series},
       volume = {12184},
        month = aug,
          eid = {121846M},
        pages = {121846M},
          doi = {10.1117/12.2627900},
       adsurl = {https://ui.adsabs.harvard.edu/abs/2022SPIE12184E..6MB}
}

@ARTICLE{2015arXiv151100737G,
       author = {{Gilbert}, James and {Dalton}, Gavin},
        title = "{Learning from history: Adaptive calibration of 'tilting spine' fiber positioners}",
      journal = {arXiv e-prints},
         year = 2015,
        month = nov,
          eid = {arXiv:1511.00737},
        pages = {arXiv:1511.00737},
          doi = {10.48550/arXiv.1511.00737},
archivePrefix = {arXiv},
       eprint = {1511.00737},
 primaryClass = {astro-ph.IM},
       adsurl = {https://ui.adsabs.harvard.edu/abs/2015arXiv151100737G}
}

@ARTICLE{2012PhRvD..86j3518P,
       author = {{Parkinson}, David and {Riemer-S{\o}rensen}, Signe and {Blake}, Chris and {Poole}, Gregory B. and {Davis}, Tamara M. and {Brough}, Sarah and {Colless}, Matthew and {Contreras}, Carlos and {Couch}, Warrick and {Croom}, Scott and {Croton}, Darren and {Drinkwater}, Michael J. and {Forster}, Karl and {Gilbank}, David and {Gladders}, Mike and {Glazebrook}, Karl and {Jelliffe}, Ben and {Jurek}, Russell J. and {Li}, I. -hui and {Madore}, Barry and {Martin}, D. Christopher and {Pimbblet}, Kevin and {Pracy}, Michael and {Sharp}, Rob and {Wisnioski}, Emily and {Woods}, David and {Wyder}, Ted K. and {Yee}, H.~K.~C.},
        title = "{The WiggleZ Dark Energy Survey: Final data release and cosmological results}",
      journal = {\prd},
         year = 2012,
        month = nov,
       volume = {86},
       number = {10},
          eid = {103518},
        pages = {103518},
          doi = {10.1103/PhysRevD.86.103518},
archivePrefix = {arXiv},
       eprint = {1210.2130},
 primaryClass = {astro-ph.CO},
       adsurl = {https://ui.adsabs.harvard.edu/abs/2012PhRvD..86j3518P}
}

@ARTICLE{2021MNRAS.500..101Z,
       author = {{Zhang}, Feifan and {Wang}, Jianping and {Liu}, Zhigang and {Zhai}, Chao and {Chu}, Jiaru},
        title = "{Collision possibility analysis and collision avoidance for multi-object fibre-fed spectrographs with theta-phi positioners}",
      journal = {\mnras},
         year = 2021,
        month = jan,
       volume = {500},
       number = {1},
        pages = {101-108},
          doi = {10.1093/mnras/staa1944},
       adsurl = {https://ui.adsabs.harvard.edu/abs/2021MNRAS.500..101Z}
}
\bibliographystyle{aasjournalv7}

\end{document}